\documentclass[floats,floatfix,showpacs,amssymb,prd,twocolumn,superscriptaddress,nofootinbib,longbibliography,reprint,preprintnumbers]{revtex4-1}
\usepackage{amssymb,amsmath,verbatim,mathtools,needspace,enumitem,etoolbox,graphicx,physics,microtype,afterpage,bigints,gensymb,tabularx,array,booktabs}

\usepackage{multirow}
\usepackage[dvipsnames, usenames]{xcolor}
\definecolor{linkcolor}{rgb}{0.0,0.3,0.5}
\definecolor{dodgerblue}{HTML}{1E90FF}
\usepackage[unicode, colorlinks=true, linkcolor=linkcolor, citecolor=linkcolor, filecolor=linkcolor,urlcolor=linkcolor, pdfusetitle]{hyperref}
\usepackage[all]{hypcap}
\usepackage[T1]{fontenc}
\usepackage[utf8]{inputenc}
\usepackage{orcidlink}
\usepackage[normalem]{ulem}
\usepackage{soul}
\usepackage{comment}
\usepackage{graphicx}

\usepackage{lipsum}

\definecolor{romared}{RGB}{142,0,28}
\usepackage{hyperref}
\hypersetup{
    pdfnewwindow=true,      %
    colorlinks=true,       %
    linkcolor=romared,          %
    citecolor=romared,        %
    filecolor=romared,      %
    urlcolor=romared        %
}

\newcolumntype{Y}{>{\centering\arraybackslash}X}

\def\bi{\begin{itemize}[noitemsep,leftmargin=*]
\setlength\itemsep{1em}
        }
\def\ei{\end{itemize}}

\newcommand{\PS}{{\sc ParSpec }}

\newcommand{\beq}{\begin{eqnarray}}
\newcommand{\eeq}{\end{eqnarray}}

\newcommand{\tn}{\textnormal}

\makeatletter
\newcommand*{\balancecolsandclearpage}{\close@column@grid \cleardoublepage \twocolumngrid}
\makeatother

\newcommand{\jhu}{William H.\ Miller III Department of Physics and Astronomy, Johns Hopkins University, \\ 3400 N. Charles Street, Baltimore, Maryland, 21218, USA}
\newcommand{\GSSI}{Gran Sasso Science Institute (GSSI), I-67100 L’Aquila, Italy}
\newcommand{\GranSasso}{INFN, Laboratori Nazionali del Gran Sasso, I-67100 Assergi, Italy}
\newcommand{\sapienza}{Dipartimento di Fisica, ``Sapienza'' Universit\'a di Roma}
\newcommand{\INFNRoma}{Sezione INFN Roma 1, P. Aldo Moro 5, 00185 Roma, Italy}
\newcommand{\pisa}{Dipartimento di Fisica, Universit\'a di Pisa}
\newcommand{\INFNpisa}{Sezione INFN Pisa, L. Bruno Pontecorvo 3, 56127 Pisa, Italy}
\newcommand{\sissa}{SISSA, International School for Advanced Studies, via Bonomea 265, 34136 Trieste, Italy}
\newcommand{\trieste}{INFN, Sezione di Trieste, via Valerio 2, 34127 Trieste, Italy}
\newcommand{\ifpu}{IFPU, Institute for Fundamental Physics of the Universe, via Beirut 2, 34014 Trieste, Italy}

\begin{document}

\title{Black hole spectroscopy beyond Kerr:\\agnostic and theory-based tests with next-generation interferometers}

\author{Andrea Maselli}
\address{\GSSI}
\address{\GranSasso}

\author{Sophia Yi}
\address{\jhu}

\author{Lorenzo Pierini}
\address{\sapienza}
\address{\INFNRoma}

\author{Vania Vellucci}
\address{\sissa}
\address{\trieste}
\address{\ifpu}

\author{Luca Reali}
\address{\jhu}

\author{Leonardo Gualtieri}
\address{\pisa}
\address{\INFNpisa}

\author{Emanuele Berti}
\address{\jhu}

\date{\today}

\begin{abstract}
  Black hole spectroscopy is a clean and powerful tool to test gravity in the strong-field regime and to probe the nature of compact objects. Next-generation ground-based detectors, such as the Einstein Telescope and Cosmic Explorer, will observe thousands of binary black hole mergers with large signal-to-noise ratios, allowing for accurate measurements of the remnant black hole quasinormal mode frequencies and damping times.  In previous work we developed an observable-based parametrization of the quasinormal mode spectrum of spinning black holes beyond general relativity ({\sc ParSpec}). In this paper we use this parametrization to ask: can next-generation detectors detect or constrain deviations from the Kerr spectrum by stacking multiple observations of binary mergers from astrophysically motivated populations? We focus on two families of tests: (i) \textit{agnostic} (null) tests, and (ii) \textit{theory-based} tests, which make use of quasinormal frequency calculations in specific modified theories of gravity. We consider in particular two quadratic gravity theories (Einstein-scalar-Gauss-Bonnet and dynamical Chern-Simons gravity) and various effective field theory-based extensions of general relativity. We find that robust inference of hypothetical corrections to general relativity requires pushing the slow-rotation expansion to high orders. Even when high-order expansions are available, ringdown observations alone may not be sufficient to measure deviations from the Kerr spectrum for theories with dimensionful coupling constants. This is because the constraints are dominated by ``light'' black hole remnants, and only few of them have sufficiently high signal-to-noise ratio in the ringdown. Black hole spectroscopy with next-generation detectors may be able to set tight constraints on theories with dimensionless coupling, as long as we assume prior knowledge of the mass and spin of the remnant black hole.
\end{abstract}

\preprint{ET-0305A-23}

\maketitle

\tableofcontents

\maketitle

\section{Introduction}
\begin{figure}[t]%
    \includegraphics[scale = 0.45]{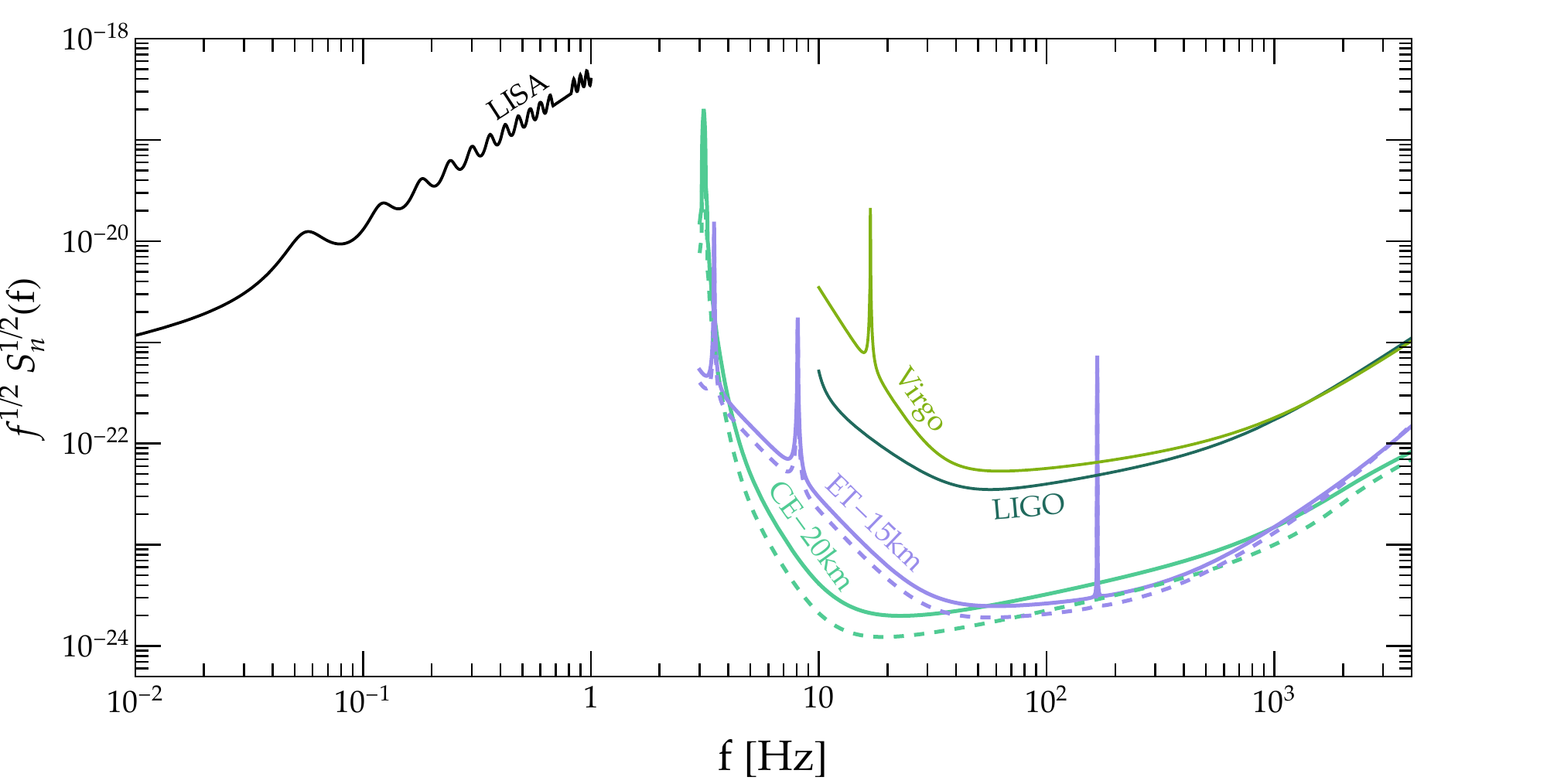}
    \caption{Characteristic strain for ET and CE, compared to LIGO and Virgo at design sensitivity, as well as to the space-based interferometer LISA. For ET (CE), solid and dashed curves correspond to detector configurations with $15$\,km  ($20$\,km) and $20$\,km ($40$\,km) armlength, respectively. 
    In this work we assume a $40$\,km CE~\cite{Evans:2023euw}, 
    and a network of two L-shaped aligned 
    $15$\,km ET~\cite{Branchesi:2023mws}.}
    \label{fig:psd}
\end{figure}

The LIGO-Virgo-KAGRA gravitational-wave (GW) detector network has already observed about $100$ events produced by merging compact binaries~\cite{LIGOScientific:2021djp}.
This number is expected to grow to several hundreds during the ongoing fourth observing run (O4), and by orders of magnitude when next-generation (XG) detectors, such as the Einstein Telescope (ET) in Europe~\cite{Punturo:2010zz} and Cosmic Explorer (CE) in the US~\citep{2015PhRvD..91h2001D}, will start taking data.
The improved noise power spectral density (PSD) and observing volume of these detectors (see Fig.~\ref{fig:psd}) will open the door to new and exciting science, ranging from studies of the astrophysical formation scenarios of compact object binaries to precise tests of general relativity (GR) and cosmology~\cite{Maggiore:2019uih,Perkins:2020tra,Branchesi:2023mws,Borhanian:2022czq,Evans:2023euw,Gupta:2023lga}. 

A promising strategy to test gravity in the strong-field regime and to check if the remnants of these mergers really are the Kerr black holes (BHs) of general relativity is based on ``black hole spectroscopy.'' The idea is that the entire quasinormal mode (QNM) frequency spectrum of Kerr BHs is uniquely determined by their mass $M$ and dimensionless spin $\chi$~\cite{Kokkotas:1999bd,Ferrari:2007dd,Berti:2009kk}. The measurement of the frequency and damping time of a single QNM yields the BH mass and spin. At least in principle, the measurement of three or more quantities (either frequencies or damping times) in the so-called ``ringdown'' signal following a binary BH merger can be a test for consistency with the predictions of GR~\cite{Detweiler:1980gk,Dreyer:2003bv,Berti:2005ys}.
In practice, precision BH spectroscopy will probably require the detection of several ringdown events with the large signal-to-noise ratios (SNRs) expected from XG detectors, or from the space-based Laser Interferometer Space Antenna (LISA)~\cite{Gossan:2011ha,Meidam:2014jpa,Berti:2016lat,Yang:2017zxs,Maselli:2017kvl,Baibhav:2017jhs,Baibhav:2018rfk,Baibhav:2020tma,Ota:2021ypb,Bhagwat:2021kwv,Baibhav:2023clw}.

Tests of modified gravity theories can be either ``theory-agnostic'' or ''theory-specific.''

Agnostic tests are similar in spirit to the Will-Nordtvedt ``parametrized-post-Newtonian'' (PPN) formalism~\cite{Will:1972zz}: the deviations of observable quantities from the GR predictions are described through a set of free parameters that can either be constrained from the data, or measured if a deviation is found.
Theory-agnostic parametrizations are useful as long as the parameters can be mapped to the predictions of specific beyond-GR theories. In the PPN expansion, the parametrization involves the spacetime metric~\cite{Will:2014kxa}. In the so-called ``parametrized post-Einsteinian'' (PPE) framework, the parametrization involves the amplitude and phase of the (inspiral) gravitational waveform~\cite{Blanchet:1994ez,Arun:2006hn,Arun:2006yw,Yunes:2009ke,Chatziioannou:2012rf,Chamberlain:2017fjl,Perkins:2020tra}: see e.g.~\cite{Yunes:2013dva,Berti:2018cxi} for reviews.

In the theory-specific approach, the goal is to compute the observable quantities (in our context, for example, the QNM frequencies) within each modified gravity theory.
This is not always possible or practical. When it is, one can place bounds (or measure) the parameters of the theory by direct comparison with the data. 

There have been several recent attempts at parametrizing the ringdown 
waveform (see e.g.~\cite{Tattersall:2017erk,Berti:2018vdi,Cardoso:2019mqo,McManus:2019ulj,Volkel:2022aca,Volkel:2022khh,Franchini:2022axs}).
The special nature of the Kerr metric in GR implies that the perturbation equations are ``miraculously'' separable~\cite{Teukolsky:2014vca}, leading to decoupled ordinary differential equations for the radial and angular dependence of the perturbations (the ``Teukolsky equations''~\cite{Teukolsky:1973ha}). Separability is usually lost in beyond-GR theories, and the calculation of QNM frequencies usually proceeds in one of two ways. The most common procedure is to first look for spherically symmetric BH solution, and then consider rotational corrections to the background metric and to the perturbation equations by working in a small-spin expansion. This approach is fairly common in GR~\cite{Pani:2013pma,Franchini:2023xhd}, and it has recently been extended to theories such as Einstein-scalar-Gauss-Bonnet (EsGB) gravity~\cite{Pierini:2021jxd,Pierini:2022eim}, dynamical Chern-Simons (dCS) gravity~\cite{Wagle:2021tam,Srivastava:2021imr,Wagle:2023fwl}, and certain classes of effective field theory (EFT) modifications of GR~\cite{Cano:2020cao,Cano:2021myl}.
Other approaches try to bypass the small-spin expansion using either numerical or analytical methods. For example, one can use spectral methods to find the QNM spectrum by solving the (nonseparable) partial differential equations that describe perturbations of the Kerr-Newman metric in GR~\cite{Dias:2015wqa,Dias:2021yju,Dias:2022oqm,Davey:2023fin} (see also~\cite{Chung:2023zdq}). More recently, various groups have developed perturbation schemes leading to ``modified Teukolsky equations'' in modified theories of gravity~\cite{Li:2022pcy,Hussain:2022ins,Cano:2023tmv,Li:2023ulk,Dey:2022pmv,Ghosh:2023etd}. These schemes have been applied to EFT modifications of GR in Ref.~\cite{Cano:2023jbk}; in this case the small-spin expansion is performed on the perturbations, but not on the background.

The \PS parametrization of Ref.~\cite{Maselli:2019mjd} is motivated, at least in part, by the small-spin approach to the problem of computing the QNM spectrum of BHs in modified gravity theories, but both approaches have found application in GW data analysis (see e.g.~\cite{Carullo:2021dui} for an application of the \PS formalism to GW data, and~\cite{Carullo:2021oxn} for constraints on the Kerr-Newman QNMs computed via spectral methods).

The \PS formalism also relies on the assumption that modified gravity corrections can be treated perturbatively: while one can contemplate the possibility of nonperturbative corrections that can dramatically alter the spectrum~\cite{Cardoso:2019rvt}, there are two strong arguments for working within a perturbative framework. The first argument is theoretical: most modified theories of gravity only make sense as EFTs, and a perturbative expansion around GR usually determines their domain of validity (see e.g. Refs.~\cite{Julie:2022huo,Hegade:2022hlf} and~\cite{Alexander:2009tp} for discussions in the context of EsGB and dCS gravity, respectively). The second argument is experimental: GR is so well constrained that the perturbative approach used in this paper is general enough for most applications (see e.g.~\cite{Berti:2015itd}). 

In this paper we focus on the observable-based \PS parametrization of the QNM spectrum developed in Ref.~\cite{Maselli:2019mjd}.
While the parametrization is theory-agnostic, specific theories can be mapped to a given subset of the \PS parameters.
In fact, in this paper we will discuss the implications of a {\sc ParSpec}-based analysis of XG data on the three specific classes of theories listed above: EsGB, dCS, and various EFT theories.
Our goal is to assess whether theory-agnostic and theory-specific BH spectroscopy tests are feasible with XG detectors.

The plan of the paper is as follows.
In Sec.~\ref{sec:ParSpec} we briefly review the \PS parametrization, our data analysis framework, and the population models used in our forecast.
In Sec.~\ref{sec:approaches} we describe our implementation of theory-agnostic and theory-specific tests.
In Sec.~\ref{sec:results} we present the results of our analysis, and in Sec.~\ref{sec:conclusions} we summarize our main conclusions.
To improve readability, we relegate to the appendices some of the most technical aspects, including the mapping between the \PS expansion and the QNM frequencies in specific theories (Appendix~\ref{app:qnmtheories});
the remnant mass and spin accuracy achievable with ringdown observations (Appendix~\ref{sec:appmasses});
differences in ringdown SNR between ET and CE (Appendix~\ref{sec:snrconf});
and a list of the Fisher matrix components used in our parameter estimation calculations (Appendix~\ref{sec:fishercomp}).
Throughout the paper we use geometrical units $(G=c=1)$.

\section{The parametrized spectroscopy pipeline}
\label{sec:ParSpec}

\subsection{ParSpec}\label{sec:parspec}
The key idea behind \PS is to express the QNM frequencies and damping times of a set of $N$ observed ringdown events in terms of a two-parameter Taylor expansion, such that the $(\ell,m)$ QNM frequency $\omega_i^{\ell m}$ and damping time $\tau_i^{\ell m}$ of the $i$-th source $(i=1,\dots,N)$ read: 
\begin{align}
&M_i \omega_i^{\ell m} = \sum_{k_1=0}^{n_1} \chi_i^{k_1} \bar{\omega}_{\ell m}^{(k_1)} +\sum_{k_2=0}^{n_2}\chi_i^{k_2} \bar{\omega}_{\ell m}^{(k_2)}\gamma_i \delta  \omega_{\ell m}^{(k_2)} \ ,\label{eq:parspec-expa1}\\
& \tau_i^{\ell m}/M_i = \sum_{k_1=0}^{n_1} \chi_i^{k_1} \bar{\tau}_{\ell m}^{(k_1)} +\sum_{k_2=0}^{n_2} \chi_i^{k_2} \bar{\tau}_{\ell m}^{(k_2)}\gamma_i \delta  \tau_{\ell m}^{(k_2)} \label{eq:parspec-expa2} \, ,
\end{align}
where $M_i$ and $\chi_i$ are the remnant BH's detector-frame mass and dimensionless spin, and $\bar{\omega}_{\ell m}^{(k_1)},\bar{\tau}_{\ell m}^{(k_1)}$ are the $k_1$-order dimensionless coefficients in a small-spin ($\chi_i \ll 1$) expansion of the mode's frequency and damping time for a Kerr BH.
Beyond-Kerr corrections at order $k_2$ in the spin expansion are specified by the \PS dimensionless parameters $\delta \omega_{\ell m}^{(k_2)}$ and $\delta \tau_{\ell m}^{(k_2)}$. Given some ``fundamental'' modification of GR, these deviations are \textit{universal}, i.e., they do not depend on the source.

We assume that the deviation from GR is characterized by a fundamental coupling constant $\alpha$, with dimensions $[\alpha]=({\rm mass})^p$ ($p\ge0$), which is considered as a small expansion parameter. The dimensionless coupling constants $\gamma_i$ are linear in the coupling. They can depend on the source $i$ (e.g., they can scale with some power of the remnant BH's mass in theories of gravity where the action is quadratic or of higher order in the curvature), but not on the specific QNM. At leading order in our perturbative scheme, they are given by
\begin{equation}
    \gamma_i=\frac{\alpha}{M_i^p}(1+z_i)^p\,,
    \label{eq:parspeccoupl}
\end{equation}
where $M_i$ is the ADM mass of the BH in the detector frame, and $z_i$ is the redshift of the source. We will focus on three representative cases: $p=0$, $p=4$ and $p=6$.

Within this expansion, GR is recovered (by definition) in the limit $\gamma_i \rightarrow 0$. 
Assuming GR corrections to be perturbative  
also requires that 
$\vert\gamma_i \delta \omega_{\ell m}^{(k_2)}\vert 
\lesssim 1$ and $\vert\gamma_i \delta \tau_{\ell m}^{(k_2)} \vert\lesssim 1$. 
Note that the order of the spin expansion of the QNM frequencies in GR, denoted by $n_1$ in Eqs.~\eqref{eq:parspec-expa1} and \eqref{eq:parspec-expa2}, can be different from the order of the spin expansion that allows for deviations from GR, $n_2$.

Consider $N$ ringdown events in which we extract from the signal the frequencies and damping times of $q$ QNMs for each source. This yields $\mathcal{O}=2N\times q$  observables that can be used, at least in principle, to determine the beyond-GR parameters of the expansion in Eqs.~\eqref{eq:parspec-expa1}--\eqref{eq:parspec-expa2}, i.e., the quantities $\vec\zeta=(\delta\omega^{(k_2)}_{\ell m},\delta\tau_{\ell m}^{(k_2)} )$. 
As discussed in Ref.~\cite{Maselli:2019mjd}, there are two possible strategies to determine these quantities: (i) we assume that the (unperturbed) masses and spins of the sources, $(M_i,\chi_i)_{i=1\ldots N}$, are known {\it a priori}; or (ii) we extract these quantities from the ringdown signal by considering them as part of the parameter set to be determined, i.e., $\vec\zeta=(\delta\omega^{(k_2)}_{\ell m},\delta\tau_{\ell m}^{(k_2)},M_i,\chi_i)$.

Of course, the first strategy is preferable whenever possible, since it generally leads to smaller errors. This approach requires {\it a priori} knowledge of the masses and spins of the remnant. In GR, these quantities be obtained (for example) by using the parameters of the binary progenitors and semianalytical fits of numerical relativity simulations~\cite{Healy:2014yta}. Going beyond GR, if we are interested in tests of a specific gravity theory, numerical relativity simulations in that theory would be necessary in order to derive similar fits. For tests not based on specific theories  -- hereafter, ``agnostic'' tests -- and as long as the deviations from GR are small, we can assume $(M_i,\chi_i)_{i=1\ldots N}$ to be the GR masses and spins, obtained for example from the inspiral signal through the analytical fits derived within GR~\cite{Healy:2014yta}. In this way, the shifts between the ``true'' masses and spins of the remnant and those obtained from the GR fits are absorbed in the unknown parameters $\delta\omega^{(k_2)}_{\ell m}$ (see the discussion in Ref.~\cite{Maselli:2019mjd}).
In summary: when performing theory-specific tests in a theory for which numerical relativity simulations are not available, we are forced to follow the second strategy, but the larger number of parameters is expected to lead to larger errors.

If we know in advance the masses and spins of the sources (strategy (i) above), we can use the $\mathcal{O}=2N \times q$ observables to constrain $2(n_2+1)q$ parameters, therefore we must have $N \geq n_2+1$.  If, instead, we use only the ringdown signal (strategy (ii) above), there are now $2(n_2+1)q+2N$ free parameters, and thus the minimum number of sources is $N > q(1+n_2)/(q-1)$. Constraining the GR modifications requires in this case at least two modes, i.e., $q>1$, since the first mode is used to determine the mass and spin.  In this paper we will investigate both of these methodologies, extending the analysis of Ref.~\cite{Maselli:2019mjd}.

To infer the beyond-Kerr corrections we adopt a hierarchical inference scheme. For a given set of $N$ astrophysical sources we compute the corresponding QNM frequencies and damping times, and assume they are observed by a next-generation (XG) detector, which can be either ET or CE.
We compute the joint posterior probability distribution associated with the mode measurements using a Fisher Information Matrix (FIM) approach.  

The observed values of $(\omega^{\ell m},\tau^{\ell m})_{i=1\ldots N}$ are interpreted in terms of the \PS template by sampling the $\vec{\zeta}$ parameters of Eqs.~\eqref{eq:parspec-expa1}--\eqref{eq:parspec-expa2} through Markov Chain Monte Carlo (MCMC) simulations.
Below we give more details about each step of the analysis.

\subsection{Quasinormal mode templates}

For a given multipolar component of the radiation with angular indices $(\ell m)$, the ``plus'' and ``cross'' polarizations of the ringdown waveform are given by
\begin{align}
h_+(t)=\frac{M{\cal A}_{\ell m}Y^{\ell m}_{+}}{r}\tn{Re}\left[e^{-t/\tau_{\ell m}+i(\omega_{\ell m}t+\phi_{\ell m}-m\varphi)}\right]\ ,\label{math:hp}\\
h_\times(t)=\frac{M{\cal A}_{\ell m}Y^{\ell m}_{\times}}{r}\tn{Im}\left[e^{-t/\tau_{\ell m}+i(\omega_{\ell m}t+\phi_{\ell m}-m\varphi)}\right]\ ,\label{math:hc}
\end{align}
where the amplitude  ${\cal A}_{\ell m}$ and the phase $\phi_{\ell m}$ are real quantities, we have defined
\begin{align}
Y^{\ell m}_{+}(\iota)=_{-2}Y^{\ell m}(\iota,0)+(-1)^\ell _{-2}Y^{\ell -m}(\iota,0)\ ,\\
Y^{\ell m}_{\times}(\iota)=_{-2}Y^{\ell m}(\iota,0)-(-1)^\ell _{-2}Y^{\ell -m}(\iota,0)\ ,
\end{align}
$\iota$ is the inclination angle of the source with respect to the detector, and $_{-2}Y^{\ell m}(\iota,\varphi)$ are the spin-weighted spherical harmonics of spin-weight $-2$~\cite{Baibhav:2018rfk}.
To compute the signal in the frequency domain we follow the Flanagan-Hughes convention, i.e., we assume that the waveform for $t<0$ is found by reflection through the origin of the waveform for $t>0$, and then we divide the amplitude by a factor $\sqrt{2}$ to compensate for the doubling~\cite{Flanagan:1997sx,Berti:2005ys}. Under this approximation the Fourier transforms of Eqs.~\eqref{math:hp} and \eqref{math:hc} can be computed analytically, and they read:
\begin{align}
\tilde{h}_+(f)&=\frac{M{\cal A}_{\ell m}}{\sqrt{2}r}\left[\hat{Y}^{\ell m}_{+}e^{i\phi_{\ell m}}b_+
+\hat{Y}^{\ell m\star}_{+} e^{-i\phi_{lm}}b_-\right]\,,\label{math:hpf}\\
\tilde{h}_\times(f)&=\frac{M{\cal A}_{\ell m}}{i\sqrt{2}r}\left[\hat{Y}^{\ell m}_{{\times}}e^{i\phi_{\ell m}}b_+
-\hat{Y}^{\ell m\star}_{{\times}}e^{-i\phi_{\ell m}}b_-\right]\,,\label{math:hcf}
\end{align}
where $\hat{Y}^{\ell m}_{+,\times}=Y^{\ell m}_{+,\times}e^{-im\varphi}$ and $b_\pm$ are the Breit-Wigner functions,
\begin{equation}\label{BreitWigner}
b_\pm=\frac{1/\tau_{lm}}{\tau_{\ell m}^{-2}+(\omega \pm \omega_{\ell m})^2}\,.
\end{equation}
The full waveform is then given by 
\begin{equation}
\tilde{h}(f)=F_{+}\tilde{h}_+(f)+F_{\times}\tilde{h}_\times(f)\ ,\label{math:totalh}
\end{equation}
where $F_{+,\times}$ are the pattern functions of the detector, which depend on the source position in the sky
and on the polarization angle~\cite{Sathyaprakash:2009xs}.

Given the Fourier-domain amplitude of Eq.~\eqref{math:totalh}, we can compute the signal-to-noise ratio (SNR) $\rho$ for a GW interferometer with noise power spectral density $S_n(f)$ from the relation
\begin{equation}
\rho^2=4\int_{f_\tn{min}}^{f_\tn{max}} \frac{\tilde{h}(f)\tilde{h}^\star(f)}{S_n(f)}df\ .\label{math:snr}
\end{equation}
Hereafter we average over the detector and BH angular variables. Making use of the identities
\begin{align}
\langle F_{+,\times}^2\rangle=\frac{1}{5}\ ,\ \langle F_+F_\times\rangle=0\ ,\ 
\langle (Y_{\ell m}^+)^2+(Y_{\ell m}^\times)^2\rangle=\frac{1}{\pi}\ ,\nonumber
\end{align}
from Eq.~\eqref{math:snr} we find
\begin{equation}
\rho^2=\frac{2}{5\pi}\frac{M^2{\cal A}_{\ell m}^2}{r^2}\int_{f_\tn{min}}^{f_\tn{max}}\frac{b_+^2+b_-^2}{S_n(f)}df\ .\label{math:snr2}
\end{equation}
The SNR can be expressed in terms of the GW energy spectrum $dE/df$,
\begin{equation}
\rho^2=\frac{2}{5\pi^2 r^2}\int_{f_\tn{min}}^{f_\tn{max}}\frac{1}{f^2S_n(f)}\frac{dE}{df}df\ ,\label{math:snr3}
\end{equation}
which in turn is related to the radiation efficiency
\begin{equation}
\epsilon_\tn{RD}=\frac{1}{M}\int_{f_\tn{min}}^{f_\tn{max}} \frac{dE}{df} df\ ,\label{math:erd}
\end{equation}
a measure of the amount of energy radiated by the $(\ell,m)$ mode.
From Eqs.~\eqref{math:snr2} and \eqref{math:erd} we find 
\begin{equation}
\epsilon_\tn{RD}=\pi M{\cal A}^2_{\ell m}\int_{f_\tn{min}}^{f_\tn{max}} df f ^2(b_+^2+b_-^2) \ .\label{math:erd2}
\end{equation}
For a given choice of the source parameters, $(M,r,\omega_{\ell m},\tau_{\ell m},\phi_{\ell m})$, and of the efficiency $\epsilon_\tn{RD}$, we can numerically solve Eq.~\eqref{math:erd2} to find the signal amplitude ${\cal A}_{\ell m}$.
We compute $\epsilon_\tn{RD}$ for a given $(\ell, m)$ multipolar component from the semi-analytic formula of Ref.~\cite{Baibhav:2017jhs}:
\begin{equation}
\epsilon_\tn{RD}=\left[a_{\ell m}+b_{\ell m} \chi_+ +c_{\ell m} \chi_-\right]^2\ ,\label{math:epsilonfit}
\end{equation}
where $a_{\ell m},b_{\ell m}$ and $c_{\ell m}$ are functions of the binary mass ratio $q=m_1/m_2\ge1$ and of two spin parameters,
\begin{equation}
\chi_\pm=\frac{m_1\chi_1\pm m_2\chi_2}{m_1+m_2}\ ,
\end{equation}
which in turn depend on the masses $m_{1,2}$ and on the dimensionless spins $\chi_{1,2}$ of the progenitors.

\subsection{Parameter estimation of quasinormal mode frequencies}\label{sec:qnmpe}

To estimate the measurement accuracy of QNM frequencies and damping times by XG detectors, we use the waveform template of Eq.~\eqref{math:totalh} and a FIM analysis~\cite{Vallisneri:2007ev}.

The interferometer output $s(t)=h(t,\vec{\theta})+n(t)$ is the sum of the detector's (stationary) noise $n(t)$ and of the actual signal. The probability distribution of the source parameters $\vec{\theta}$, up to a normalization constant, is
\begin{equation}
p(\vec{\theta}\vert s)\propto p^{(0)}(\vec{\theta})e^{-\frac{1}{2}(h(\vec{\theta})
-s\vert h(\vec{\theta})-s)}\ ,\label{math:prob}
\end{equation}
where $p^{(0)}(\vec{\theta})$ is the prior on the source parameters~\cite{Cutler:1994ys}, and we have introduced the inner product between two waveforms
\begin{equation}
(h_1\vert h_2)=2\int_{f_\tn{min}}^{f_\tn{max}}\frac{\tilde{h}_1(f) \tilde{h}_2^{\star}(f)+
\tilde{h}_1^{\star}(f) \tilde{h}_2(f)}{S_{n}(f)}df\ .\label{math:inner}
\end{equation} 
For GW signals with large SNRs, we expect $p(\vec{\theta}\vert s)$ to be peaked around the \textit{true} values $\vec{\xi}$. Therefore, we can expand Eq.~\eqref{math:prob} as a Taylor series up to second order in $\Delta \vec{\theta}=\vec{\theta} -\vec{\xi}$, such that 
\begin{equation}\label{prob1}
p( \vec{\theta}\vert s)\propto p^{(0)}(\vec{\theta})e^{-\frac{1}{2}\Gamma_{ij}\Delta \theta^{i}\Delta \theta^{j}}\ ,
\end{equation}
where 
\begin{equation}
\Gamma_{ij}=\left(\frac{\partial h}{\partial \theta^i}\bigg\vert \frac{\partial h}{\partial \theta^j}\right) \label{math:FisherM}
\end{equation}
is the FIM, computed at $\vec{\theta}=\vec{\xi}$. 
The covariance matrix $\Sigma^{ij}$ of the waveform parameters is then given by 
\begin{equation}
\Sigma^{ij}=\left(\Gamma^{-1}\right)^{ij}\ ,\label{math:cov}
\end{equation}
where $\Gamma^{-1}$ is the inverse of the FIM.
In the single-mode analysis, the waveform $\tilde{h}(f)$ is completely specified by four parameters: $\vec{\theta}=(\bar{{\cal A}}_{\ell m},\phi_{\ell m},\omega_{\ell m},\tau_{\ell m})$, where we have defined the effective amplitude $\bar{{\cal A}}_{\ell m}=M{\cal A}_{\ell m}/r$. 
Herefter, following the conventions of Ref.~\cite{Berti:2005ys}, we fix $\phi_{\ell m}=0$. The components of $\Gamma_{ij}$ are listed in Appendix~\ref{sec:fishercomp}.
 
For the two-mode study, we replace the polarizations~\eqref{math:hp}--\eqref{math:hc} with a sum over two multipolar components of the form~\cite{Berti:2005ys}:
\begin{align}
h_+(t)=\sum_{\ell m}\frac{M{\cal A}_{\ell m}Y^{\ell m}_{+}}{r}\tn{Re}\left[e^{-\frac{t}{\tau_{\ell m}}+i(\omega_{\ell m}t+\phi_{\ell m})}\right]\ ,\\
h_\times(t)=\sum_{\ell m}\frac{M{\cal A}_{\ell m}Y^{\ell m}_{\times}}{r}\tn{Im}\left[e^{-\frac{t}{\tau_{\ell m}}+i(\omega_{\ell m}t+\phi_{\ell m})}\right]\ ,
\end{align}

Here we consider two scenarios in which, along with the fundamental $\ell=m=2$ mode, we include either the $(\ell, m)=(21)$ or the $(\ell, m)=(33)$ component (note that we do not include overtones in our analysis~\cite{Baibhav:2023clw,Nee:2023osy}).
In this two-mode model, the Fisher matrix~\eqref{math:FisherM} becomes an $8\times8$ matrix in the variables $\vec{\theta}=(\bar{{\cal A}}_{\ell_1 m_1},\phi_{\ell_1 m_1},\omega_{\ell_1 m_1},\tau_{\ell_1 m_1},
\bar{{\cal A}}_{\ell_2 m_2},\phi_{\ell_2 m_2},\omega_{\ell_2 m_2},\tau_{\ell_2 m_2})$. 
Since the two modes have different harmonic indices $(\ell,\,m)$, by angular averaging and using the orthogonality of the spherical harmonics we can decouple the two modes. This property simplifies our calculations, since the FIM is block-diagonal and can be written as a combination of the matrices $\Gamma^{(1,2)}_{ij}$ corresponding to each mode:
\begin{equation}
\Gamma_{ij}=
\begin{pmatrix}
\Gamma^{(1)}_{ij} & 0 \\
0 & \Gamma^{(2)}_{ij}
\end{pmatrix}\,.\label{Gamma_bd}
\end{equation} 
Similarly, the total SNR is given by the sum in quadrature of the two modes $\rho=\sqrt{\rho^2_{(1)}+\rho^2_{(2)}}$. 
These considerations are easily generalized to $q$ modes with different values of $(\ell,\,m)$, as long as we do not consider overtones: in this case, the FIM is simply the block-diagonal combination of $q$ matrices corresponding to the different QNMs.

In the evaluation of SNRs and Fisher matrix components, we fix the integration limits to $f_\tn{min}=3$~Hz and $f_\tn{max}=5$~kHz.
For ET we consider a network of two L-shaped aligned detectors with $15$\,km armlength~\cite{Branchesi:2023mws}, while for CE we focus on single $40$\,km interferometer~\cite{Evans:2023euw}.

\subsection{Sampling the beyond-Kerr parameters}\label{sec:beyondGRan}

From the covariance matrices for $\omega_{\ell m}$ and $\tau_{\ell m}$ we can infer the probability distribution of the beyond-Kerr parameters using a Bayesian approach based on MCMC simulations. 
In the most general case, given the set of parameters
\begin{equation}
\vec{\zeta}=(\delta\omega_{\ell_j m_j}^{(k)}, \delta\tau_{\ell_j m_j}^{(k)},M_i,\chi_i)\qquad
{\small
\begin{cases}
i=1,\ldots,N\\
j=1,\ldots,q\\
k=1,\ldots,n_2
\end{cases}}\ ,
\end{equation} 
for a set $\vec{d}$ of $N$ ringdown observations consisting of $2N\times q$ frequencies and damping times, the posterior probability distribution of $\vec{\zeta}$ is given by: 
\begin{equation}
p(\vec{\zeta}\vert\vec{d}) \propto \mathcal{L}(\bar{d}| \vec{\zeta})p_0(\vec{\zeta})\ ,
\end{equation}
where $\mathcal{L}(\vec{d}| \vec{\zeta})$ is the likelihood, which we 
choose to be a Gaussian  for each event:
\begin{equation}
\mathcal{L}(\vec{d}| \vec{\zeta})= \mathcal{N}(\vec{\mu_i},\Sigma_i)\ .
\label{eq:likel}
\end{equation}
The vector $ \vec{\mu_i}$ is defined as
\begin{equation}
\vec{\mu_i}=(\vec{\mu_i}^{(1)},...,\vec{\mu_i}^{(q)})^T\ .
\end{equation}
Each $\vec{\mu_i}^{(j)}$ is a two-component vector that depends on the difference between the observed mode frequencies ($j=1,...,q$) and the parametrized templates in Eqs.~\eqref{eq:parspec-expa1} and \eqref{eq:parspec-expa2}:
\begin{align}
\vec{\mu}_i^{(j)}= 
\begin{bmatrix}
\omega_i^{\ell_j m_j}-\omega_{i,\tn{obs}}^{\ell_j m_j}\\
\tau_i^{\ell_j m_j}-\tau_{i,\tn{obs}}^{\ell_j m_j}\\
\end{bmatrix}
\label{eq:def-mu-parspec} \, .
\end{align}
In Eq.~\eqref{eq:likel} $\Sigma_i$ denotes the covariance matrix \eqref{math:cov}, which includes errors on (and correlations between) the frequencies and damping times of the $i$-th source.
For each value of $i$, $\Sigma_i$ is a block-diagonal $(4q\times 4q)$ square matrix.
Because the covariance matrix is block-diagonal, the likelihood function for a $q$-mode analysis (neglecting the overtones) can be recast as a product of Gaussian distributions:
\begin{equation}
\mathcal{N}( \vec{\mu_i}, \Sigma_i)= \prod_{j=1}^q \mathcal{N}( \mu_i^{(j)}, \Sigma_i^{(j)})\ .
\end{equation}
Finally, since the $N$ ringdown observations are all independent, the combined likelihood function of the  parameters can be  
factorized as
\begin{equation}
\mathcal{L}(\vec{d} | \vec{\zeta})= \prod_{i=1}^N \mathcal{L}_i(\vec{d} | \vec{\zeta})= \prod_{i=1}^N\prod_{j=1}^q \mathcal{N}( \mu_i^{(j)}, \Sigma_i^{(j)})\ .
\end{equation}
We sample the posterior distribution using the \textit{emcee} algorithm with stretch move~\cite{emcee}. For each data set, we run $n\gg N$ walkers of $2\cdot 10^6$ samples, discarding the first half as burn-in and applying a thinning factor of $0.2$ for the remaining samples. 

\subsection{Astrophysical population models}\label{sec:models}

We construct our binary black hole (BBH) population by sampling the masses from the \textsc{POWER LAW+PEAK} phenomenological model favored by the latest LIGO/Virgo/KAGRA catalog, GWTC-3~\cite{KAGRA:2021duu}.

The primary mass $m_1$ follows a truncated power law with the addition of a Gaussian peak and an exponential tapering at low masses:
\begin{align}
P(m_1) & \propto [ (1-\lambda)P_{\rm law}(m_1|\gamma_1,m_{\rm max}) \notag \\
 & + \lambda G(m_1|\mu_m,\sigma_m) ] \, S(m_1|m_{{\rm min}}, \delta_m)\,.
\end{align}
Here, $P_{\rm law}(m_1|\gamma_1,m_{\rm max})$ is a power-law distribution with slope $\gamma_1=-3.40$ and cut-off at $m_{\rm max}=86.85\, M_\odot$, $G(m_1|\mu_m,\sigma_m)$ is a Gaussian distribution with mean $\mu_m=33.73\, M_\odot$ and standard deviation $\sigma_m=3.36\, M_\odot$, and $S(m_1|m_{\rm min},\delta_m)$ is a smoothing function that rises monotonically from $0$ to $1$ within $[m_{\rm min},m_{\rm min}+\delta_m]$, with $m_{\rm min}=5.08\, M_\odot$ and $\delta_m=4.83\, M_\odot$.

The secondary mass $m_2$ is obtained from the mass ratio $q=m_2/m_1$, sampled from a smoothed power law
\begin{equation}\label{eq:qdistr}
P(q) \propto q^{\gamma_q} S(m_1 q|m_{\rm min}, \delta_m) \,,
\end{equation}
with $\gamma_q=1.08$.  

We sample the BBH population up to redshift $z=10$. The BBH redshift distribution is assumed to follow the Madau-Dickinson cosmic star-formation rate (SFR)~\cite{Madau:2014bja}, with parameters taken from the phenomenological fit of Ref.~\cite{Ng:2020qpk}. The normalization is set by the local BBH merger rate $R_{\rm m}=28.3~\rm Gpc^{-3}yr^{-1}$, which is favored by the GWTC-3 catalog for the \textsc{POWER LAW+PEAK} model~\cite{KAGRA:2021duu}.

We assume BH spins aligned or antialigned with the binary orbital angular momentum, and we sample their magnitudes from two different distributions. In \texttt{model I}, the dimensionless spin components $\chi_{1,2}$ follow a Beta distribution with parameters $\alpha=2$ and $\beta=5$: this model is meant to qualitatively reproduce the \textsc{Default} spin model of the GWTC-3 catalog~\cite{KAGRA:2021duu}. %
In \texttt{model II}, the individual dimensionless spin magnitudes $\chi_{1,2}$ are uniformly sampled within the range $[-1,1]$.

\section{Case studies}\label{sec:approaches}

We consider here different and complementary case studies. The main purpose of these studies is to compare ``agnostic'' null tests of the Kerr QNM spectrum against theory-based approaches, which exploit direct calculations of the frequencies and damping times of spinning BH solutions in beyond-GR theories.  

\subsection{Agnostic tests}\label{sec:agnostic}

We focus first on agnostic null-hypothesis tests of GR, with the goal of understanding the largest deviation in the QNM spectrum compatible with Kerr predictions.  
In this case we assume the injected values of $(\omega_{i,\tn{obs}}^{\ell_j m_j},\tau_{i,\tn{obs}}^{\ell_j m_j})$ in Eq.~\eqref{eq:def-mu-parspec} to be computed in GR, and interpret them in terms of the \PS template. We consider two families of beyond-Kerr corrections, specified by different mass dimensions of the fundamental coupling of the theory, $\alpha$. 

The simplest case we consider is that of a dimensionless coupling constant, i.e., $p=0$ in Eq.~\eqref{eq:parspeccoupl}. In this case, $\gamma_i$ is the same for all sources ($\gamma=\alpha$), and it can be reabsorbed in the definitions of $\delta  \omega_{\ell_j m_j}^{(k_2)}$ and $\delta  \tau_{\ell_j m_j}^{(k_2)}$. This case includes certain families of scalar-tensor theories of gravity~\cite{Barausse:2013nwa}, and it was investigated in Ref.~\cite{Maselli:2019mjd} under the assumption of a random distribution of the mass and SNR of the ringdown events (rather than the astrophysically motivated population models used in this work). 

As a second scenario we consider the case of a coupling constant with dimensions $({\rm mass})^4$, i.e., $p=4$~\eqref{eq:parspeccoupl}:
\begin{equation}
\gamma_i=\frac{\alpha}{M_i^4}(1+z_i)^4\,.
\end{equation}
Again, $\alpha$ can be absorbed in the definitions of the frequency and damping time deviation coefficients, so the number of parameters that must be constrained is unchanged. 
This case is particularly relevant because it includes notable examples of beyond-GR theories, such as Einstein-scalar-Gauss-Bonnet (EsGB) gravity~\cite{Kanti1995,PaniCardoso2009,Salcedo2016,Pierini:2021jxd,Pierini:2022eim} and dynamical Chern-Simons (dCS) gravity~\cite{Alexander:2009tp,Yunes:2009hc,Molina:2010fb,Srivastava:2021imr,Wagle:2021tam,Wagle:2023fwl}, as discussed in Sec.~\ref{sec:gravityth} below. 

For both parametrizations ($p=0$ and $p=4$), we perform the Bayesian analysis discussed in Sec.~\ref{sec:beyondGRan}. We compare the inference of the \PS parameters in the ``optimistic'' case in which the masses and spins of the remnant BHs are inferred from the inspiral-merger phase against the ``pessimistic'' case in which we only use the ringdown phase, and therefore we need (at a minimum) two modes to perform a test. For the latter case, we will investigate in detail how the analysis depends on the specific choice of the ``secondary'' QNM used in the test.

In the $p=0$ case, we assume that the \PS parameters $(\delta\omega_{\ell m}^{(k_2)},\,\delta\tau_{\ell m}^{(k_2)})$ are drawn from uniform priors in the range $[-1,\,1]$. 
In the $p=4$ case, we use flat priors for $(\delta\omega_{\ell m}^{(k_2)},\,\delta\tau_{\ell m}^{(k_2)})$ in the range $[-3\times10^4,\,3\times10^4]\,\tn{km}^4$. We also impose the conditions $\vert\gamma_i(\delta\omega_{\ell m}^{(k_2)},\,\delta\tau_{\ell m}^{(k_2)})\vert\lesssim1$, as required by the perturbative character of the \PS expansion.
Finally, when masses and spins are allowed to vary in the MCMC, they are sampled from uniform distributions in the ranges $[20,\,500]M_{\odot}$ and $[0.5,\,1]$, respectively.

\subsection{Theory-specific tests}\label{sec:gravityth}

As a top-down approach, we study deviations from Kerr ringdown waveforms focusing on three families of modified gravity theories: 
EsGB gravity, dCS gravity, and the same class of effective field theory (EFT) models considered in Ref.~\cite{Cano:2023jbk}. These theories are particularly appealing in the context of our study for two reasons: (i) the BH solutions in all of these theories are different from those of GR (i.e., the no-hair theorems do not apply), and (ii) the QNM frequencies of these BH solutions have been explicitly computed in the slow-rotation approximation at different orders in the slow-rotation expansion, as we discuss below.

In EsGB and dCS gravity, strong-field, large-curvature modifications are induced by a scalar field $\varphi$ that is non-minimally coupled to the gravitational sector. In both classes of theories, the deviations from GR depend on a fundamental coupling constant ($\alpha_\tn{GB}$ for EsGB gravity, $\alpha_\tn{CS}$ for dCS gravity) with dimensions of $({\rm mass})^2$.

The action of EsGB gravity is
\begin{equation}
{\cal S}_\tn{GB}=\int d^4x \sqrt{-g}\left(R-\frac{1}{2}\partial_\mu\varphi\partial^\mu\varphi
+\frac{\alpha_\tn{GB}}{4}e^{\varphi}{\cal R}_\tn{GB}^2\right)\ ,\label{action_EsGB}
\end{equation}
where ${\cal R}^2_\tn{GB}=R_{\mu\nu\rho\sigma}R^{\mu\nu\rho\sigma}-4 R_{\mu\nu}R^{\mu\nu}+R^2$ is the Gauss-Bonnet invariant,
and $R_{\mu\nu\rho\sigma},R_{\mu\nu},R$ are the Riemann tensor, the Ricci tensor and the Ricci scalar, respectively~\cite{Kanti1995}. 

The action of dCS gravity is
\begin{equation}{\cal S}_\tn{CS}=\int d^4x\sqrt{-g} \left(R-\frac{1}{2}\partial_\mu\varphi\partial^\mu\varphi
+\frac{\alpha_\tn{CS}}{4}\phantom{a}^{\star}RR\,\varphi \right)\ ,\label{action_dCS}
\end{equation}
where $\phantom{a}^{\star}RR=\phantom{a}^{\star}R^{\mu}{_{\nu}}^{\kappa\delta}R^\nu{_{\mu\kappa\delta}}$ is the Pontryagin density, and $\phantom{a}^{\star}R^{\mu\nu\kappa\delta}$ is the dual of the Riemann tensor~\cite{Alexander:2009tp}. 
In Eqs.~\eqref{action_EsGB} and \eqref{action_dCS} we do not consider the matter sector of the theory, because we are interested in astrophysical processes involving BHs. 
For simplicity, we also neglect the possibility of a nonzero scalar field potential.

In both theories, corrections with respect to the BH solutions in GR scale as $\beta_\tn{GB,CS}=\alpha_\tn{GB,CS}/{M^2}$. In the limit $\beta_\tn{GB,CS}\rightarrow0$ we recover the Kerr metric. The shifts in the QNM frequencies are proportional to the {\it square} of $\beta_\tn{GB,CS}$, i.e., to the square of the fundamental coupling constant. Therefore, in both cases the parameter $\alpha$ appearing in Eq.~\eqref{eq:parspeccoupl} -- say, $\alpha=\alpha_{\rm GB}^2$ for EsGB gravity, and $\alpha=\alpha_{\rm CS}^2$ for dCS gravity -- has dimensions of $({\rm mass})^4$.

The gravitational QNM spectrum in EsGB gravity was computed in Ref.~\cite{Blazquez-Salcedo:2016enn} for static BH solutions, and in Ref.~\cite{Pierini:2022eim} for slowly rotating BHs at second order in a small-spin expansion. For dCS gravity, the QNMs of static BHs coincide with those of GR. The QNM spectrum of slowly rotating BHs was computed in Refs.~\cite{Wagle:2021tam,Srivastava:2021imr,Wagle:2023fwl} at first order in a small-spin expansion. These calculations also assume a perturbative expansion in the coupling constant $\alpha_\tn{GB,CS}$. In the spirit of the \PS framework, we truncate this weak-coupling expansion at leading order.

Following Ref.~\cite{Cano:2023jbk}, for EFT gravity we consider the most general extension of GR specified by the action:
\begin{align}
{\cal S}_\tn{EFT}=\frac{1}{16\pi}\int d^4x\sqrt{-g}
&\left(R +\ell^4[\lambda_\tn{evn}{\cal R}^3+
\lambda_\tn{odd}\tilde{{\cal R}}^3]\right .\nonumber\\
&+\left .\ell^6[\epsilon_1{\cal C}^2+
\epsilon_2\tilde{{\cal C}}^2+
\epsilon_3\tilde{{\cal C}}{\cal C}]\right)\ .
\end{align}

The action contains up to eight derivatives, encoded within the curvature scalars
\begin{align}
{\cal R}^3=R{^{\rho\sigma}}_{\mu \nu}
R_{\rho\sigma}{^{\delta\gamma}}
R_{\delta\gamma}{^{\mu\nu}}\quad \ ,&\quad 
\tilde{{\cal R}}^3=R{^{\rho\sigma}}_{\mu \nu}
R_{\rho\sigma}{^{\delta\gamma}}
\tilde{R}_{\delta\gamma}{^{\mu\nu}}\ ,\nonumber\\
{\cal C}=R_{\mu\nu\rho\sigma}R^{\mu\nu\rho\sigma}\quad\ ,&\quad 
\tilde{{\cal C}}=R_{\mu\nu\rho\sigma}\tilde{R}^{\mu\nu\rho\sigma}\ ,
\end{align}
where 
$\tilde{R}_{\alpha\beta\mu\nu}=\epsilon_{\alpha\beta\delta\sigma}R^{\delta\sigma}{_{\mu\nu}}$. 
Corrections to GR are controlled by the length-scale $\ell\sim\Lambda^{-1}_\tn{cut}$, which is related to the EFT cutoff, and by the dimensionless coefficients $\{\lambda_\tn{evn,odd},\epsilon_{1,2,3}\}$.

The QNM frequencies for slowly rotating BH solutions in these theories have recently been computed at high order in a small-spin expansion, including terms up to $\mathcal{O}(\chi^{12}$)~\cite{Cano:2023jbk}. For cubic and quartic theories, the deviations from the Kerr spectrum depend on the coupling constants
\begin{align}
\beta^{\tn{cubic}}_q&=\ell^4\lambda_q/M^4=\alpha^4_q/M^4\quad q=\{\tn{evn},\tn{odd}\}\ ,\nonumber\\
\beta^{\tn{quartic}}_q&=\ell^6\lambda_q/M^6=\alpha^6_q/M^6\quad q=\{1,2,3\}\ .
\end{align}
The validity of the EFT approach requires that these coupling constants satisfy the conditions $\beta^\tn{cubic,quartic}_{q}\ll 1$.  Within the \PS framework, the parameter $\alpha$ appearing in the coupling \eqref{eq:parspeccoupl} is mapped to $\alpha=\alpha_q^4$ and $\alpha=\alpha_q^6$ for the cubic and quartic models, respectively.  In this work we consider eight different EFT gravity theories, for which QNMs have been explicitly computed in Ref.~\cite{Cano:2023jbk}. In particular, we focus on: (i) two parity-violating models, identified by $q=\{\tn{odd}^+,3^{+}\}$; and (ii) six parity-preserving theories, labeled by $q=\{\tn{evn}^+,\tn{evn}^-,1^{+},1^{-},2^{+},2^{-}\}$.\\

At variance with the agnostic, null test approach of Sec.~\ref{sec:agnostic}, for this analysis we assume that the injected values of $(\omega_{i,\tn{obs}}^{\ell_j m_j},\tau_{i,\tn{obs}}^{\ell_j m_j})$ are given by the \textit{true} values obtained in EsGB, dCS or EFT gravity for a given combination of the BH mass, the BH spin, and the coupling constant.

We map the numerical frequencies and damping times computed in Refs.~\cite{Pierini:2022eim}, \cite{Wagle:2021tam,Srivastava:2021imr,Wagle:2023fwl} and~\cite{Cano:2023jbk} 
to the template \eqref{eq:parspec-expa1}--\eqref{eq:parspec-expa2}, i.e., 
we determine the values of $\delta  \omega_{\ell_j m_j}^{(k_2)}$ 
and $\delta  \tau_{\ell_j m_j}^{(k_2)}$ for each theory (and for 
each GW source) by setting 
 \begin{align}\label{eq:couplingGBCS}
 \gamma^\tn{cubic}_{i}= \beta^\tn{cubic}_q=\frac{\alpha_q^4}{M_i^4}\quad\ ,&\quad
 \gamma^\tn{quartic}_{i}= \beta^\tn{quartic}_q=\frac{\alpha_q^6}{M_i^6}\ ,\nonumber\\
  \gamma_{\tn{GB},i}=\beta_\tn{GB}^2= \frac{\alpha_\tn{GB}^2}{M_i^4}\quad\ ,&\quad
 \gamma_{\tn{CS},i}= \beta_\tn{CS}^2=\frac{\alpha_\tn{CS}^2}{M_i^4}\ .
 \end{align}
\par\noindent
The technical details of the mapping can be found in Appendix~\ref{app:qnmtheories}.
With the shifts $(\delta  \omega_{\ell_j m_j}^{(k_2)},\delta  \tau_{\ell_j m_j}^{(k_2)})$ fixed by the theory, we recover the injected QNMs using the \PS template by MCMC sampling on the coupling $\alpha_\tn{GB,CS}$, and also on $(M_i,\chi_i)$. 
This is because, at variance with the  ``agnostic'' case -- where $M_i$, $\chi_i$ in Eqs.~\eqref{eq:parspec-expa1}--\eqref{eq:parspec-expa2} are the GR masses and spin, and the shifts between those and the ``true'' masses
and spins of the remnant  are absorbed in the parameters $(\delta\omega_{\ell m}^{(k_2)},\delta\tau_{\ell m}^{(k_2)})$ -- in theory-specific tests the BH mass and spin enter in  Eqs.~\eqref{eq:parspec-expa1}--\eqref{eq:parspec-expa2}  with their \textit{physical} values.

For this reason we must infer $M_i$ and $\chi_i$ directly from each ringdown observation, through a two-mode analysis. In other words, we are forced to follow the second strategy discussed in Sec.~\ref{sec:parspec}.
Fortunately, in this case the inclusion of the secondary QNM introduces a smaller set of parameters to be constrained compared to the agnostic approach. This is because the shifts are fixed, and the only free parameter beyond $(M_i,\,\chi_i)$ is the coupling constant $\alpha_{\tn{GB,CS},q}$.
We impose flat priors on the coupling constants of the three theories in the range $\alpha_\tn{GB,CS}\sim{\cal U}_{[0,\,100]}$, and $\alpha_q\sim{\cal U}_{[0,\,50]}$. Masses and spins are sampled from uniform distributions within $[3,\,150]M_{\odot}$ and $[0.5,\,1]$, respectively.

\subsection{Numerical setup}\label{sec:setup}

With all the ingredients described in the previous sections, we can now forecast the constraints on beyond-Kerr QNMs according to the following procedure:
\begin{enumerate}
\item We compute the mass and the spin of the remnant for the binary systems of our catalog exploiting semi-analytic relations derived from NR~\cite{Healy:2014yta}.  For each BH we derive the fundamental mode ($\ell=m=2$) and the secondary modes (either $\ell=m=3$ or $\ell=2, m=1$) through the templates of Eqs.~\eqref{eq:parspec-expa1} and \eqref{eq:parspec-expa2}, both in GR and in the gravity theories described in Sec.~\ref{sec:gravityth}. We choose independently the order of the spin expansion for the GR ($n_1$) and beyond-GR ($n_2$) coefficients. In particular we set $n_1=12$, as this is sufficient to reproduce the QNMs of the full Kerr solution for a typical merger remnant ($\chi\sim 0.7$) with an accuracy better than $1\%$~\cite{Maselli:2019mjd}: see also the discussion in Appendix~\ref{app:qnmtheories}.  In the agnostic case we vary $n_2$ to study the constraints that can be inferred on different spin-dependent \PS coefficients. In the theory-specific study, $n_2$ is necessarily limited by the accuracy of the QNM calculations available so far, i.e., $n_2=2$, $n_2=1$ and $n_2=12$ for EsGB, dCS and EFT gravity, respectively.
\item We compute the radiation efficiency of all mode/binary configurations (and hence the corresponding amplitudes ${\cal A}_{\ell m}$) using Eq.~\eqref{math:epsilonfit}.
  We then determine the SNR of each multipolar component $(\ell\,m)$, and select only events for which the SNR of the $(22)$ QNM is larger than the detectability threshold $\rho=12$, as measured by ET or CE. From now on, we will refer to this subset as the ``detectable'' events. We repeat this procedure for both spin distributions (``\texttt{model I}'' and ``\texttt{model II}'' of Sec.~\ref{sec:models}).
\item We compute the FIM for each of the selected ringdown events, evaluating the joint likelihood distribution of frequencies and damping times, in both the single-mode and two-mode configurations.  For theory-based tests we apply the FIM approach to QNMs computed in EsGB, dCS and EFT gravity, which are uniquely determined by the coupling constant of the theory, together with the remnant's masses and spins. For simplicity, we select the same ensemble of events -- i.e., those for which the SNR of the dominant mode {\em in GR} is larger than $12$. This is justified, as the SNR computed in these theories of gravity typically differs by less than a percent from the values computed in GR.
\item We use the likelihood functions as seeds for our Bayesian analysis to sample the \PS parameters for the agnostic tests (with $p=0$ and $p=4$), and the fundamental couplings $\alpha_\tn{GB/CS}$ for the theory-based tests.
The remnant masses and spins are kept fixed in the single-mode analysis, while they can be allowed to vary together with the beyond-GR parameters in the two-mode approach.
\end{enumerate}

When quoting constraints on all parameters of interest, we refer to the 90\% highest posterior density intervals, except for distributions bounded by the prior, for which we report the one-side 90\% probability interval.
We show posterior distributions as histograms of the samples inferred from the MCMC (or Kernel Density Estimations of the latter, computed in {\sc Mathematica} assuming a Gaussian Kernel).

\begin{figure}[t]
    \includegraphics[scale = 0.5]{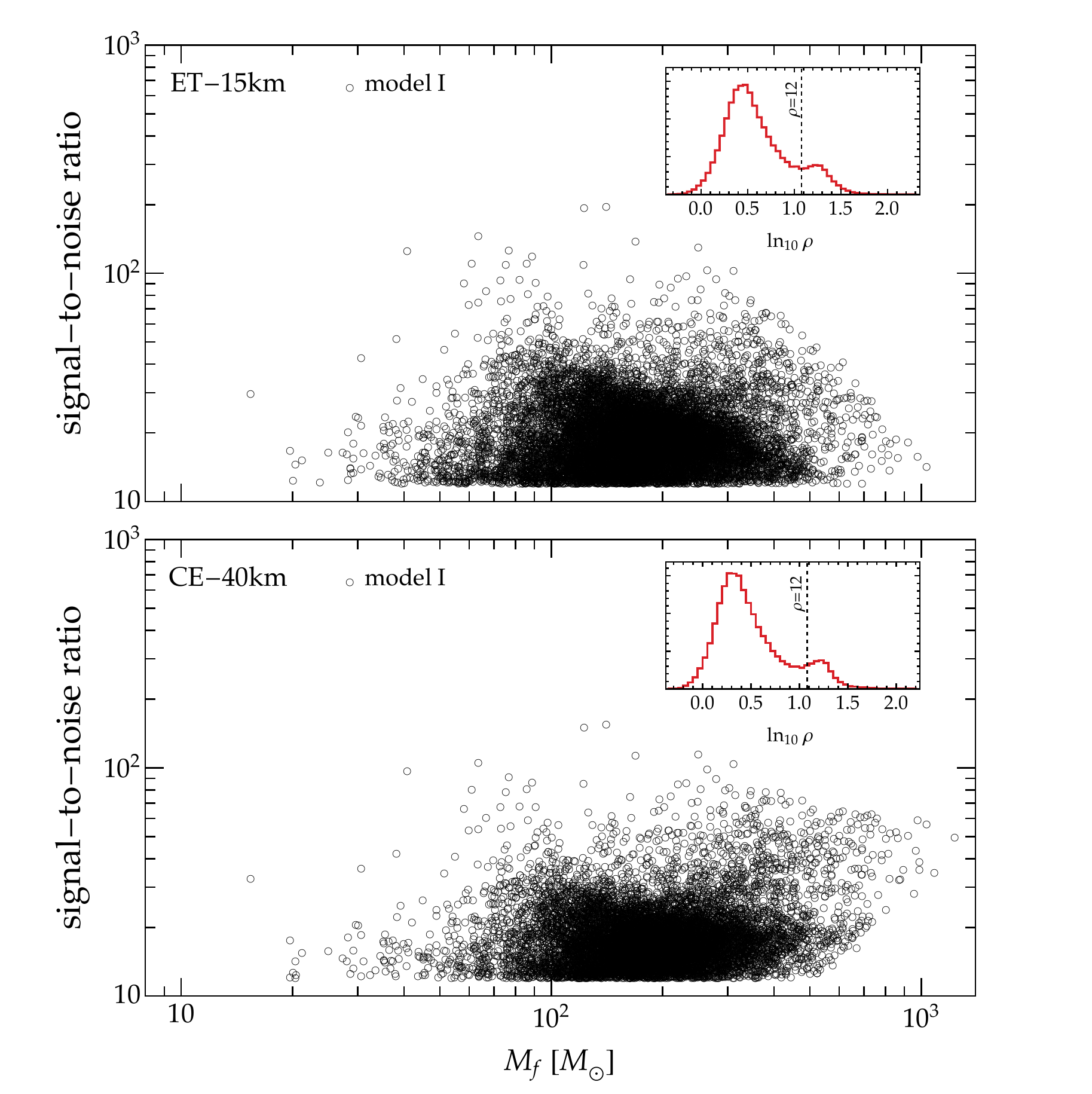}
    \caption{SNR of the $(22)$ mode as a function of the detector-frame remnant mass for detectable events in ET (top panel) and CE (bottom panel), assuming the \texttt{model I} population.
      ``Detectable'' events are those for which the SNR of the $(22)$ QNM is larger than the detection threshold, $\rho\ge 12$.
      The inset in each panel shows the SNR distribution of the \textit{whole} catalog, before applying the detection threshold cut. The SNR--$M_f$ distribution for \texttt{model II} is similar.}
    \label{fig:snr}
\end{figure}

\begin{figure}[htbp!]
     \includegraphics[scale = 0.4]{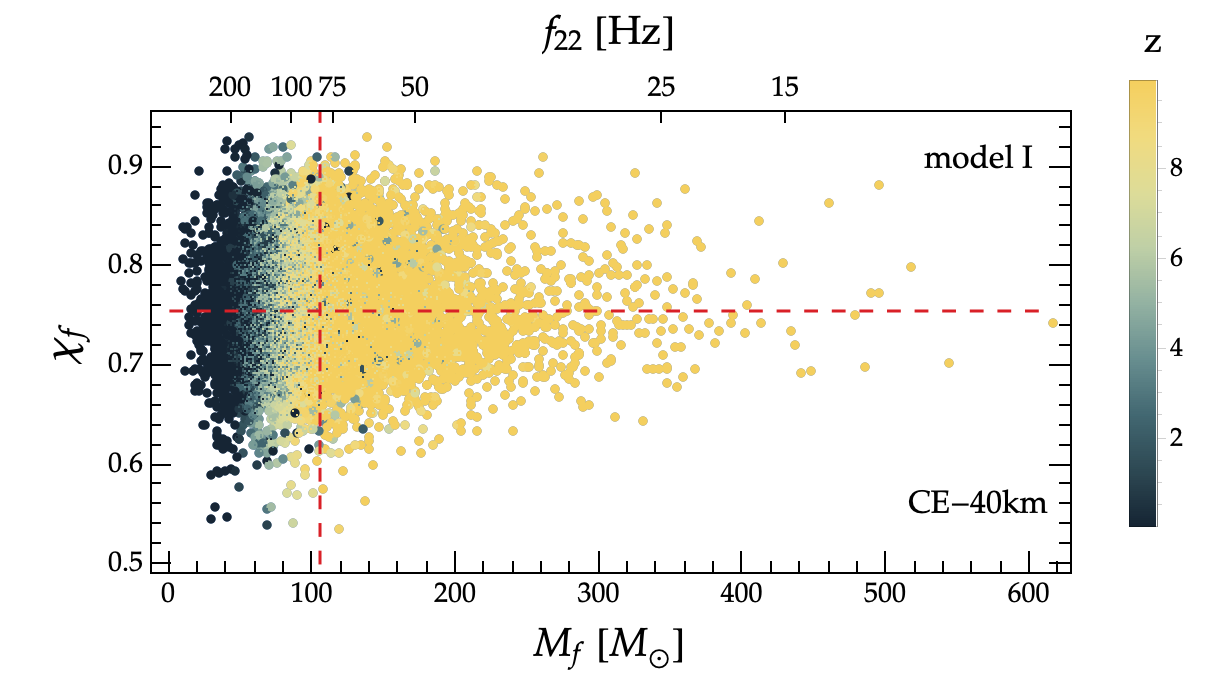}
     \includegraphics[scale = 0.4]{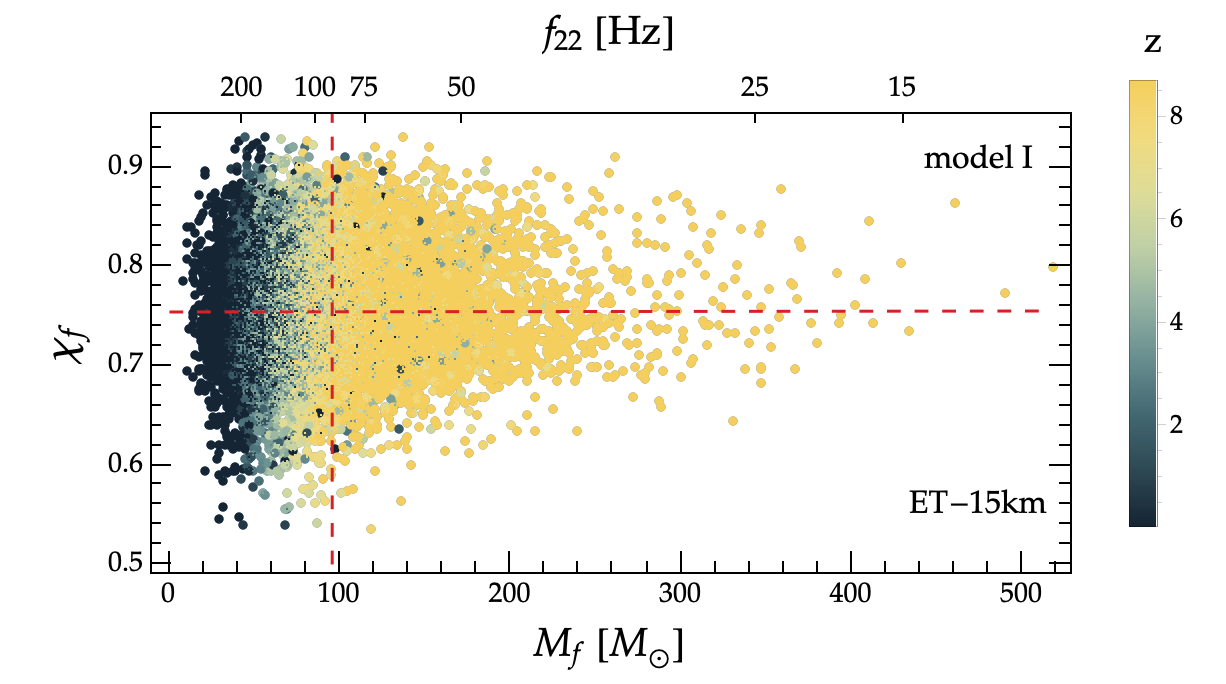}
    \includegraphics[scale = 0.4]{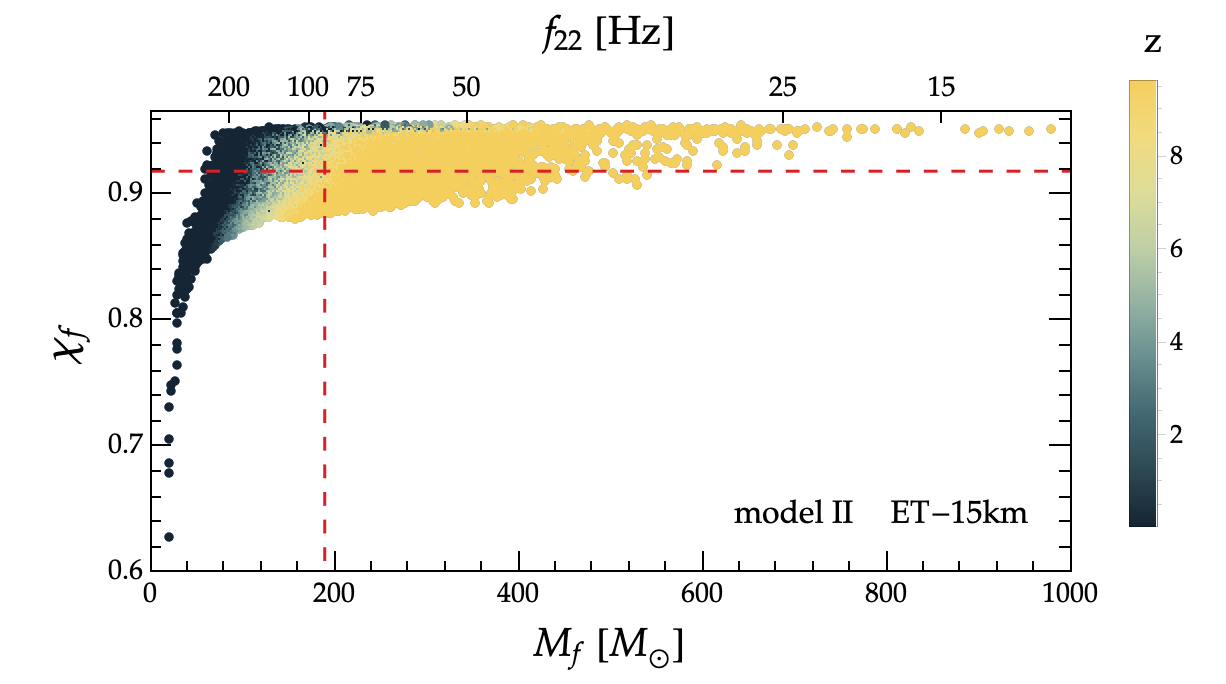}
     \includegraphics[scale = 0.4]{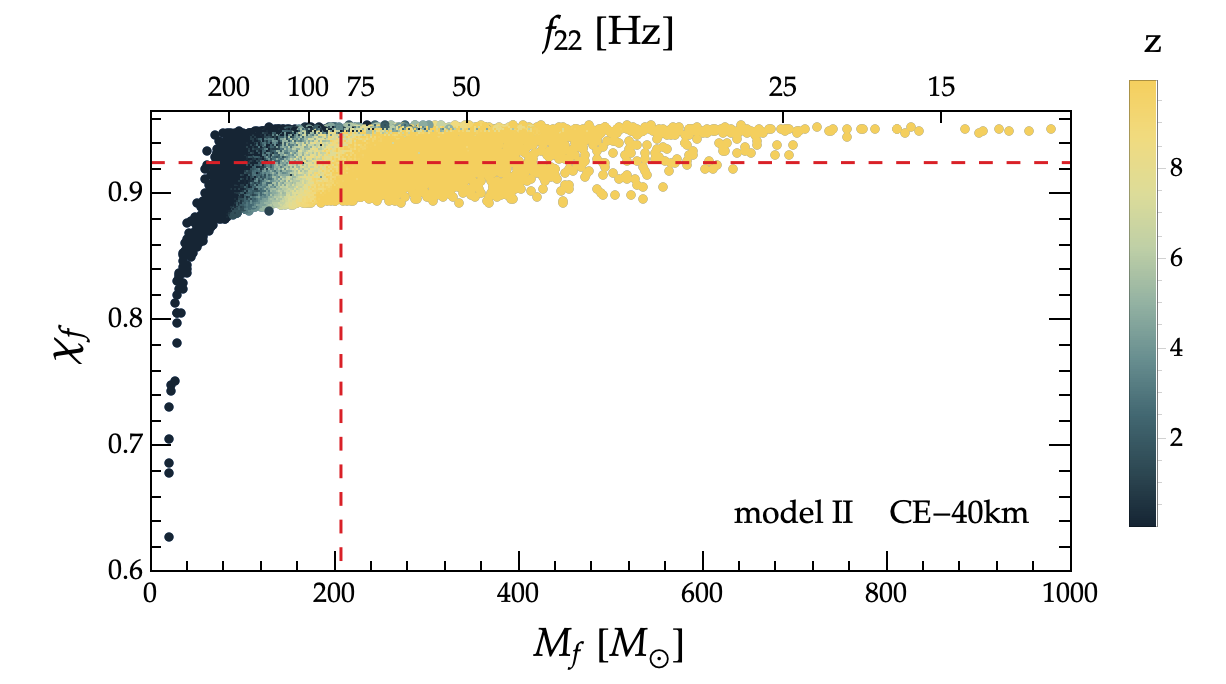}
    \caption{Distribution of the remnant detector-frame masses and spins obtained from the binary population catalog for detectable events. The color code identifies the redshift associated to the binary, for a specific detector. Dashed lines correspond to the mean values of the distributions. For reference, the top x-axis in each panel shows the frequency of the fundamental $(22)$ mode for a remnant BH with dimensionless spin $\chi_f=0.8$, and mass corresponding to the bottom x-axis.}    \label{fig:finalmasses}
\end{figure} 

\section{Results}
\label{sec:results}

The two panels of Fig.~\ref{fig:snr} show the SNR distribution as a function of the remnant mass for the events with a detectable ringdown, assuming the spin distribution of \texttt{model I}. Within the $10^5$ BBH binaries of the catalog we find $N=13547$ and $N=11126$ detectable events for ET and CE, respectively. The mean SNR of observable events is $\rho\simeq21$ ($\rho\simeq19$) for ET (CE).
The qualitative behavior for the \texttt{model II} population is similar, although the number of selected events decreases: we find $N=12487$ for ET and $N=9021$ for CE.

As shown in Fig.~\ref{fig:finalmasses}, different assumptions on the spins (referred to as ``\texttt{model I}'' and ``\texttt{model II}'' above) lead to significantly different mass-spin distributions of the BH remnants, with mean spin values clustering around $\chi_f\sim0.75$ for \texttt{model I} and $\chi_f\sim0.9$ for \texttt{model II}. 

\begin{table}[h]
	\centering
	\begin{tabular}{ c | c c c | c}
        \hline
		\hline
		  & $\rho\geq 12$ & $\rho\geq 25$ 
		& $\rho\geq 100$ & ${\rm max} \ \rho$ \\
		\hline
		ET - \texttt{model I}& $13547$ & $2554$ & $14$ & $196$ \\
		CE - \texttt{model I}& $11126$ & $1454$ & $6$ & $155$ \\
   		ET - \texttt{model II}& $12487$ & $1662$ & $7$ & $164$ \\
		CE - \texttt{model II}& $9021$ & $754$ & $2$ & $119$ \\
		\hline
        \hline
	\end{tabular}
		\caption{Number of ringdown observations with signal-to-noise ratio of the fundamental mode larger than $(12,25,100)$ for the two spin distributions we consider. The last column shows the maximum SNR of each population.}
	\label{tab:snr}
\end{table}

\begin{figure}[t]%
    \includegraphics[scale = 0.62]{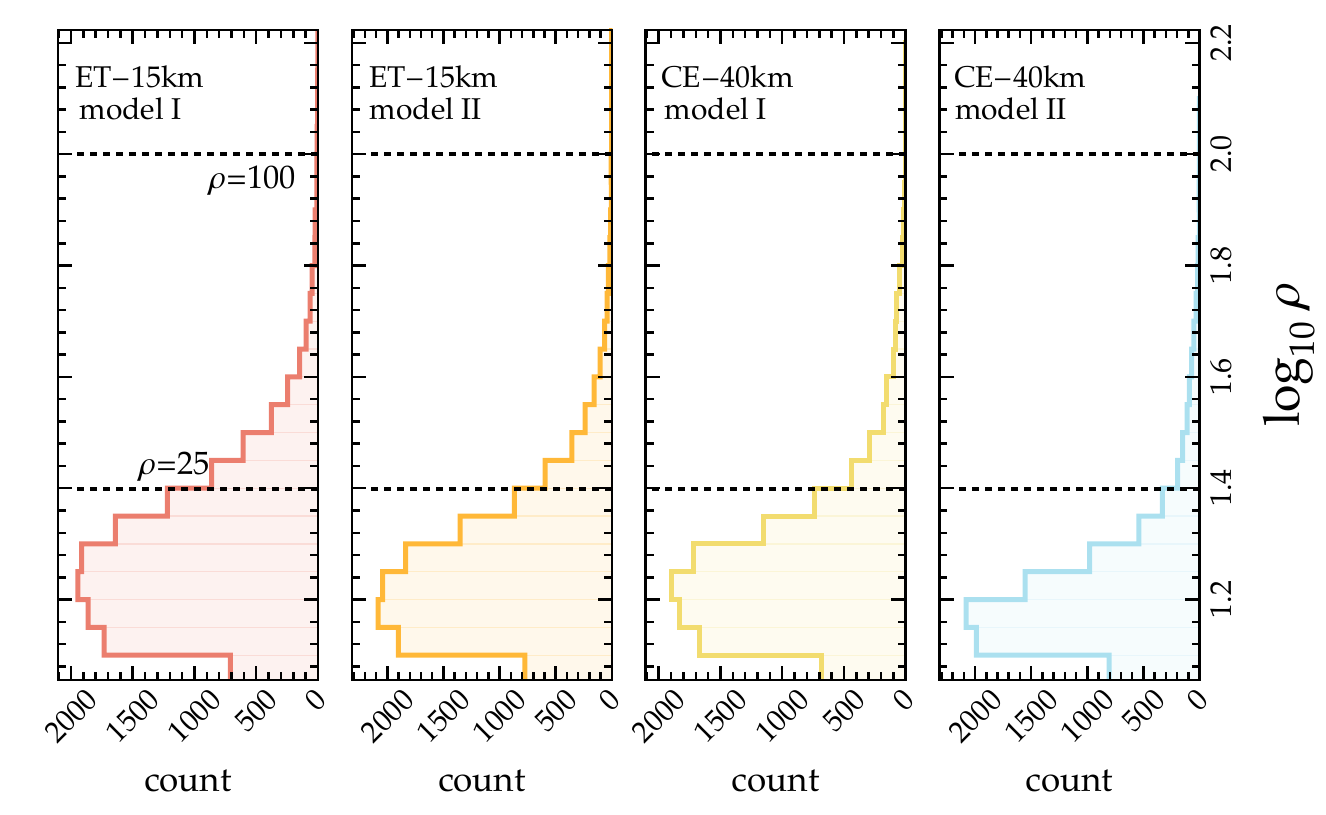}
    \caption{Distribution of the SNR for the fundamental $\ell=m=2$ mode, for the selected events observed by ET and CE above the detectability threshold $\rho=12$.}
   \label{fig:snr2}
\end{figure}

In Fig.~\ref{fig:snr2} we show the distribution of the SNR for the two detectors and for the two spin models.
Note that \texttt{Model I} yields slightly larger values of the SNR for both ET and CE.
In the frequency range which is most relevant for ringdown detection, CE and ET (in a single-detector configuration) have similar SNR (see the discussion in Appendix~\ref{sec:snrconf}).  The difference in the number of events above the threshold is mostly due to the fact that we consider a two-detector network configuration for ET. This boosts the ET SNR by a factor of $\sqrt{2}$.
In Table~\ref{tab:snr} we summarize some salient features of these SNR distributions: for both detectors we estimate thousands of ringdown events with $\rho\geq 25$, but only a few events with $\rho\geq 100$.

\begin{figure}[htbp!]
    \includegraphics[width=8cm]{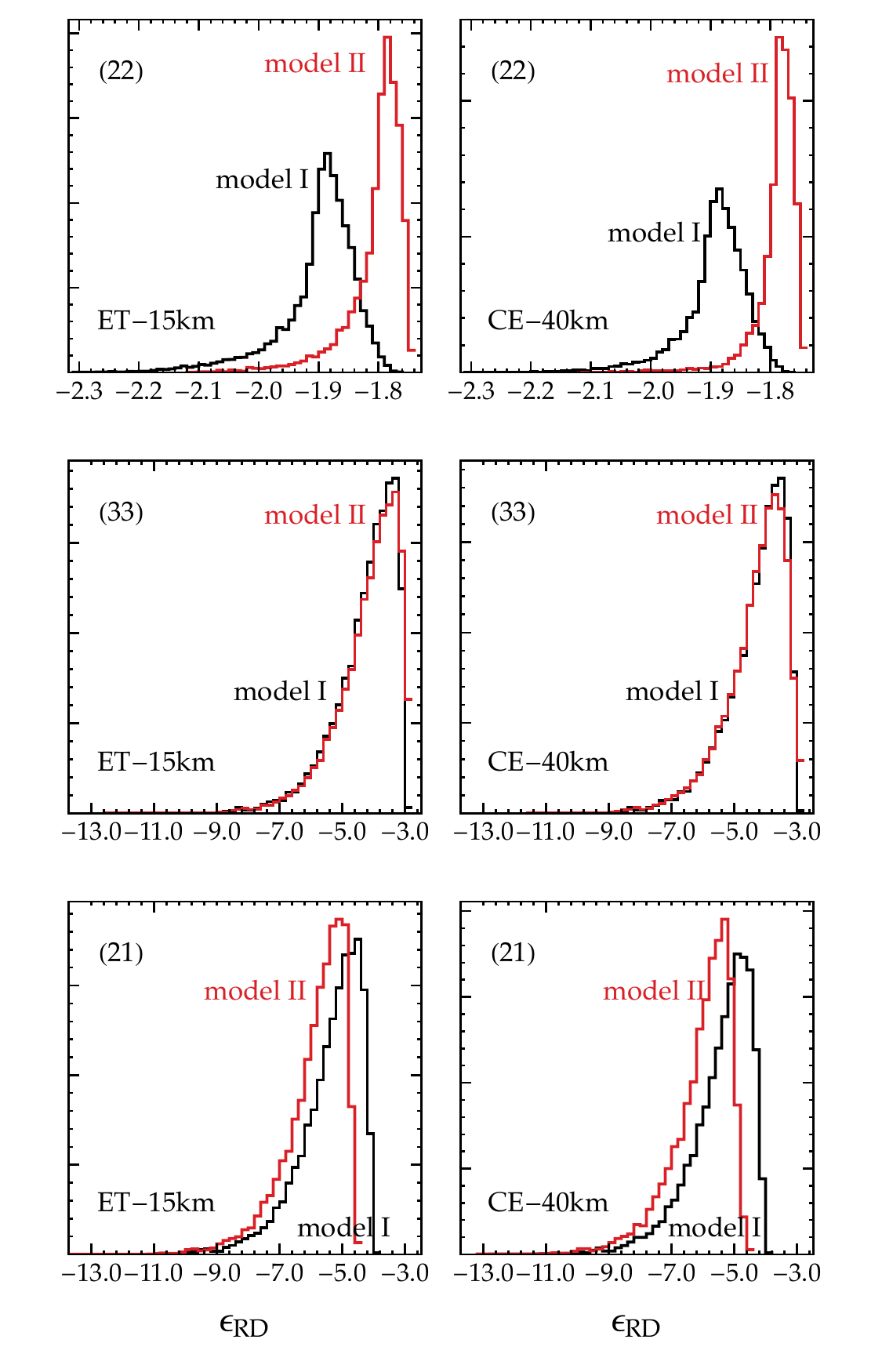}
    \caption{Distribution of the energy released in the $(22)$, $(33)$ and $(21)$ modes for the ringdown events with signal-to-noise ratio of the fundamental QNM larger than 12, as observed by ET and CE, assuming both population models described in Sec.~\ref{sec:models}.}\label{fig:hist_amp}
\end{figure} 

\begin{figure}[htbp!]
    \includegraphics[scale = 0.34]{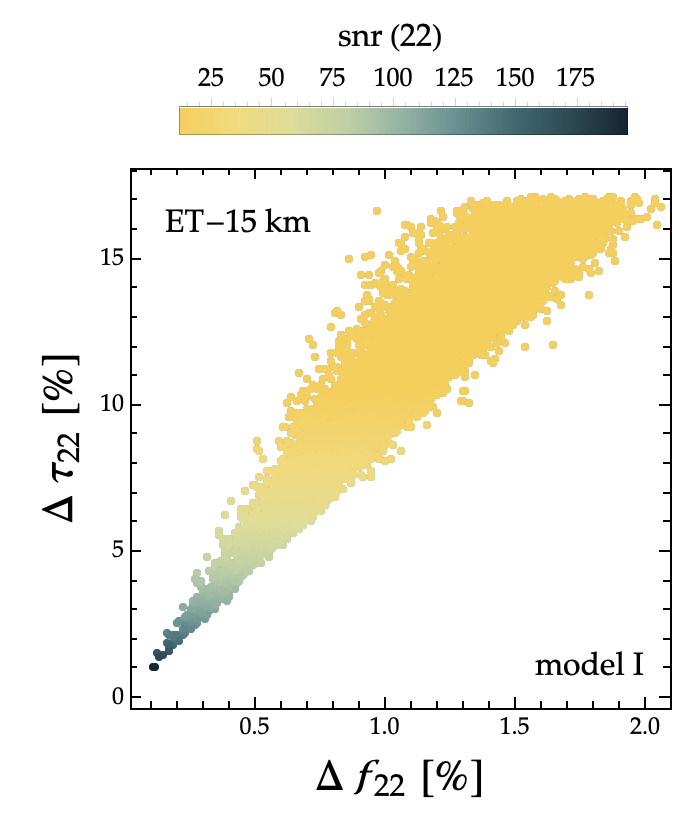}
\includegraphics[scale = 0.34]{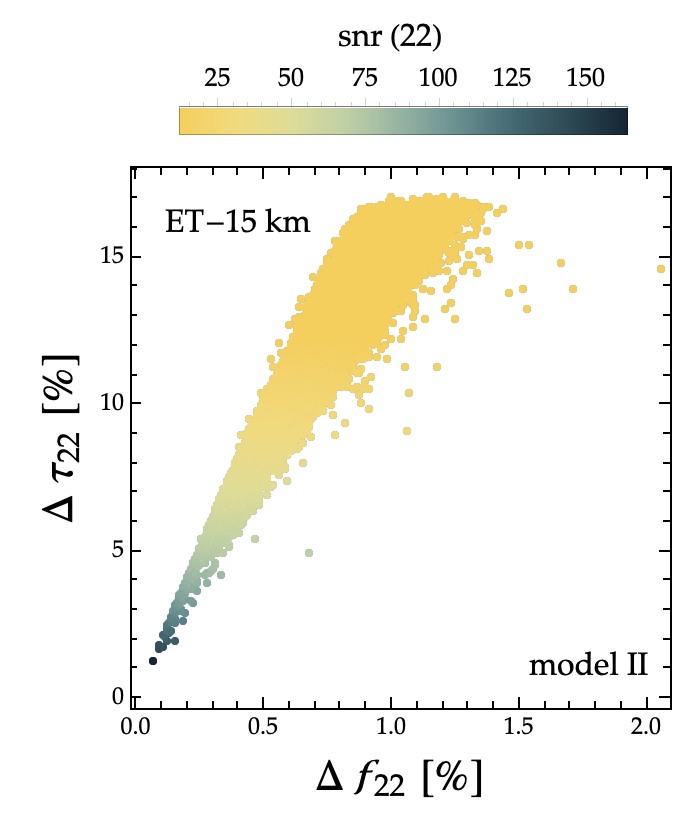}
 \includegraphics[scale = 0.34]{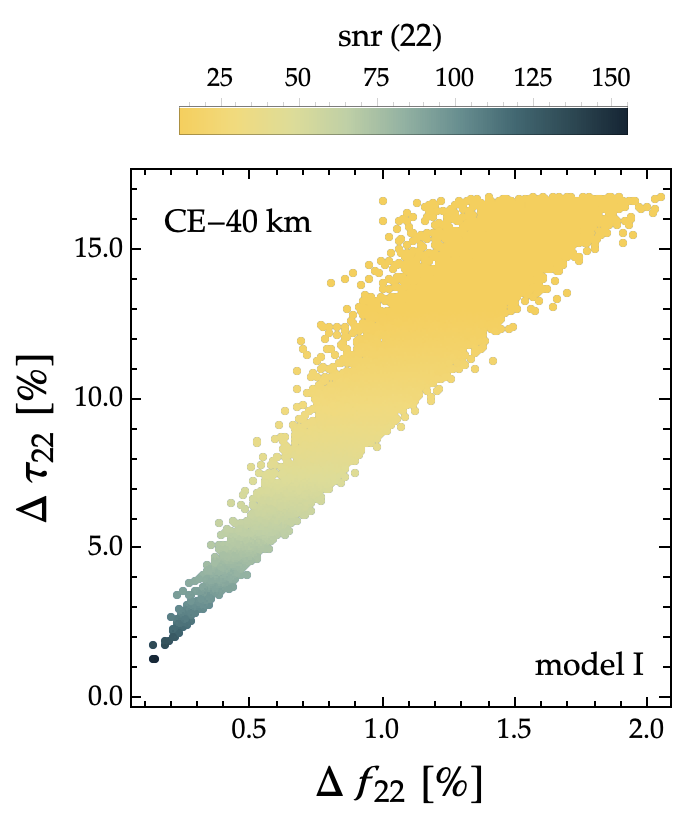}
\includegraphics[scale = 0.34]{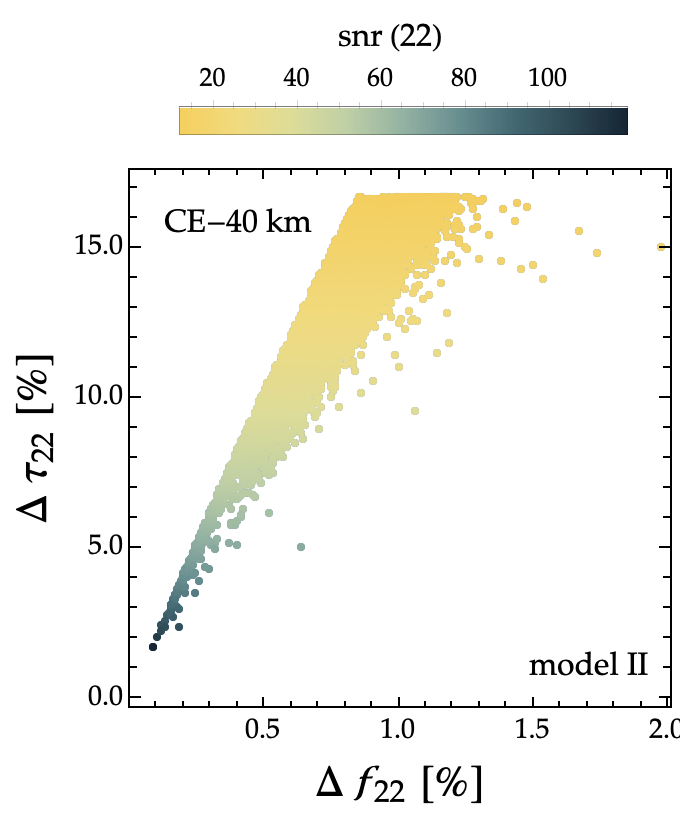}
    \caption{Relative percentage errors on the frequency and the damping time of the fundamental QNM with $\ell=m=2$ %
    for the binary systems  described in Sec.~\ref{sec:setup}. The color scheme identifies the SNR of the $(22)$ mode as detected by ET (top row) and CE (bottom row).}
    \label{fig:errors-fisher-app-fund}
\end{figure}

\begin{figure*}[htbp!]
    \includegraphics[scale = 0.65]{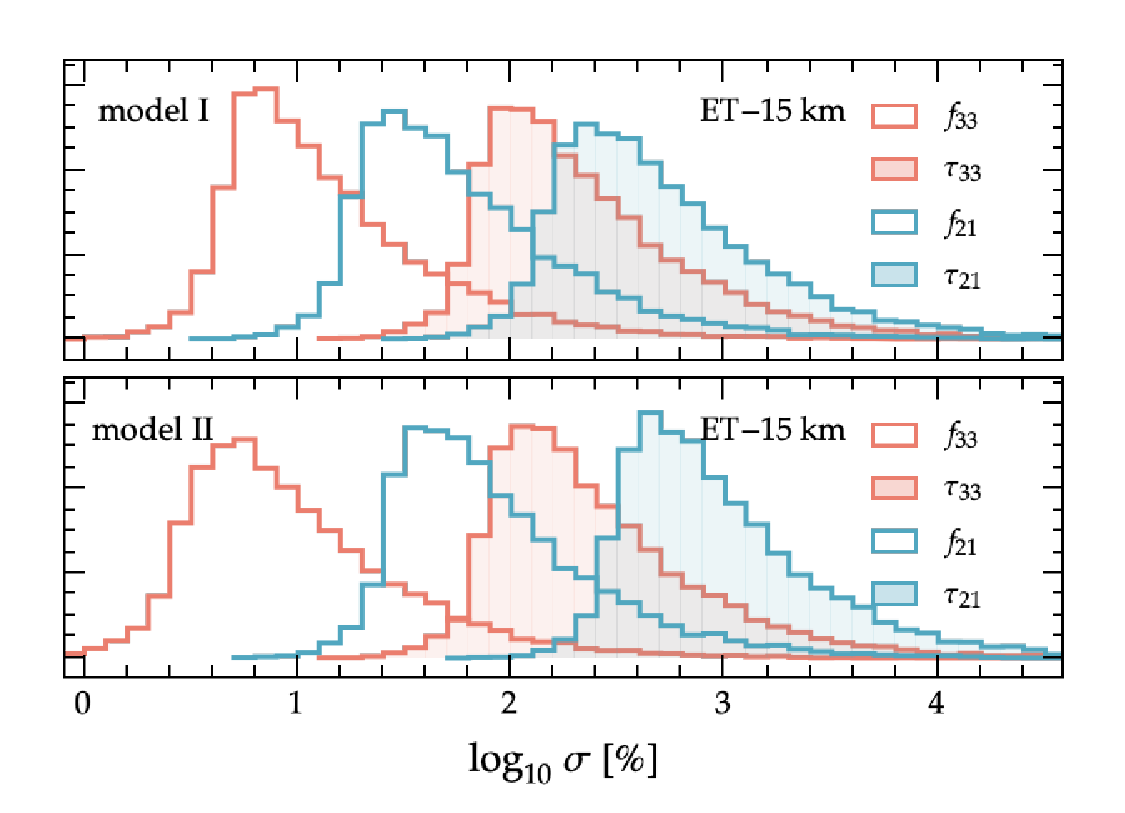}
     \includegraphics[scale = 0.65]{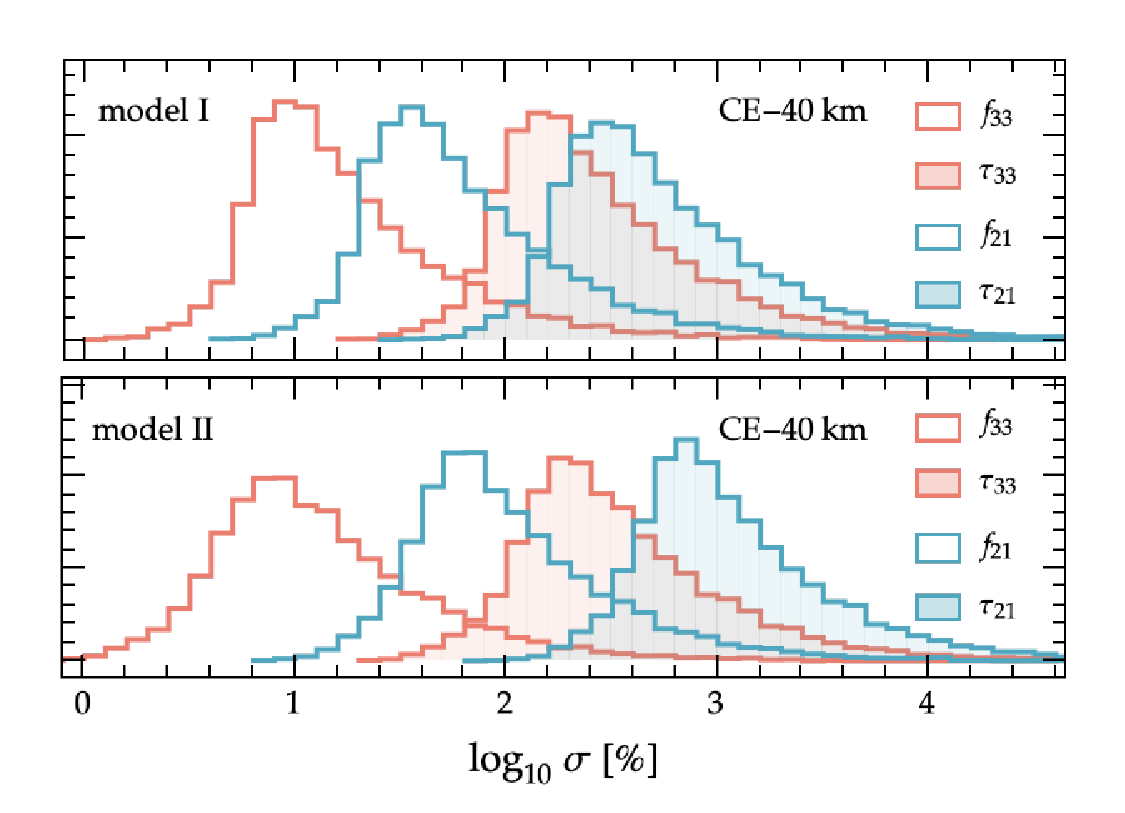}
    \caption{Relative percentage errors on the frequency and the damping time of the secondary modes, $(\ell,m)=(33)$ and $(\ell,m)=(21)$, for ET and CE.The top and bottom panels refer to the two population models we considered.}
    \label{fig:errors-fisher-app-sec}
\end{figure*}

The GW energy released in the different QNMs considered in our analysis is shown in Fig.~\ref{fig:hist_amp}. The component spins directly affect the spin of the remnant, leading to sensibly different results for the $\ell=m=2$ mode amplitude in \texttt{model I} and \texttt{model II} (top panel).

The dependence of the effective amplitude on the spin is quite mild for the $(33)$ and the $(21)$ QNM amplitudes, because our astrophysical population is dominated by comparable-mass binaries -- cf. Eq.~\eqref{eq:qdistr} -- and odd-$m$ multipoles of the radiation vanish by symmetry for equal-mass binaries~\cite{Berti:2007fi,Kidder:2007rt}.  Unlike the $(33)$ mode, the $(21)$ mode is quite sensitive to the binary component spins: see e.g. Eq.~(3) of Ref.~\cite{Berti:2007nw}.  This explains why the $(21)$ mode amplitudes for \texttt{model I} and \texttt{model II}, shown in the bottom panel of Fig.~\ref{fig:hist_amp}, are quite different.  However, for our population model the $(33)$ mode generally yields larger values of the ringdown efficiency $\epsilon_\tn{RD}$.

The four panels of Fig.~\ref{fig:errors-fisher-app-fund} show the percentage relative errors on the frequency and on the damping time of the fundamental mode with $\ell=m=2$ inferred through the FIM approach described in Sec.~\ref{sec:qnmpe}, for all detectable ringdown events. For both ET and CE the mode frequency can be measured with an accuracy better than $1\%$, while the damping time is usually measured with less accuracy. Only few observations (those with $\rho\gtrsim 100$) have relative errors on $\tau_{22}$ at the level of $2\%$-$3\%$.

In Fig.~\ref{fig:errors-fisher-app-sec} we show the corresponding errors for the secondary modes.
As expected from our discussion of the effective amplitudes, the error histograms are almost independent of the underlying astrophysical spin distribution, with the frequency and damping time of the $(33)$ mode being better measured than the frequency and damping time of the $(21)$ mode.

\subsection{Agnostic inference}

We start by exploring agnostic tests of GR, using simple templates at first and then progressively complicating the analysis. 
We follow the two strategies discussed in Sec.~\ref{sec:parspec}: (i) we assume that the (unperturbed) masses and spins of the sources, $(M_i,\chi_i)_{i=1\ldots N}$, are known {\it a priori}; or (ii) we extract these quantities from the ringdown signal by considering them as part of the parameter set to be determined, i.e., $\vec\zeta=(\delta\omega^{(k_2)}_{\ell m},\delta\tau_{\ell m}^{(k_2)},M_i,\chi_i)$.

\subsubsection{Strategy (i): Mass and spin are known}
\label{sec:res1}

\noindent
{\bf \em No spin-dependent corrections ($n_2=0$), dimensionless coupling ($p=0$).}
We consider QNM modifications at zeroth order in the spin ($n_2=0$), assume that the underlying theory has a dimensionless coupling ($p=0$), and investigate the constraints on this class of beyond-GR modifications with the \PS template of Eqs.~\eqref{eq:parspec-expa1} and \eqref{eq:parspec-expa2}. We perform a single-mode analysis, i.e., we use only observations of the $(22)$ mode, with both mass and spin of the remnant being set equal to their GR values (see the discussion in Sec.~\ref{sec:parspec} and in Ref.~\cite{Maselli:2019mjd}).
\begin{figure*}[t]%
    \includegraphics[width=\columnwidth]{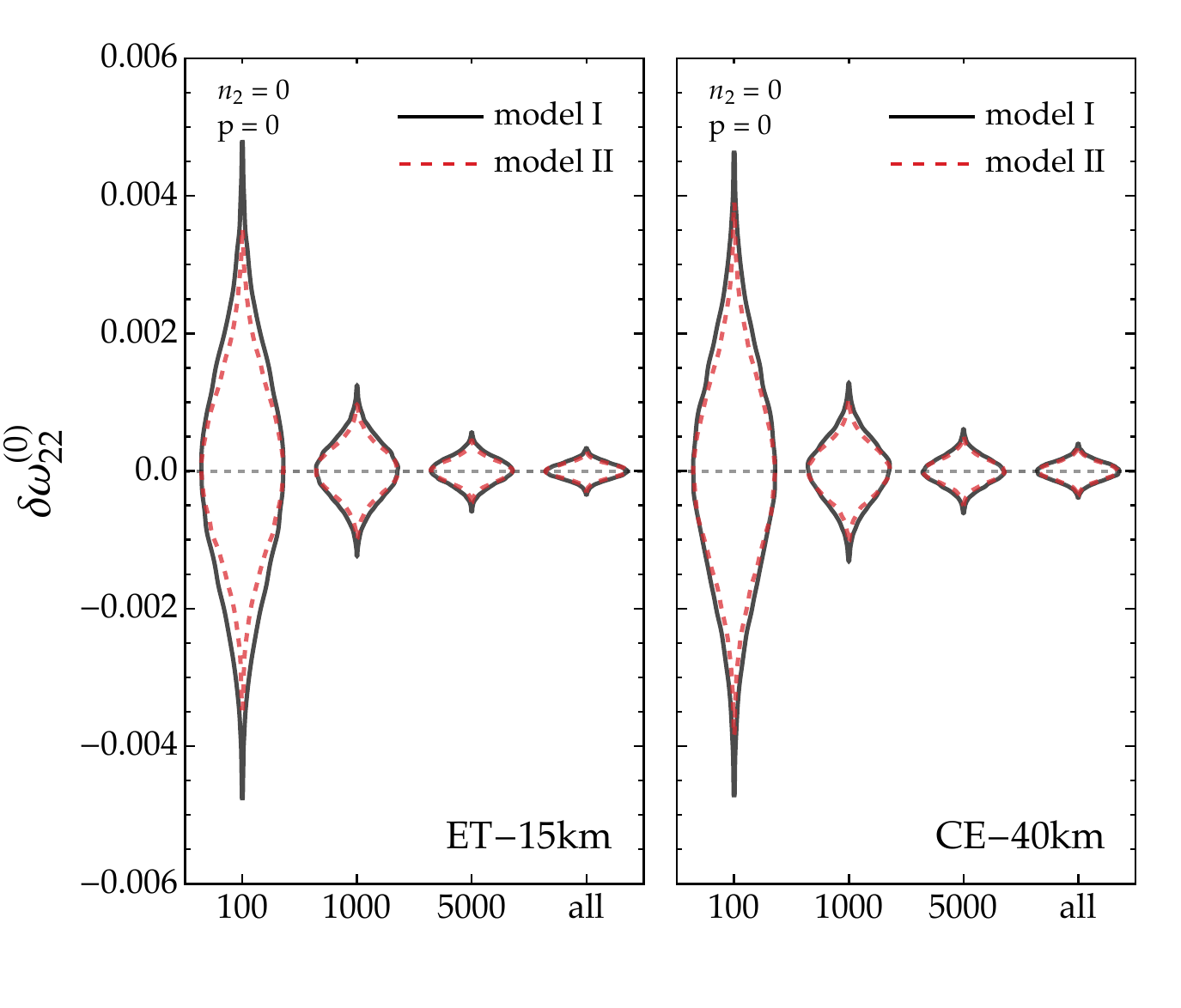}
    \includegraphics[width=\columnwidth]{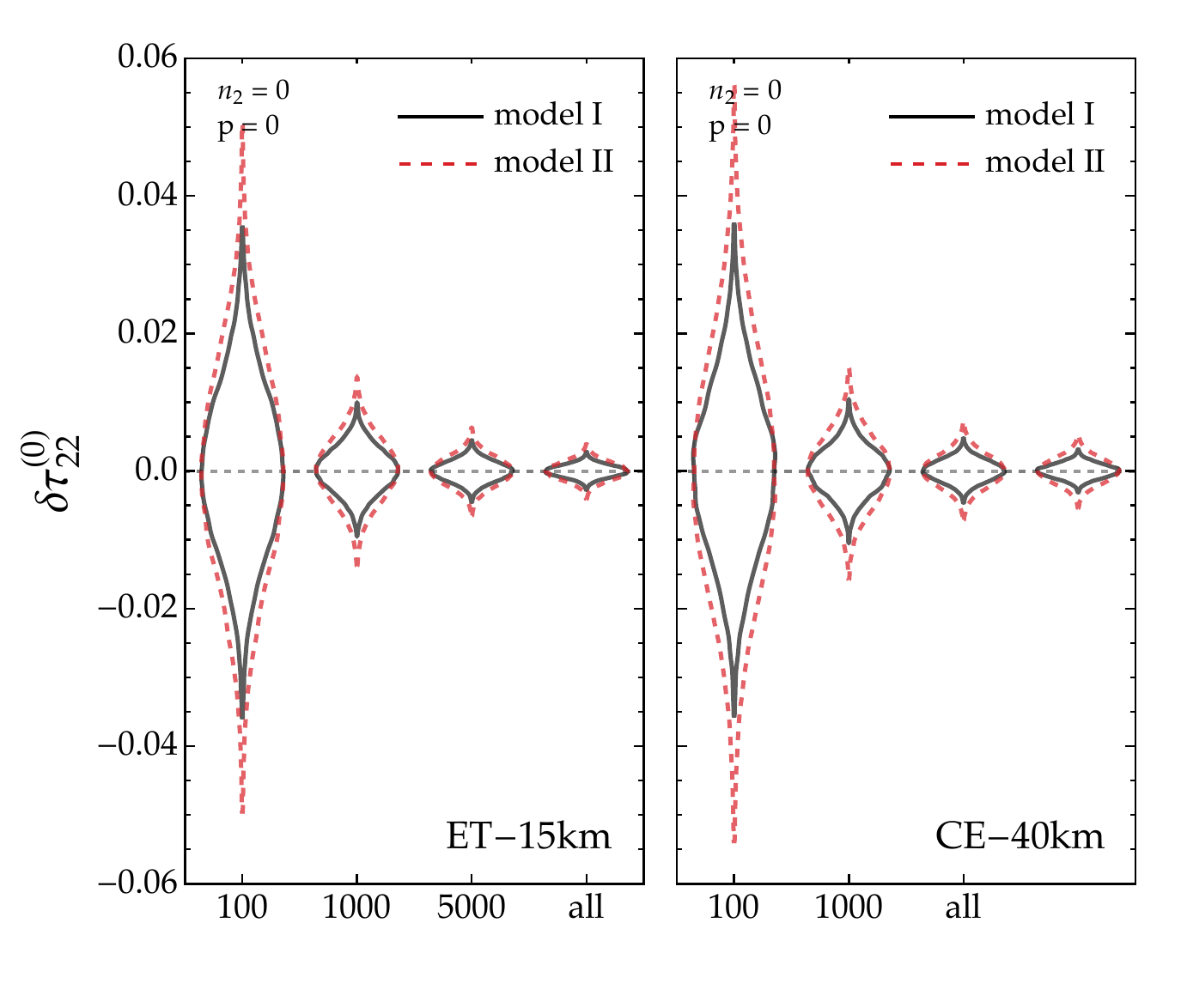}
    \caption{Posterior probability densities for the nonrotating corrections $(\delta\omega_{22}^{(0)},\delta\tau_{22}^{(0)})$, inferred in the agnostic analysis assuming strategy (i), as functions of the number of sources we analyzed.
    Results are derived following a single-mode analysis, in which only the fundamental $(22)$ mode is taken into account.
    Solid and dashed contours refer to the two population models adopted in this paper.}
  \label{fig:p0spin0pdf}
\end{figure*} 

In Fig.~\ref{fig:p0spin0pdf} we show how the posteriors of $\delta\omega_{22}^{(0)}$ and $\delta\tau_{22}^{(0)}$ get tighter by stacking an increasing number of sources.
The probability distributions for beyond-GR corrections in both the frequency and damping time are nearly symmetrical around zero (the GR value). For a given number of observations, ET provides slightly stronger bounds than CE, mainly because the ringdown SNR is typically larger. 
The posterior densities get narrower as the number of events, $N$, grows. For large $N$, the width of the distribution at $90\%$ credible level is well approximated by a $\sim N^{-1/2}$ scaling for both spin models (\texttt{model I} and \texttt{model II}) and for both detectors (ET and CE). 

The joint posterior distributions (that we do not display for brevity) show that the \PS parameters $\delta\omega_{22}^{(0)}$ and $\delta\tau_{22}^{(0)}$ are mostly uncorrelated with each other, and that $\delta\omega_{22}^{(0)}$ is always better constrained than $\delta\tau_{22}^{(0)}$.
The constraints are almost identical for \texttt{model I} and \texttt{model II}, i.e., they are insensitive to the spin prescription. This will be discussed in greater detail below, when we will include spin-dependent corrections in the analysis.
By stacking all of the events for \texttt{model I} we find $\vert \delta\omega_{22}^{(0)}\vert \lesssim1.8\cdot10^{-4}$, $\vert \delta\tau_{22}^{(0)}\vert \lesssim 1.4\cdot10^{-3}$ for ET and $\vert \delta\omega_{22}^{(0)}\vert \lesssim2.1\cdot10^{-4}$, $\vert \delta\tau_{22}^{(0)}\vert \lesssim 1.7\cdot10^{-3}$ for CE. For \texttt{model II} the bounds are $\vert \delta\omega_{22}^{(0)}\vert \lesssim1.5\cdot10^{-4}$, $\vert \delta\tau_{22}^{(0)}\vert \lesssim 2.2\cdot10^{-3}$ for ET and $\vert \delta\omega_{22}^{(0)}\vert \lesssim1.9\cdot10^{-4}$, $\vert \delta\tau_{22}^{(0)}\vert \lesssim 2.9\cdot10^{-3}$ for CE, respectively.

\begin{figure}[htbp!]
    \includegraphics[width=8.5cm]{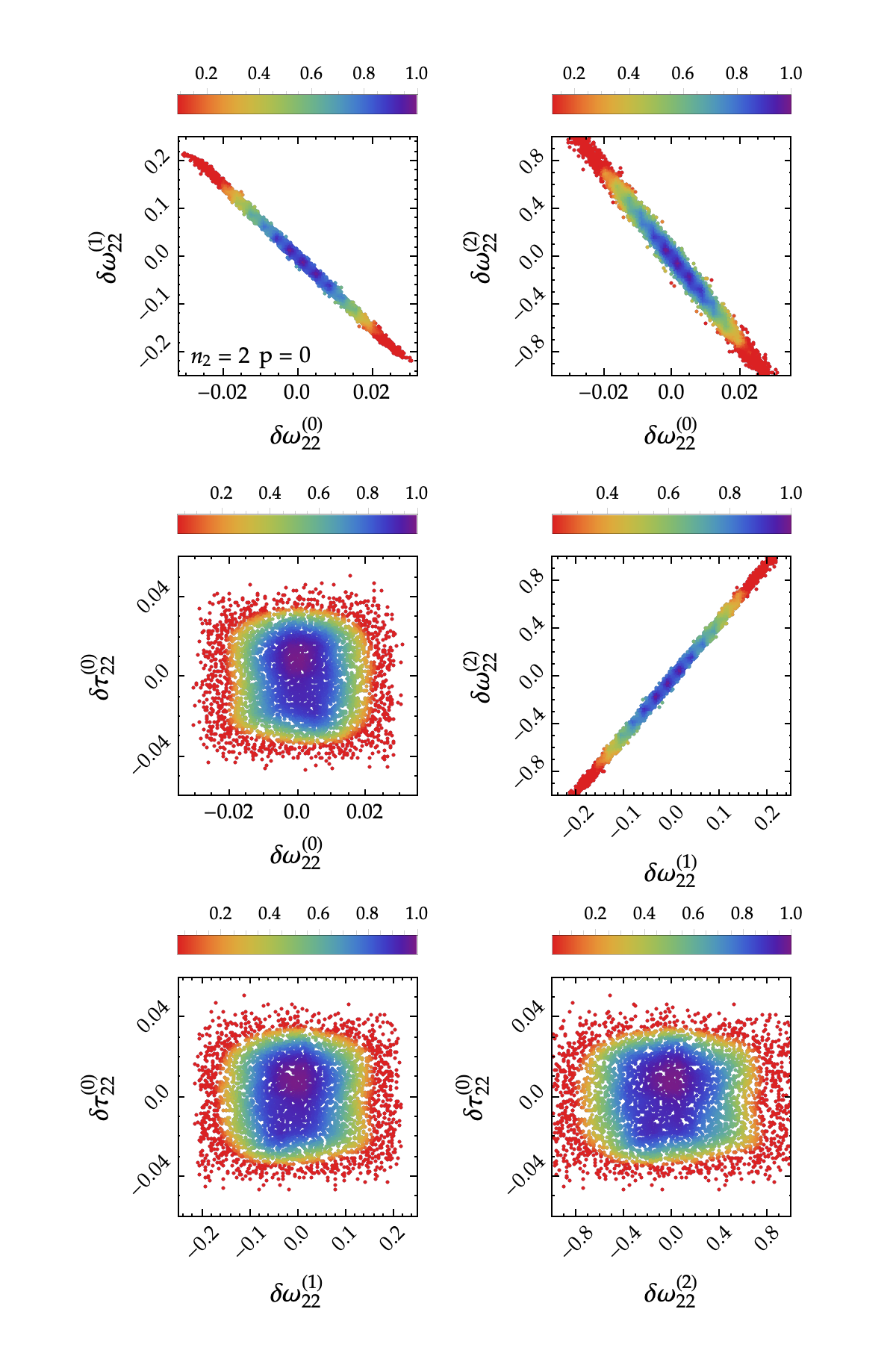}
    \caption{Joint posterior samples for the \PS parameters in the agnostic analysis using strategy (i), including quadratic spin corrections ($n_2=2$) and dimensionless coupling ($p=0$). The color scheme identifies the probability of the samples, with red (blue) regions corresponding to low (high) probabilities. We stack the full set of events observable by ET and assume the spin distribution of \texttt{model I}, but results for \texttt{model II} are qualitatively similar.} %
    \label{fig:jp0spin0pdf}
\end{figure} 

\noindent
{\bf \em Spin-dependent corrections ($n_2=2$), dimensionless coupling ($p=0$).}
We now augment the \PS template by including spin corrections up to second order in $\chi_f$, i.e., we set $n_2=2$ in Eqs.~\eqref{eq:parspec-expa1}--\eqref{eq:parspec-expa2}. In Fig.~\ref{fig:jp0spin0pdf} we show the 68\% and 90\% credible regions of the joint posteriors of the beyond-GR parameters that are constrained by the data. 
For brevity we focus on ET, and we stack the full set of detectable BH events. The results in Fig.~\ref{fig:jp0spin0pdf} confirm that there is a definite hierarchy among the \PS parameters: the posterior distributions of  $\delta \omega_{22}^{(0)},\delta\omega_{22}^{(1)}$ and $\delta \omega_{22}^{(2)}$ are informative with respect to the prior support, while all spinning corrections to the damping time (with the exception of $\delta \tau^{(0)}_{22}$) are unconstrained. 
The coefficients that control the frequency deviations are strongly correlated with each other, while they are almost uncorrelated with $\delta \tau_{22}^{(0)}$. 

These broad conclusions hold regardless of the detector and population models, as shown by the first two columns of Table~\ref{tab:CLp0spin2}, where we list the 90\% credible intervals of the parameters for both ET and CE. 
\texttt{Model II} yields in general looser constraints on all the \PS coefficients, because the typical SNRs and the number of stacked events are lower than those for \texttt{model I}.  
While the nonrotating corrections $(\delta \omega_{22}^{(0)},\delta \tau_{22}^{(0)})$ are constrained with the tightest accuracy, their bounds increase by more than two orders of magnitude compared to the case shown in Fig.~\ref{fig:p0spin0pdf}, due to correlations among the spin-dependent \PS coefficients.

We have also studied how the constraints change with the total number of stacked ringdown observations. In Table~\ref{tab:CLp0spin2} we show the $90\%$ credible levels for $N=1000$ and $N=500$ events. The bounds on the parameters broaden (as expected) as $N$ decreases,
and the posterior on some of the frequency modifications becomes uninformative, or almost as large as the prior.
\begin{table*}
\centering
\begin{tabular}{cc|cc|cc|cc}
\hline
\hline
&  & \texttt{model I} & \texttt{model II} & \texttt{model I} & \texttt{model II} & \texttt{model I} & \texttt{model II} \\
&  & all & all & $N=1000$ & $N=1000$ & $N=500$ & $N=500$ \\
\hline
CE-40km & $\delta \omega_{22}^{(0)} $ &[-0.0234, 0.0237] &[-0.0341, 0.0363] &[-0.0282, 0.0273] 
&[-0.0485, 0.0509] &[-0.0312, 0.0312] &  [-0.0602, 0.0586] \\
& $\delta \omega_{22}^{(1)} $ &[-0.178, 0.175] &[-0.222,0.213] &[-0.193, 0.196] &[-0.257, 0.246] &[-0.204, 0.205] & [-0.276, 0.277] \\
& $\delta \omega_{22}^{(2)} $ &[-0.853, 0.834] &[-0.978, 0.780] & - & - & - & - \\
&$ \delta \tau_{22}^{(0)} $ &[-0.0343, 0.0337] &[-0.0667,  0.0709] &[-0.0534, 0.0529] &[-0.210, 0.201] 
&[-0.0725, 0.0746] &  [-0.279, 0.277] \\
\hline
\hline
ET-15km &  $\delta\omega _ {22}^{(0)} $ &[-0.0217, 0.0225] &[-0.034, 
        0.0366] &[-0.0266, 0.0281] &[-0.0448, 0.0464] &[-0.0311,   0.0316] &
   [-0.0541, 0.057] \\
&   $\delta\omega _ {22}^{(1)} $ &[-0.169, 0.161] 
&[-0.229, 0.212] &[-0.197, 0.188] &[-0.255, 0.243] &[-0.204, 0.203] & [-0.273, 0.258] \\
& $\delta\omega _ {22}^{(2)} $ &[-0.807, 0.771] &[-1.00, 0.741] & - & - & - & - \\
& $\delta\tau _ {22}^{(0)} $ &[-0.0323, 0.0324] &[-0.0561,0.0569] &[-0.0526, 0.0508] 
&[-0.169, 0.159] &[-0.0751,   0.0742] & [-0.25, 0.249] \\
\hline
\hline
\end{tabular}
\caption{We list the 90\% credible intervals for the marginalized distribution of the \PS parameters with quadratic spin corrections ($n_2=2$) and dimensionless coupling ($p=0$). We stack a different number $N$ of  sources for the two models. Dashed entries correspond to parameters for which the posterior is not informative with respect to the prior.}
\label{tab:CLp0spin2}
\end{table*}

\begin{figure}[t]%
    \includegraphics[width=4.2cm]{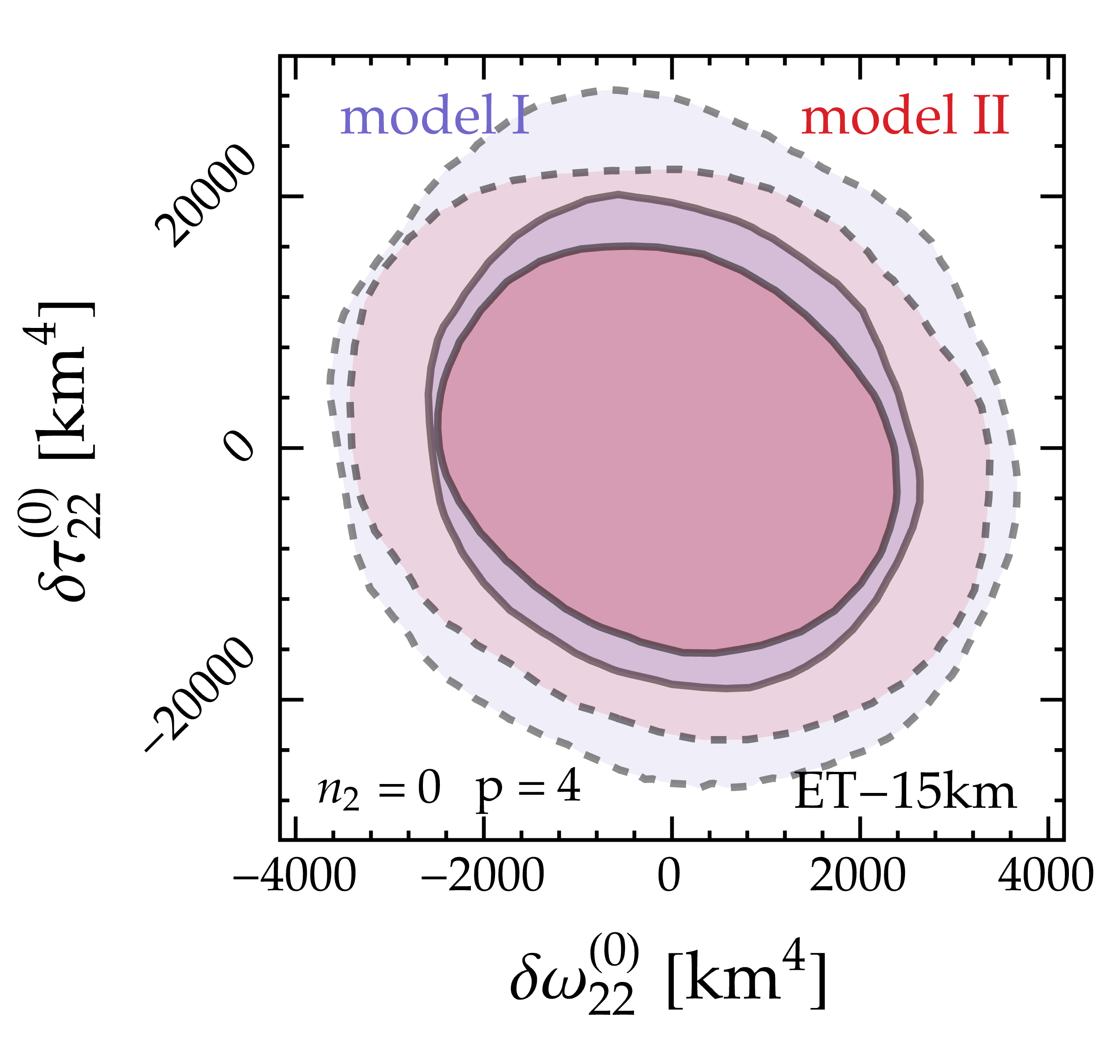}
    \includegraphics[width=4.2cm]{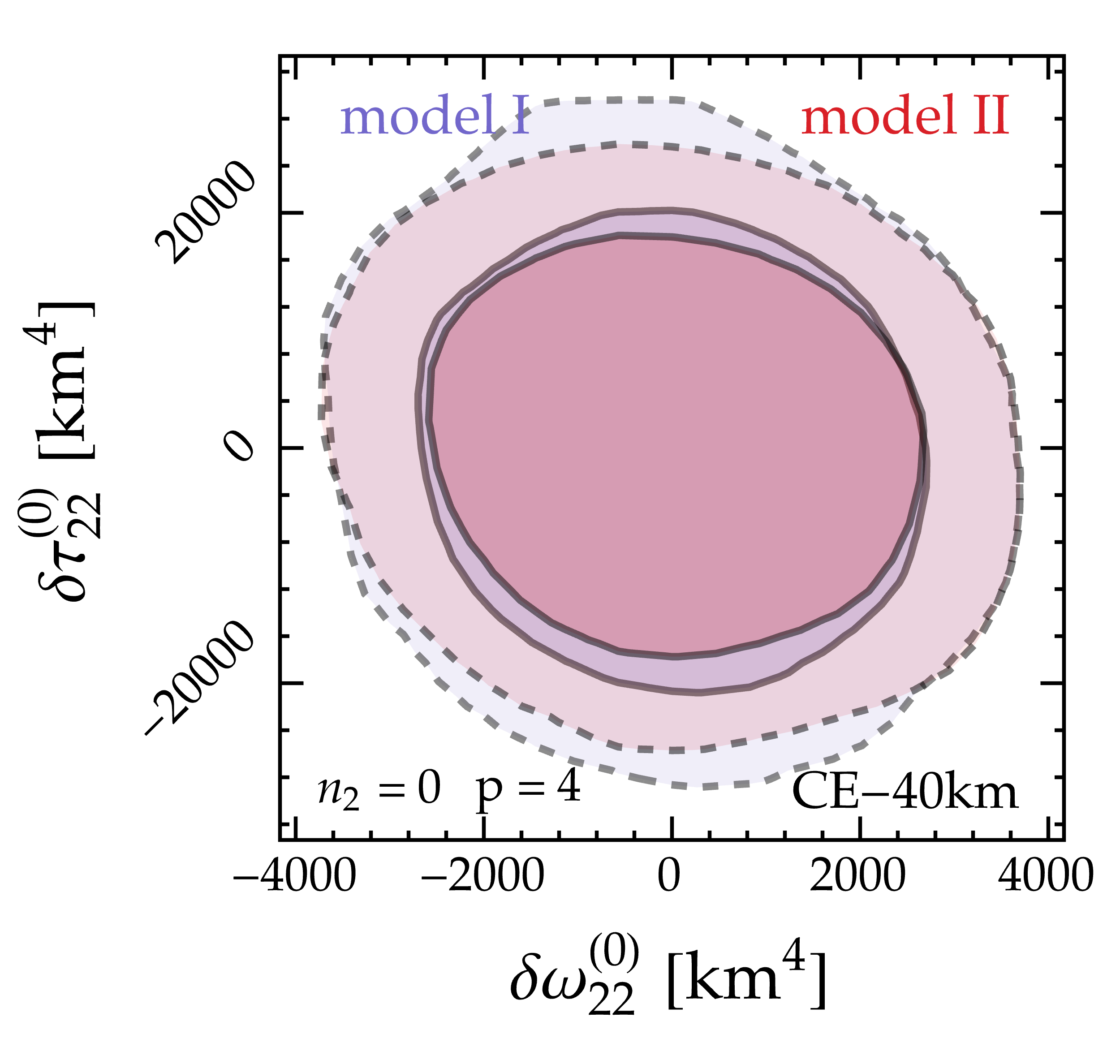}
    \caption{2D posterior distribution of the spin-independent \PS parameters $(\delta\omega_{22}^{(0)},\delta\tau_{22}^{(0)})$, assuming that Kerr deviations are controlled by a dimensionful coupling with $p=4$. We perform an agnostic analysis using strategy (i). Solid and dashed lines identify the $68\%$ and $90\%$  credible intervals, while light-purple and red regions correspond to different population models.}
    \label{fig:jp4spin0pdf}
\end{figure} 

\begin{figure*}[t]%
    \includegraphics[width=18cm]{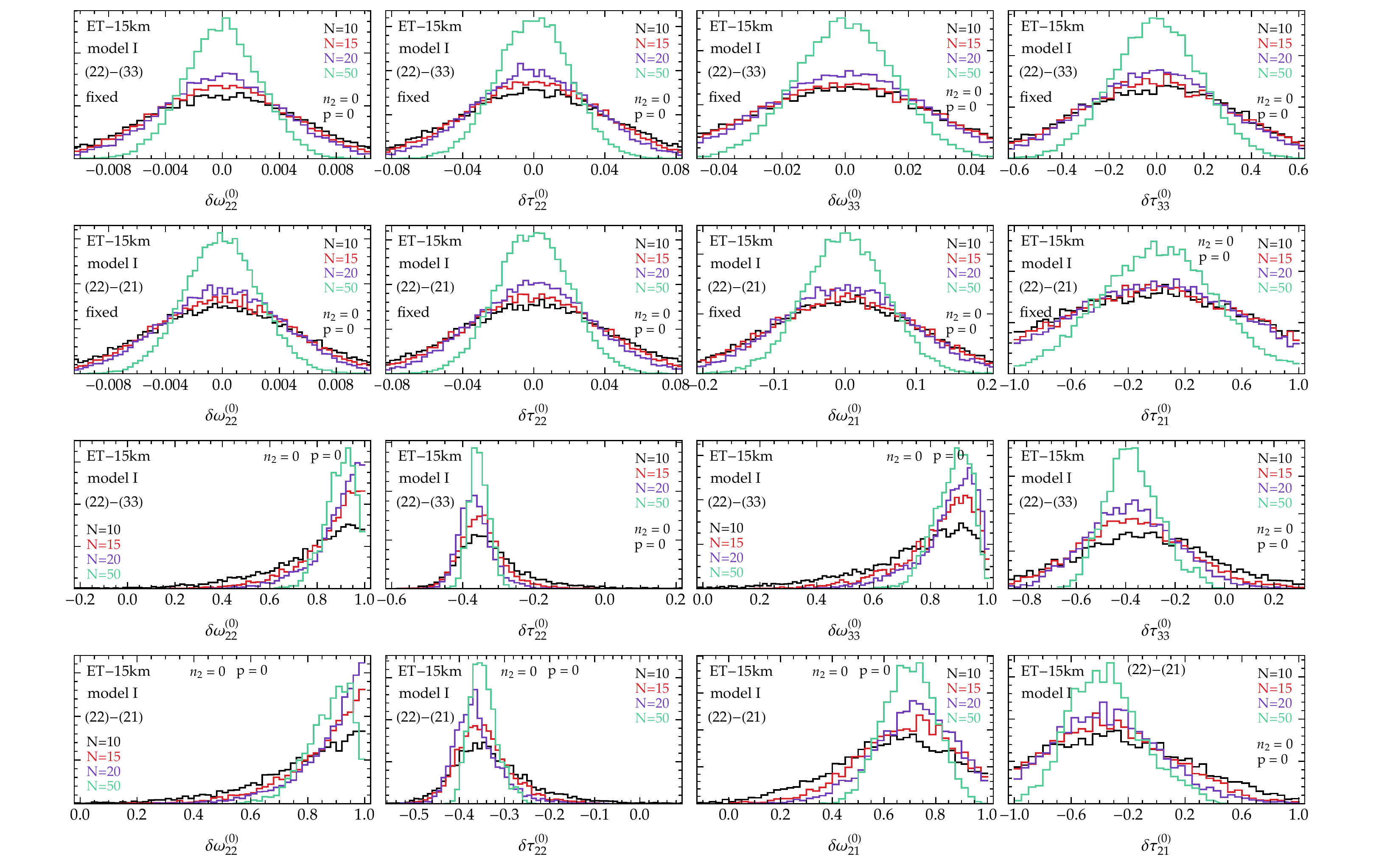}
    \caption{Distribution of the \PS  deviation parameters as a function of the number of observed events in the agnostic analysis. We assume a two-mode template which includes the (22) and either the (33) or the (21) QNM, and we use the \texttt{model I}  population. The top two rows (labeled by ``fixed'') correspond to posteriors obtained by keeping the BH masses and spins fixed to their injected values in the MCMC. The bottom rows are derived by sampling the \PS parameters as well as the mass and spin of the remnant.}  
    \label{fig:p0MChi}
\end{figure*} 

\noindent
{\bf \em Dimensionful coupling ($p=4$).}
We now consider agnostic tests of GR for  a coupling constant with dimensions $({\rm mass})^4$, i.e., $p=4$ in Eq.~\eqref{eq:parspeccoupl}.
In Fig.~\ref{fig:jp4spin0pdf} we show the joint posterior credible intervals for the nonspinning \PS parameters  $(\delta \omega_{22}^{(0)},\delta \tau_{22}^{(0)})$, computed by assuming $p=4$ in Eqs.~\eqref{eq:parspec-expa1} and \eqref{eq:parspec-expa2}. 
The results are rather insensitive to the chosen spin distribution model, and they mostly depend on the mass distribution of the remnant. Since the QNM corrections are (to a first approximation) proportional to the coupling \eqref{eq:parspeccoupl}, which scales as $\sim M_f^{-4}$, they are strongly suppressed for heavy BHs, so the constraints shown in Fig.~\ref{fig:jp4spin0pdf} are dominated by the $\sim 400$  ($\sim 200$) ET (CE) events having $M_f\lesssim 50M_\odot$. As a rough estimate, our results suggest that ET and CE may set bounds on the fundamental coupling of a putative beyond-GR gravity theory of the order $\alpha\sim[\delta\omega_{22}^{(0)}]^{1/4}\sim\mathcal{O}(10\,\tn{km})$, at 90\% credible level. 

We have also studied how the constraints are affected by the inclusion of spin-dependent terms in the \PS template. We find that the bounds on the spin-dependent corrections are not informative with respect to the prior, and that the credible intervals on the nonspinning terms $(\delta\omega_{22}^{(0)},\delta\tau_{22}^{(0)})$ deteriorate by about an order of magnitude, possibly violating the small-coupling assumption implicit in the \PS approach.

\subsubsection{Strategy (ii): Mass and spin are unknown}
\label{sec:res2}

\noindent
{\bf \em No spin-dependent corrections ($n_2=0$), dimensionless coupling ($p=0$).}
So far we have investigated the ability of XG detectors to constrain beyond-Kerr deviations, assuming that the remnant BH mass and spin are known (e.g, because they can be inferred from the full inspiral-merger-ringdown waveform). This procedure (that we called ``strategy (i)'' in Sec.~\ref{sec:parspec}) drastically improves the bounds, because the observation of a single QNM -- i.e., the dominant $(22)$ mode -- is sufficient to perform BH spectroscopy tests. 
We now turn to the more conservative and pessimistic approach (called ``strategy (ii)'' in Sec.~\ref{sec:parspec}) in which we use information from the ringdown only. We now need at least {\it two modes} in order to simultaneously constrain the \PS parameters, as well as the final BH mass and spin of each observed ringdown event. 

For reasons that will be apparent soon, we perform the analysis only for dimensionless couplings ($p=0$) and for nonrotating corrections ($n_2=0$). We analyze the $(22)$ mode in combination with either a $(33)$ or a $(21)$ secondary QNM.

In Fig.~\ref{fig:p0MChi} we show the posterior distributions of the \PS parameters as functions of the observed events for \texttt{model I} and \texttt{model II}, respectively.  Different histograms correspond to posteriors inferred by stacking a different number of observations.  In the first two rows of Fig.~\ref{fig:p0MChi} we assume that $M_f$ and $\chi_f$ are \textit{fixed} to their true values.  While in this case the posteriors of $\delta\omega_{22}^{(0)}$ and $\delta\tau_{22}^{(0)}$ are correctly centered around zero, and become narrower as $N$ increases, the panels in the third and fourth rows, where $M_f$ and $\chi_f$ are allowed to vary, change dramatically, and we observe a \textit{significant bias} in all parameters. In the majority of cases, the GR null hypothesis is excluded at more than 90\% of the marginalized posterior distributions.
Both the fundamental and the secondary mode are affected: their probability distribution can peak around the wrong value, and rally against the prior.
These results are independent of the choice of the secondary mode.

The systematic bias builds up due to correlations with $M_f$ and $\chi_f$, which are partially degenerate with the \PS parameters, and it can become more significant as $N$ grows. Note that for practical purposes here we consider only 50 events: this already requires more than 100 parameters -- i.e., $2\times 50$ masses and spins -- to be sampled along with the $4$ \PS parameters for the fundamental and secondary QNM.

In order to clarify the impact of the remnant BH properties on the reconstruction of $\delta\omega^{(0)}_{\ell m}$ and $\delta\tau^{(0)}_{\ell m}$, we repeat the analysis, but this time we replace the ``true'' mass and spin by ``mock'' values that are shifted by some fixed amount:
\begin{equation}
 \bar{M}_f=M_f+\Delta M_f\quad\ ,\quad  \bar{\chi}_f=\chi_f+\Delta \chi_f\ .
\end{equation}
For simplicity, we assume $\Delta M_f/M_f=\Delta \chi_f/\chi_f=\epsilon$.
We then sample over the \PS parameters only.

\begin{figure}[t]%
    \includegraphics[width=8cm]{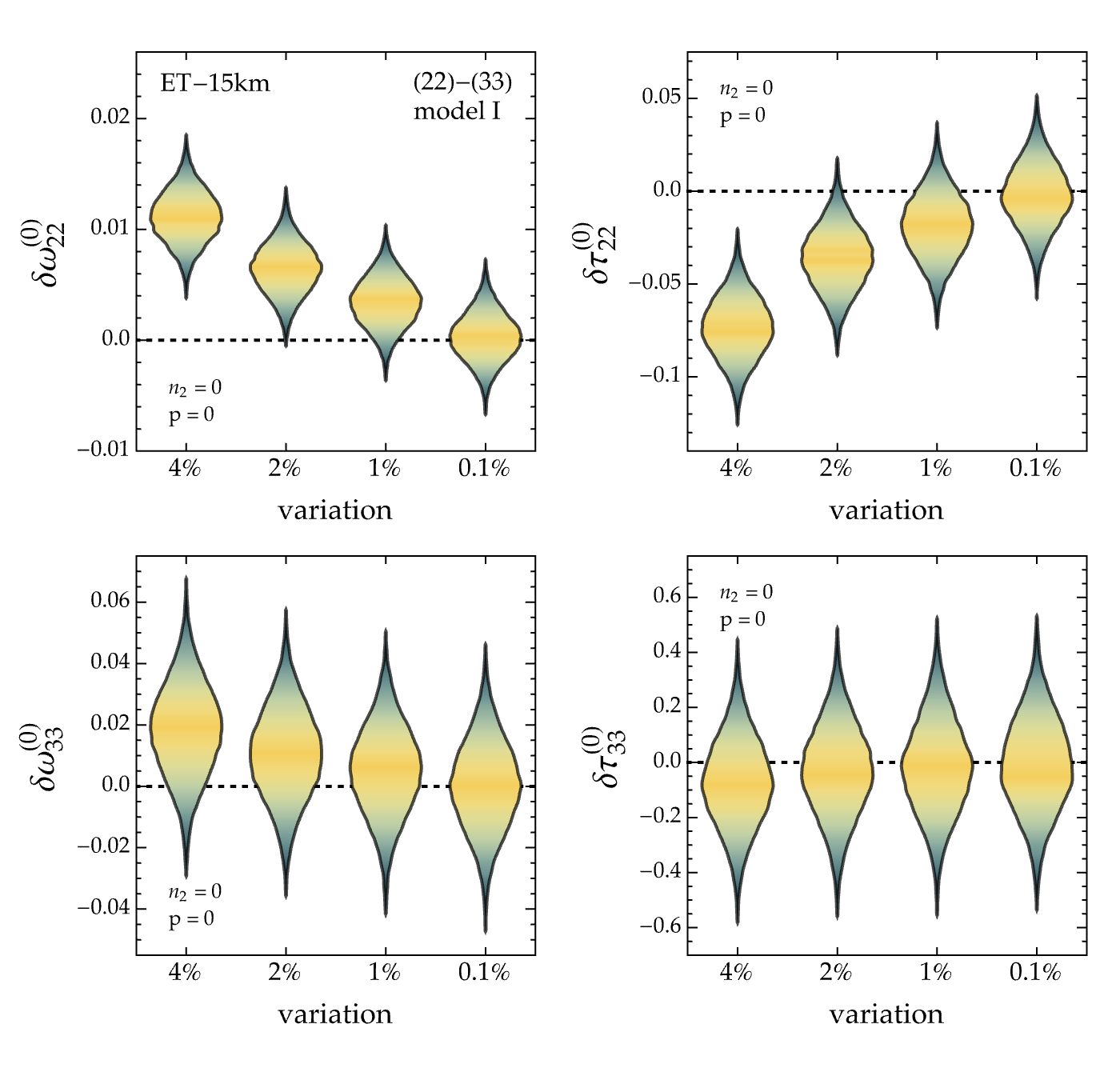}
    \caption{Posterior density distributions of the \PS parameters for the $p=0$ case, inferred in the agnostic analysis with strategy (ii), assuming that the BH masses and spins, fixed within the MCMC simulation, are shifted by a fixed amount (indicated on the x-axis) with respect to their ``true'' values. Colors closer to yellow (green) correspond to higher (lower) probability. Results are obtained by stacking $N=50$ events of the \texttt{model I} catalog.}
    \label{fig:Mchivar}
\end{figure}

In Fig.~\ref{fig:Mchivar} we show the posterior distributions inferred with the $(22)-(33)$ multi-mode analysis for ET and \texttt{model I}. We stack $N=50$ events and we consider values of $\epsilon$ ranging between $0.001$ and $0.04$.
The probability densities drawn in the four panels demonstrate how the bias evolves as a function of the relative shift in mass and spin. An accuracy of $\sim 1\%$ in the masses and spins seems to be required in order for the GR limit to return within the posterior support. 

The masses and spins of the remnant inferred from XG ringdown observations {\it alone} are expected to have $\sim 10\%$ accuracy~\cite{Bhagwat:2023jwv}, about an order of magnitude worse than the $\sim1\%$ accuracy estimated above. However, the uncertainty in the remnant's properties could be improved using different data analysis techniques (for example, by using the full inspiral-merger-ringdown information).

The origin of this bias can be traced back to the degeneracy between $(M_f,\chi_f)$ and the fundamental mode frequency and damping time~\cite{Pacilio:2023mvk}. In the case of the \PS  expansion, for small changes $\Delta M_f$, $\Delta \chi_f$, this degeneracy leads to a shift in the beyond-Kerr parameters of the form:
\begin{align}
    \delta \omega_{\ell_j m_j}^{(k_2)}\rightarrow 
\delta\omega_{\ell_j m_j}^{(k_2)}-\Delta M_f+
k_2 \Delta \chi_f\ ,\label{eq:scal1} \\
    \delta  \tau_{\ell_j m_j}^{(k_2)}\rightarrow 
\delta  \tau_{\ell_j m_j}^{(k_2)}+\Delta M_f+
k_2 \Delta \chi_f\ ,\label{eq:scal2}
\end{align}
Equations~\eqref{eq:scal1} and \eqref{eq:scal2} suggest that to avoid large systematics in the inference of $\delta \omega_{\ell_j m_j}^{(k_2)}$ and $\delta  \tau_{\ell_j m_j}^{(k_2)}$, the deviations in the mass and spin should be small enough to satisfy the condition
\begin{align}
    \vert\delta\omega_{\ell_j m_j}^{(k_2)}\vert\gg \vert\Delta M_f-
k_2 \Delta \chi_f\vert\ ,\label{eq:shiftcond1}\\
    \vert\delta  \tau_{\ell_j m_j}^{(k_2)}\vert\gg \vert\Delta M_f+k_2 \Delta \chi_f\vert\ .\label{eq:shiftcond2}
\end{align}

Note that the two-mode analysis discussed above focuses on the $p=0$ case. In general, we expect the bias to be worse for dimensionful corrections (e.g., $p=4$). This is because the mismodeling of the mass would also affect the coupling parameter $\gamma_i$, which also controls deviations from the Kerr spectrum. For this reason we do not show results for the case $p=4$, and we do not include spin-dependent corrections ($n_2>0$) in this part of the analysis.

\subsection{Theory-specific inference}\label{sec:theoryspec}

We first focus on the quadratic theories of gravity discussed in Sec.~\ref{sec:gravityth}. Both theories (EsGB and dCS) are characterized by a dimensionful coupling constant, bounded by astrophysical observations either in the gravitational or electromagnetic band. The maximum values currently allowed for the parameters $\alpha_\tn{GB,CS}$ are $\sqrt{\alpha_\tn{GB}}\simeq 6.3$~km and $\sqrt{\alpha_\tn{CS}}\simeq 8.5$~km~\cite{Lyu:2022gdr} (due to different normalizations, the EsGB coupling constant in this work is larger than the one defined in Ref.~\cite{Lyu:2022gdr} by a multiplicative factor of $4\,\pi^{1/4}$). Note however for such values of the couplings, deviations from GR can become non-perturbative, with $\beta_\tn{GB,CS}=\alpha_\tn{GB,CS}/m^2_{1,2}\sim 1$ for low-mass BHs of our catalogues.
This is problematic because both EsGB and dCS gravity should be seen as effective field theory, valid in the limit $\beta_\tn{GB,CS}\ll1$.
To be consistent with the perturbative character of \PS, in this work we inject signals for both EsGB and dCS gravity assuming a coupling such that $\beta_\tn{GB,CS}=\alpha_\tn{GB,CS}/\min[m_1,\,m_2]^2=0.1$, 
with $\min[m_1,\,m_2]$ being the smallest mass of the 
binary BHs within our population models that yield an observable ringdown signal. For both \texttt{model I} and \texttt{model I}, this leads to $\sqrt{\alpha_\tn{GB,CS}}\sim2.5$ km.

\begin{figure}[t]%
    \includegraphics[width=8.5cm]{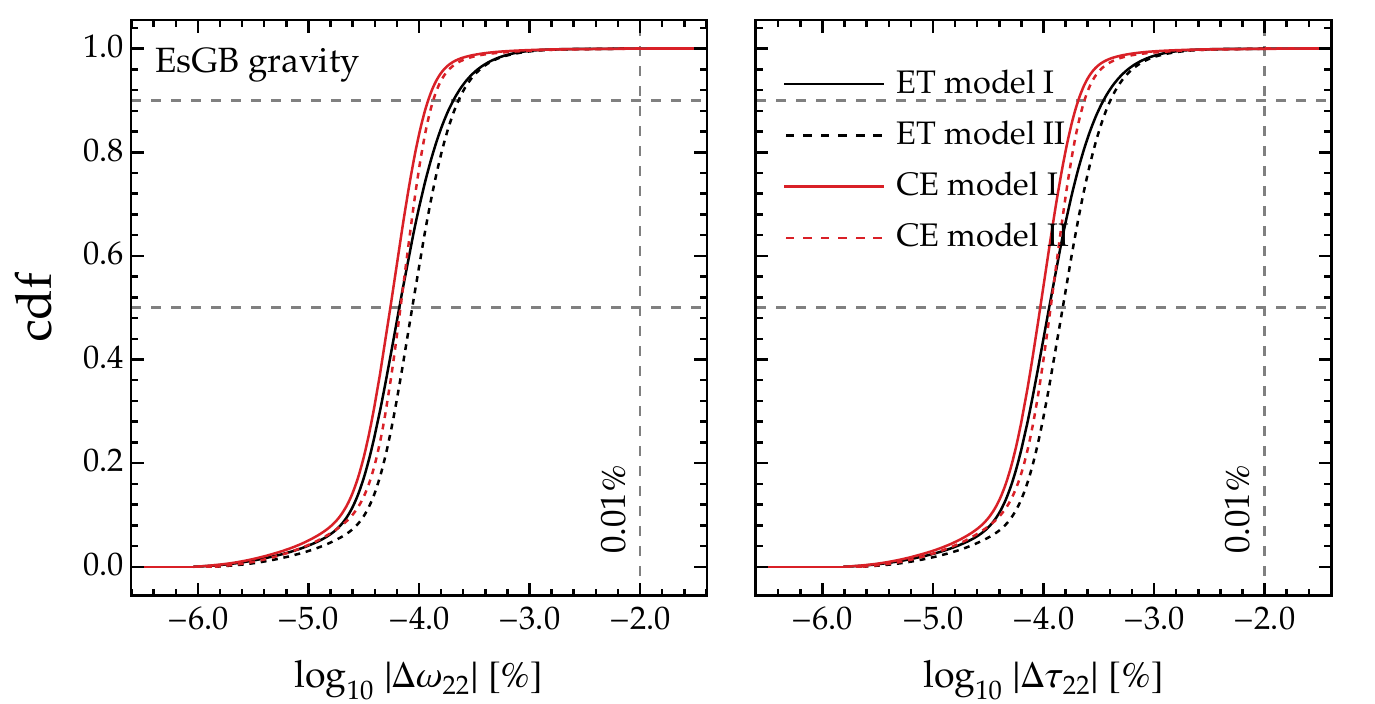}
    \includegraphics[width=8.5cm]{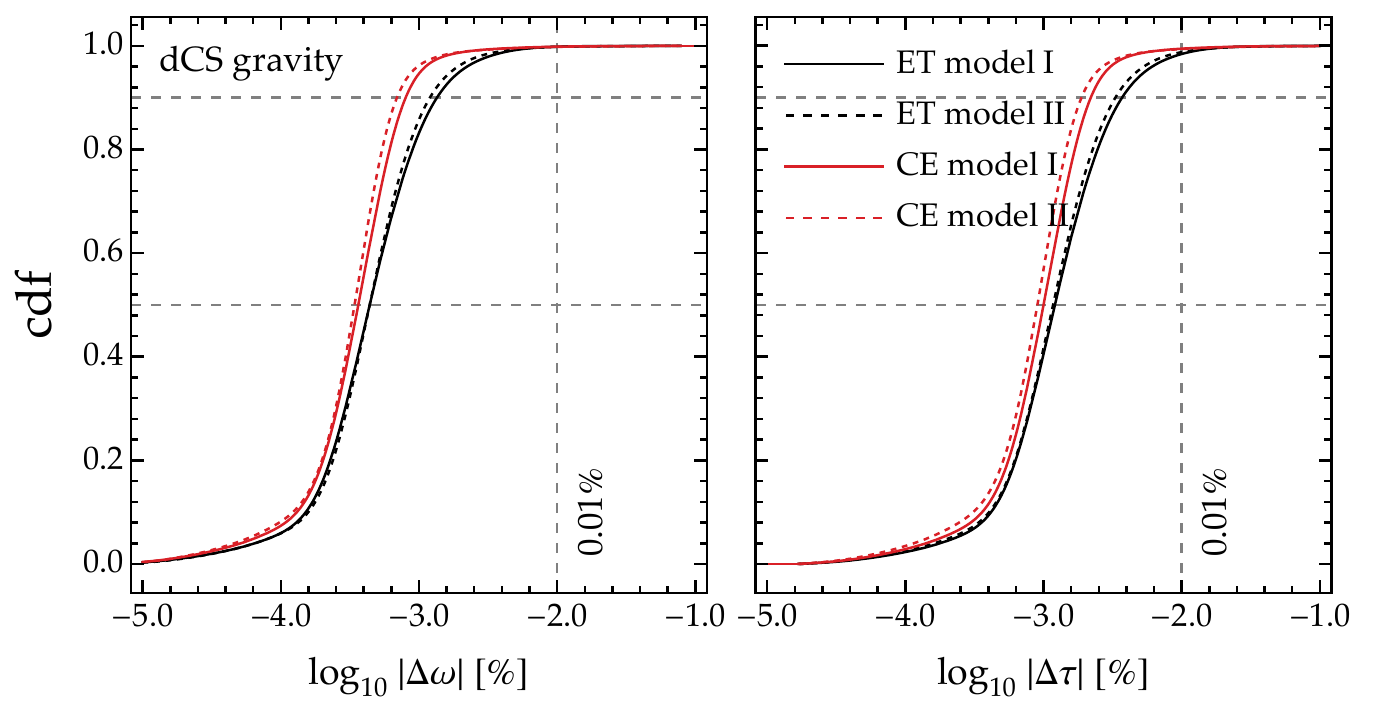}
    \caption{Cumulative distribution for the changes in the frequencies and damping times of the $(22)$ mode with respect to their Kerr values predicted by EsGB gravity (top row) and dCS gravity (bottom row), for detectable events. QNMs beyond GR are computed by assuming coupling constants $\alpha_{\tn{GB},\tn{CS}}$ that saturate the current upper bounds from astrophysical observations~\cite{Lyu:2022gdr}. The vertical lines mark $0.01\%$ changes in frequency and damping time, while the horizontal lines mark the $50\%$ and $90\%$ confidence intervals of the distribution.}
    \label{fig:alphaGBpop}
\end{figure} 

Before assessing the capability of XG detectors to constrain the coupling parameter, it is instructive to estimate the order of magnitude of the expected deviations from the Kerr predictions for the frequencies and damping times.
In Fig.~\ref{fig:alphaGBpop} we show the cumulative distributions of the changes in the $(22)$ mode predicted by EsGB gravity (top row) and dCS gravity (bottom row) relative to the (slowly rotating) Kerr predictions.
These corrections are computed using the ``beyond GR'' part of the sum in Eqs.~\ref{eq:parspec-expa1} and \ref{eq:parspec-expa2}, truncated at the order in the slow-rotation expansion at which the corrections are currently known: second order for EsGB and first order for dCS (see Appendix~\ref{app:qnmtheories}).
The cumulative distribution function is computed by considering all of the events that have an observable ringdown signal.
For completeness, in Figs.~\ref{fig:app_EsGB} and \ref{fig:app_dCS} of Appendix~\ref{app:qnmtheories} we compare the relative changes in the dominant mode to the relative changes in the subdominant $(33)$ and $(21)$ mode frequencies.

For EsGB gravity, all systems lead to changes in the frequency smaller than $\sim0.01\%$, regardless of the adopted spin distribution model.
Events observed by CE have slightly  
smaller corrections compared to ET, because the catalog observed by CE includes a smaller number of 
light BHs.
The maximum expected shifts in the damping times are of the same order of magnitude as the frequency shifts.

\begin{figure}[t]%
    \includegraphics[width=8cm]{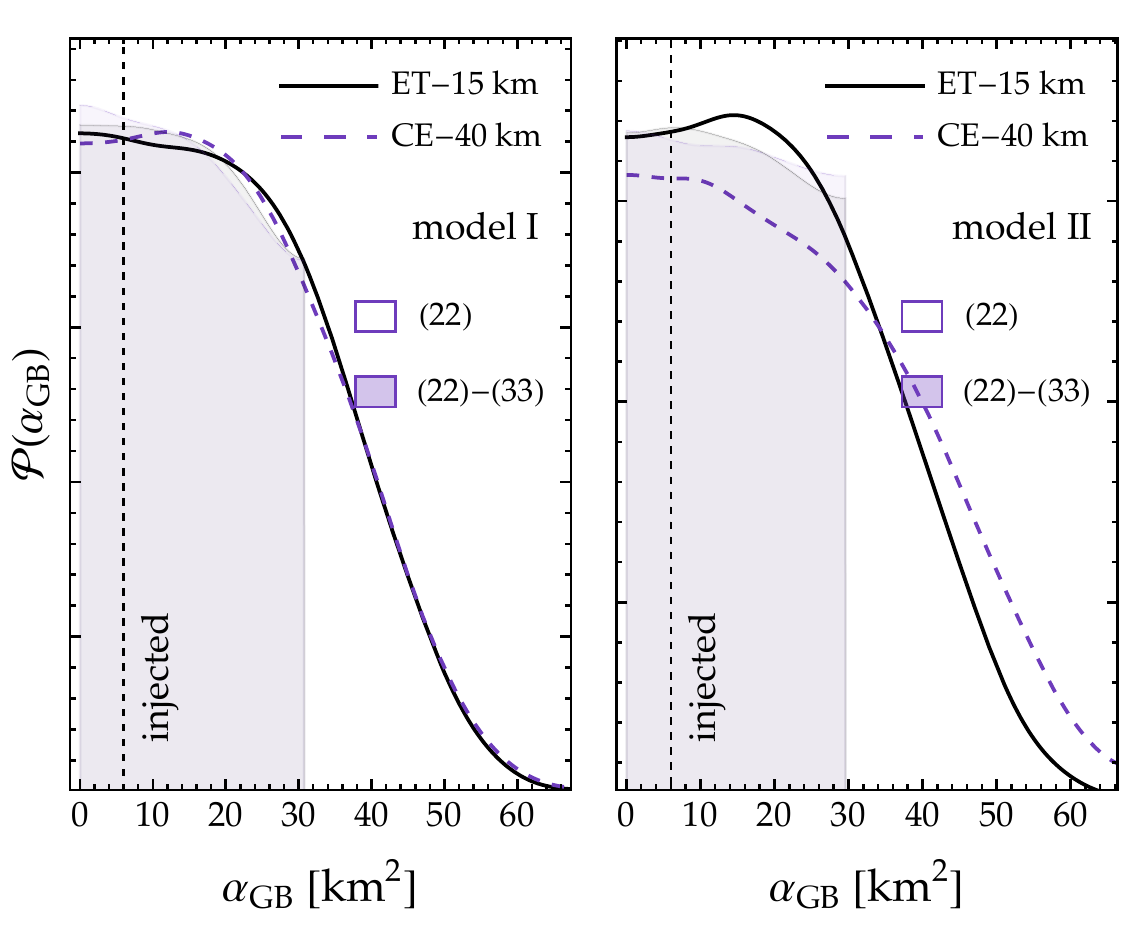}    
    \caption{Posterior distributions of the EsGB coupling constant $\alpha_\tn{GB}$. Bounds on $\alpha_\tn{GB}$ are obtained by stacking the whole set of observations for ET and CE, and by fixing both the masses and the spins of the BH remnants. The vertical dashed lines mark the injected value of the coupling. Unshaded (shaded) regions are the bounds from a one-mode (two-mode) analysis, respectively.}
    \label{fig:GBbounds}
\end{figure} 

For dCS gravity, deviations with respect to Kerr are generally larger by one order of magnitude.  Interestingly, the shifts in the damping time are bigger than the frequency shifts.

To understand how these estimates translate into actual constraints on the coupling constants, we first run our MCMC pipeline assuming that the mass and spin of the remnants are fixed to their injected values. We consider a single-mode and a two-mode analysis.  In the two-mode analysis we only use the frequency (and not the damping time) of the secondary QNM. The main reason behind this choice is that the damping time of the secondary mode is never well measured.  

\subsubsection{Einstein-scalar-Gauss-Bonnet gravity}\label{sec:gbresults}

In Fig.~\ref{fig:GBbounds} we plot the posterior of the EsGB coupling constant $\alpha_\tn{GB}$ inferred by stacking the entire set of events observed by ET and CE. Unshaded and shaded curves correspond to the $(22)$ and to the $(22)$-$(33)$ templates, respectively. In both cases our results do no show any particular preference for the injected value of the coupling: the injected value does fall within the 90\% credible intervals of the distributions, but the posteriors are fully consistent with the GR hypothesis, and only allow to infer an upper bound on $\alpha_\tn{GB}$. As expected, ET provides tighter constraints, mainly due to the larger number of observable ringdown events.
The posteriors in the $(22)$-$(33)$ case have a sharp cutoff around $\alpha_\tn{GB}\sim 30\tn{km}^2$. This cutoff is imposed by the requirement that the \PS parameters must be perturbative in nature, i.e., $\vert\gamma_i\delta\omega_{\ell m}^{(k_2)}\vert\lesssim1$ and $\vert\gamma_i\delta\tau_{\ell m}^{(k_2)}\vert\lesssim1$  (see also the discussion in Appendix~\ref{fig:app_EsGB}).  This choice imposes a tight upper bound on the frequency of the $(33)$ mode, because the coefficients of the dCS expansion (as shown in Appendix~\ref{app:qnmtheories}) are such that deviations from Kerr are larger as the BH mass decreases.  We have also run our Bayesian pipeline assuming the $(22)$-$(21)$ combination (not shown here): in this case the results are very close to the single-mode analysis, confirming that most of the information comes from the dominant $(22)$ multipole. The numerical values of the \PS coefficients for the $(21)$ QNM make it possible to sample a wider range of values for $\alpha_\tn{GB}$, so the $(22)-(21)$ posteriors do not have the sharp cutoff observed in the $(22)$-$(33)$ case.

\begin{figure}[t]%
    \includegraphics[width=8.5cm]{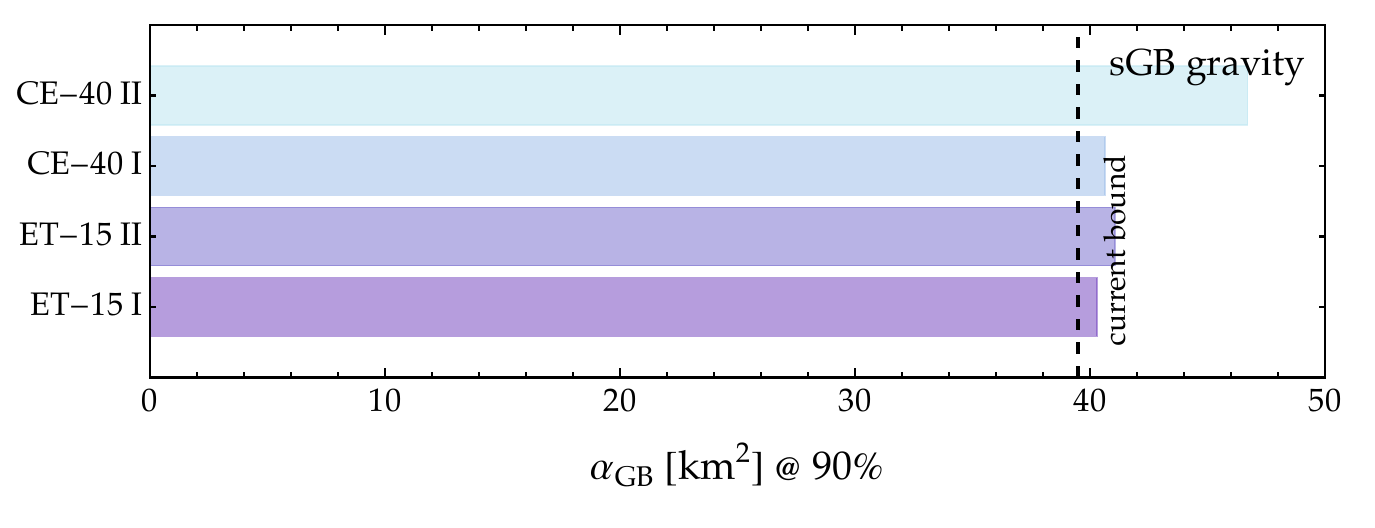}
    \includegraphics[width=8.5cm]{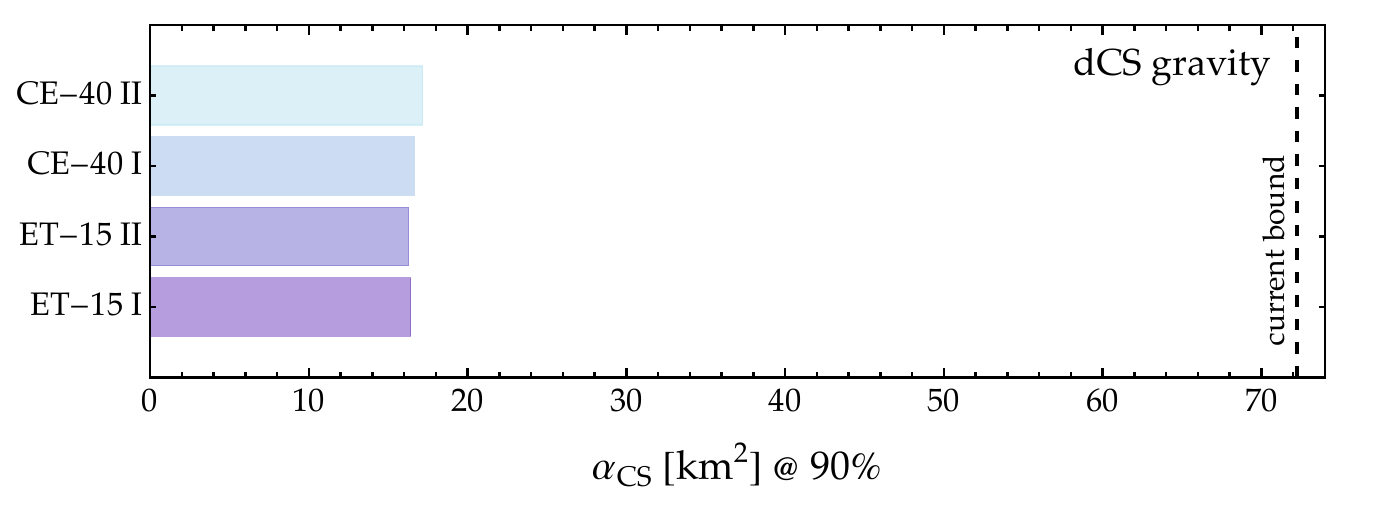}
    \caption{90\% credible intervals around the median for the coupling constants of EsGB and dCS gravity inferred using different combinations of detectors and population models. The bounds are obtained by stacking all events in a single-mode analysis, and assuming that the masses and spins of the remnants are fixed to their injected values. Vertical lines correspond to the current upper bound for $\alpha_\tn{GB,CS}$.}
    \label{fig:CSGBbounds}
\end{figure}

Our results are only mildly dependent on the chosen model for the binary spins.  Focusing on the single-mode strategy, the $90\%$ credible interval derived using ET is $\alpha_\tn{GB}\sim[0, 40]$\,km$^2$ and $\alpha_\tn{GB}\sim[0, 41]$\,km$^2$ for \texttt{model I} and \texttt{model II}, respectively.
The $90\%$ credible interval derived using CE
is $\alpha_\tn{GB}\sim[0, 41]$\,km$^2$ and $\alpha_\tn{GB}\sim[0, 47]$\,km$^2$ for \texttt{model I} and \texttt{model II}, respectively.
The top panel of Fig.~\ref{fig:CSGBbounds} summarizes the 90\% credible intervals on the EsGB coupling for different detector configurations and spin models. All constraints are close to the current upper bound for the EsGB coupling.
Given the mass dependence of the EsGB corrections (see Eq.~\eqref{eq:couplingGBCS}) we expect our results to be dominated by low-mass systems. As a check, we have repeated the analysis by considering only BH remnants with $M_f\lesssim 50M_\odot$, and the results are very similar to those shown in Fig.~\ref{fig:GBbounds}.

\begin{figure}[t]
    \includegraphics[width=8cm]{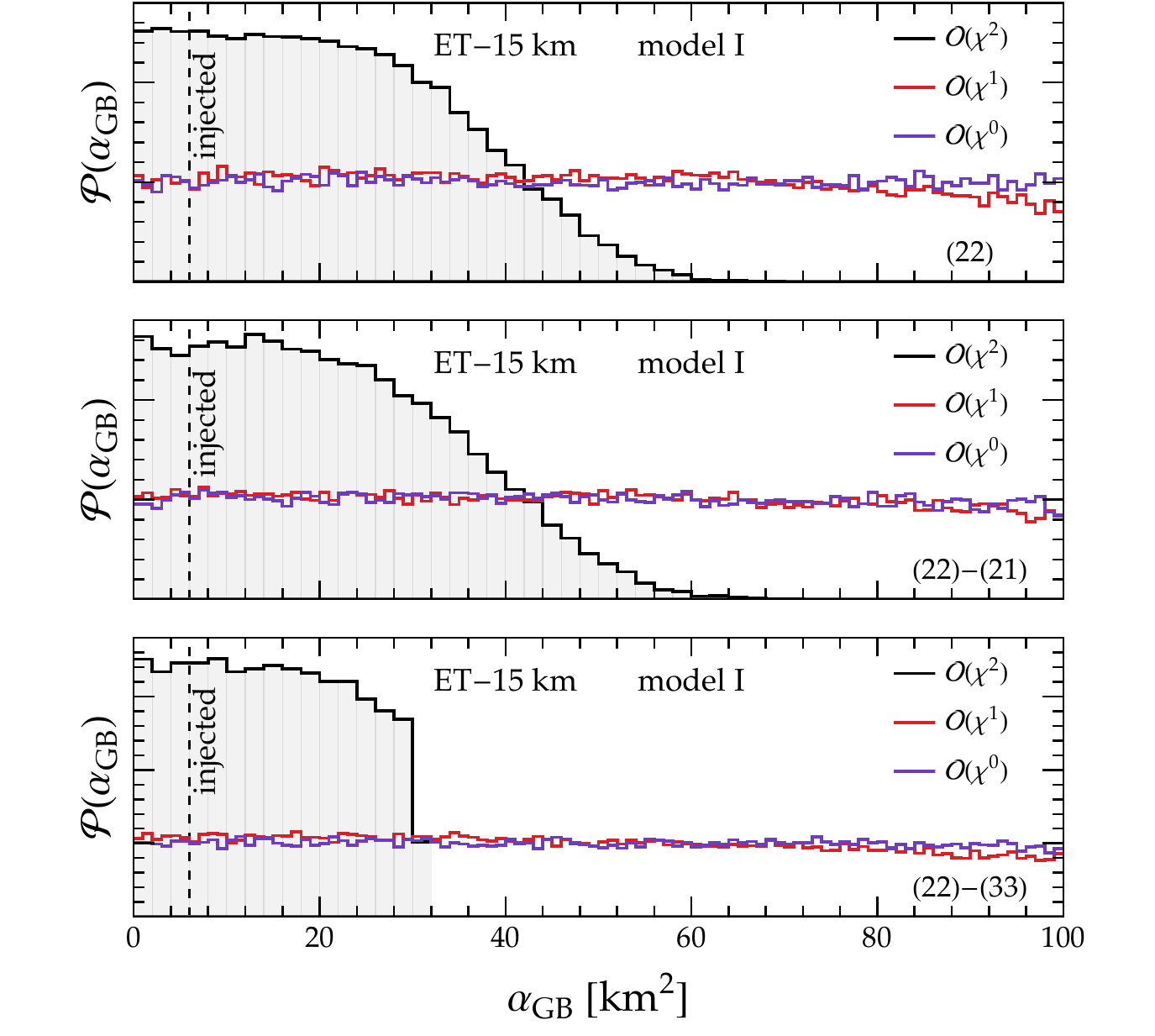}
    \caption{Posterior distribution of the EsGB coupling constant inferred from ET observations with a single-mode analysis. Samples are obtained by injecting QNM signals in EsGB gravity that include second-order corrections in the spin.  The black histogram corresponds to recovery templates which include $\mathcal{O}(\chi^2)$ spin terms in the beyond-Kerr parametrization. Red and purple histograms refer to a template that includes only nonrotating terms or first-order spin corrections, respectively.}
    \label{fig:GBbounds2}
\end{figure}

Spin-dependent corrections to the QNM frequencies in EsGB gravity are known up to quadratic order in the spin. In Fig.~\ref{fig:GBbounds2} we investigate how these corrections affect the posteriors on $\alpha_\tn{GB}$. For simplicity we focus on a single-mode analysis of ET observations.
We show the posterior probabilities obtained by fixing $n_2=0$ (purple), $n_2=1$ (red) and $n_2=2$ (black) in the recovering templates of Eqs.~\eqref{eq:parspec-expa1}--\eqref{eq:parspec-expa2}, while the injected QNM signals always include all corrections in the spin (up to and including the second-order terms).
Including only first-order or zeroth-order spin corrections in the template flattens the posterior distribution, which becomes uninformative with respect to the prior. Results for the two-mode analysis, assuming either the $(33)$ or the $(21)$ component as the secondary QNM, are qualitatively similar.
These results suggest that spin corrections play a dominant role in the Bayesian inference. This can be understood by looking at the magnitude of quadratic terms in the frequency expansion, which are about one order of magnitude larger than the zeroth-order term for $\chi_f\sim 0.7$ (see Appendix \ref{app:qnmtheories}, and in particular Fig.~\ref{fig:app_esgborder}). Therefore pushing the spin expansion to higher orders is very important to perform tests that are not dominated by systematic errors.

\begin{figure}[t]%
    \includegraphics[width=8.5cm]{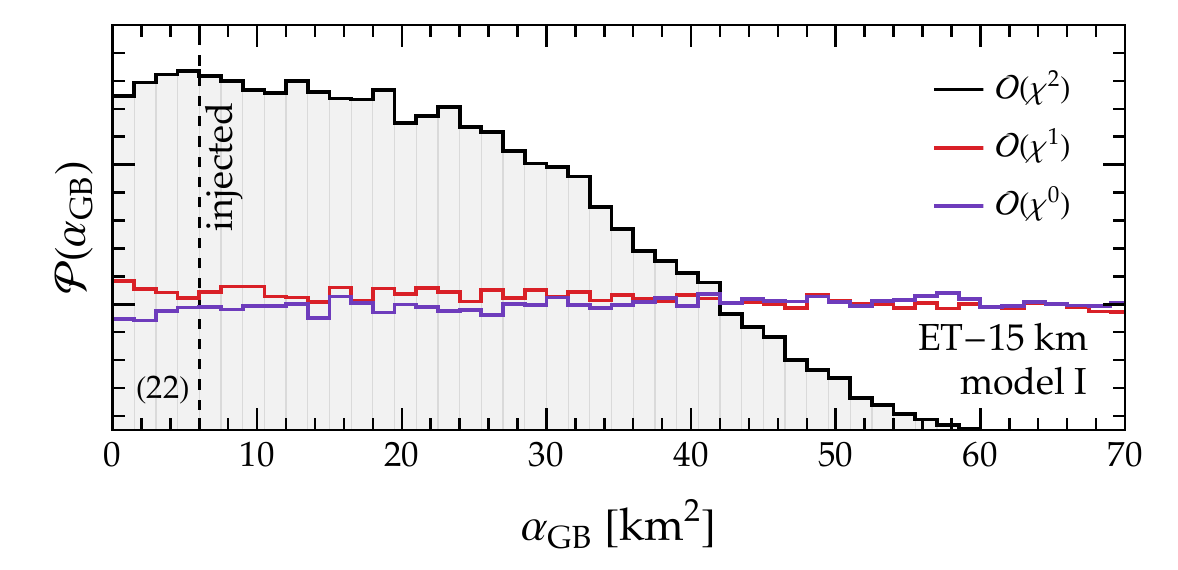}
    \caption{Posterior of the EsGB coupling constant for ET observations of BHs with $M_f\le 50M_\odot$. Black histograms are obtained by assuming that both the injected frequencies and the recovery template include quadratic beyond-GR corrections in the spin. Red and purple histograms are obtained by MCMC simulations in which both the injected frequencies and recovery templates contain only nonrotating terms or only first-order spin corrections, respectively.}
    \label{fig:GBbounds3}
\end{figure}

To further clarify the role of rotational corrections, we repeat the analysis by assuming that both the \textit{injected signal} and the \textit{recovery template} contain either nonrotating or linear-in-spin corrections beyond-Kerr corrections. We basically follow the procedure described in Sec.~\ref{sec:setup}, but now we compute FIM errors by omitting higher-order in spin corrections to the EsGB frequencies and damping times. The results are shown in Fig.~\ref{fig:GBbounds3} and support our previous analysis.  When we neglect quadratic terms in the spin the posterior distribution of the coupling is fully consistent with the uniform prior,
suggesting that $\mathcal{O}(\chi^2)$ terms dominate the reconstruction of $\alpha_\tn{GB}$. It is possible, and even likely, that bounds inferred by truncating at quadratic order are affected by a bias due to ignoring higher-order coefficients in the QNM expansion, which are currently unknown.
We remark that, even in the most optimistic case, the ringdown-only EsGB upper bounds of Fig.~\ref{fig:GBbounds} are weaker than the bounds expected from XG observations of binary inspirals, which could be as small as $\alpha_\tn{GB}\sim 10^{-3}$~km$^2$ in the absence of systematic errors (see e.g. Fig.~20 of Ref.~\cite{Perkins:2020tra}).

\begin{figure}[t]%
    \includegraphics[width=8.5cm]{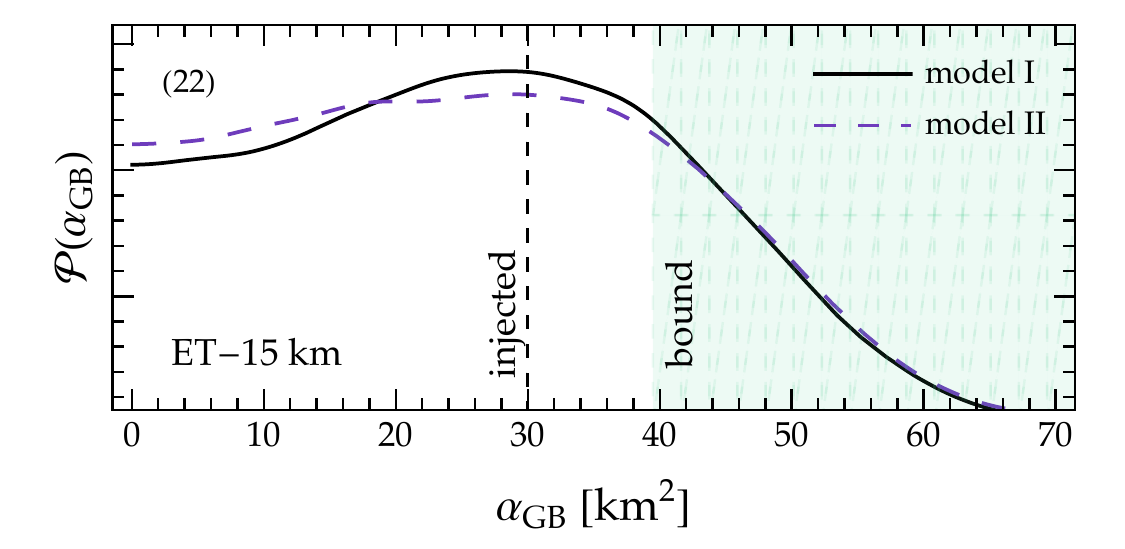}
    \caption{Posterior distributions of the EsGB coupling parameter found by stacking all events observed by ET with a single-mode strategy.  The injected signals assume a value of the coupling $\alpha_\tn{GB}/\min[m_1,\,m_2]^2=0.5$ that violates the perturbative assumption. The shaded region identifies the parameter space ruled out by astrophysical observations.}
    \label{fig:GBboundsmax}
\end{figure}

It is interesting to investigate what constraints can be placed on the EsGB coupling if we allow $\alpha_\tn{GB}$ to be nonperturbatively large.
In Fig.~\ref{fig:GBboundsmax} we show the posteriors of $\alpha_\tn{GB}$ found with ET by injecting signals such that $\beta_\tn{GB}=\alpha_\tn{GB}/\min[m_1,\,m_2]^2=0.5$, which yields $\alpha_\tn{GB}\simeq 30\ \tn{km}^2$. As expected, the detectability of beyond-Kerr deviations improves significantly: the distributions now have a clear peak around the injected value of the coupling. Note however that, even in this extreme case, the posteriors also support the GR hypothesis. Similar conclusions apply also to CE.

It is worth asking how the bounds would change if we allowed the BH masses and spins to be sampled together with $\alpha_\tn{GB}$. To limit the computational cost coming from the inclusion of a larger number of parameters in the MCMC, we restrict our analysis to ringdown events such that: (i) the remnants have masses $M_{f}\le 50M_\odot$, and (ii) the relative error on the damping time of the secondary QNM is $\sigma_{\omega_{33}}/\omega_{33}\leq 50\%$ (and similarly for the (21) component).
For concreteness we focus on ET.  Within all observable events in \texttt{model I}, we find 100 and 69 events which satisfy the two requirements above for the $(22)$-$(33)$ and $(22)$-$(21)$ templates, respectively (for \texttt{model II}, there are 92 and 50 events satisfying these conditions). Among these events, we select the 50 lightest-mass BHs, which are expected to yield the largest deviations from the Kerr spectrum.

\begin{figure}[t]
    \includegraphics[width=8.5cm]{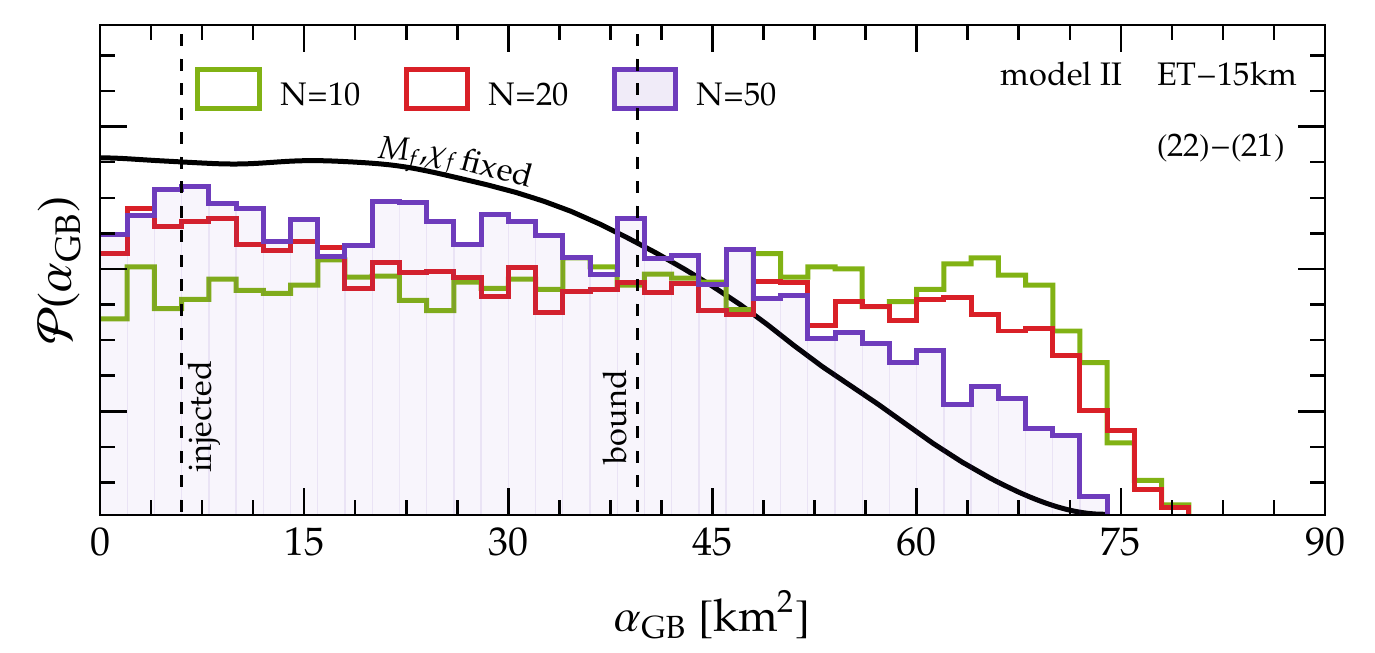}
    \caption{Posterior histograms for the GB coupling parameter inferred by stacking different sets of events observed by ET. Here we use a two-mode analysis in which the mass and spin of the remnant are sampled together with $\alpha_\tn{GB}$. The vertical dashed lines identifies the injected value of the coupling, and the current upper bound. For sake of comparison we show as solid black curve the posterior distribution obtained by fixing mass and spin to their injected values, and stacking $N=50$ events. The normalization constant of such distribution has been scaled for visualization purposes.}
    \label{fig:GBboundsMchi}
\end{figure}

In Fig.~\ref{fig:GBboundsMchi} we focus for simplicity on \texttt{model II} and on the $(22)$-$(21)$ combination (other configurations lead to similar results).
As expected, the ability of XG detectors to constrain $\alpha_\tn{GB}$ worsens dramatically.
For comparison we also plot ${\cal P}(\alpha_\tn{GB})$ when $M$ and $\chi$ are kept fixed.
When the masses and spins are sampled together with $\alpha_\tn{GB}$ the posteriors become less informative.
However the MCMC can still correctly infer the injected values of $(M_f,\chi_f)$, and masses are in general reconstructed with better accuracy than spins.

\subsubsection{Dynamical Chern-Simons gravity}\label{sec:csresults}

\begin{figure}[t]
    \includegraphics[width=7.5cm]{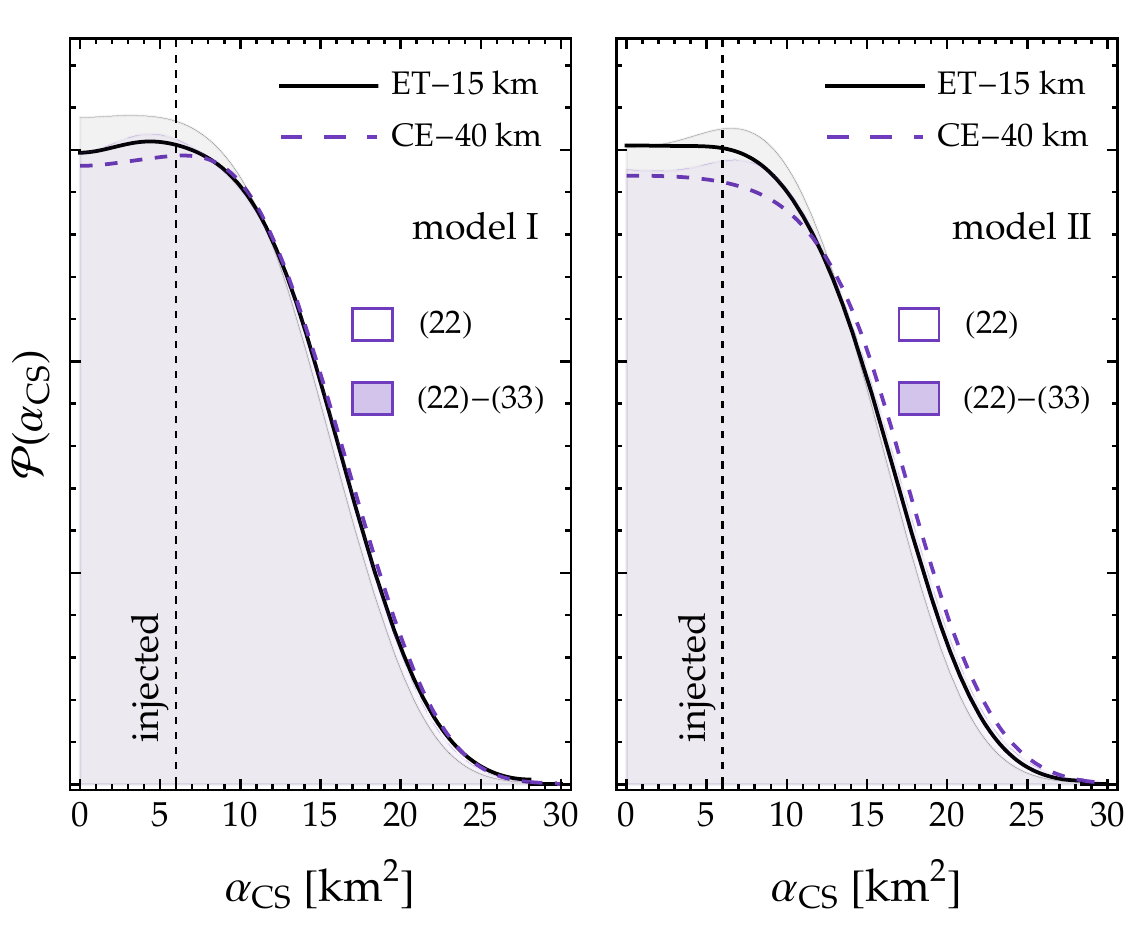}
    \caption{Same as Fig.~\ref{fig:GBbounds}, but for dCS gravity.}
    \label{fig:CSbounds}
\end{figure}

The constraining power of XG detectors improves only mildly for dCS gravity compared to EsGB, as shown in the two panels of Fig.~\ref{fig:CSbounds}.  Even in this case the posteriors of $\alpha_\tn{CS}$, obtained by stacking the full sets of BH remnants and keeping first-order spin corrections (the only corrections available so far), show full support for the GR value.

From a single-mode analysis based on the $(22)$ dominant mode we infer that ET may be able to bound the dCS coupling to the range $\alpha_\tn{CS} \sim[0,16.4]\,$km$^2$ and $\alpha_\tn{CS} \sim[0,16.3]\,$km$^2$ ($90\%$ confidence level) for \texttt{model I} and \texttt{model II}, respectively.  For CE, the forecast values have a slightly larger spread: $\alpha_\tn{CS} \sim[0,16.6]\,$km$^2$ and $\alpha_\tn{CS} \sim[0,17.2]\,$km$^2$ for \texttt{model I} and \texttt{model II}, respectively.  For completeness, we have repeated the analysis with the $(22)$-$(21)$ combination of QNMs. We find that the inferred bounds are only mildly dependent on the choice of the secondary mode.

A comparison of the 90\% credible intervals on $\alpha_\tn{CS}$ for different detector configurations and spin population models is shown in the bottom panels of Fig.~\ref{fig:CSGBbounds}.  Unlike EsGB gravity, forecasts from both detectors are tighter than the current upper bound by a factor $\sim 5$. As in the case of EsGB, the constraints on $\alpha_\tn{CS}$ are dominated by observations of low-mass sources with $M_f\le 50M_\odot$.

\begin{figure}[b]
    \includegraphics[width=8cm]{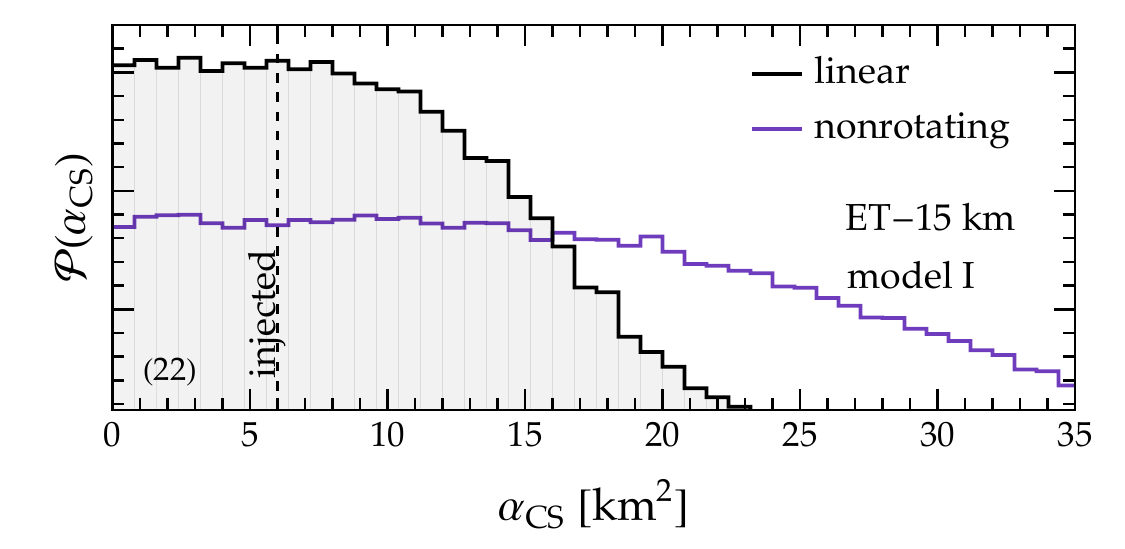}
    \caption{Posteriors of the dCS coupling constant inferred from ET observations by injecting single-mode QNM signals. Black histograms correspond to recovery templates which include $\mathcal{O}(\chi)$ spin terms, while purple histograms refer to templates which include only nonrotating terms in the beyond-Kerr parametrization.}
    \label{fig:CSbounds2}
\end{figure}

We have further assessed the relevance of spin corrections by studying how bounds on the dCS coupling change if we adopt a recovery template that only includes nonrotating coefficients in the beyond Kerr expansion. The results are shown in Fig.~\ref{fig:CSbounds2} for ET observations and \texttt{model I}. At variance with EsGB gravity, spin corrections seem to play a less dominant role for dCS: the posteriors obtained using zeroth-order terms in $\chi$ are roughly a factor of two broader, and still informative over the prior. Results for other detector configurations and BH populations are similar. While the analysis is limited by our current knowledge of higher-order spin terms in the dCS QNM spectrum, our findings seem to suggest that quadratic and higher-order corrections in $\chi$ may not dramatically affect our conclusions.

\begin{figure*}[htbp!]
    \includegraphics[width=8cm]{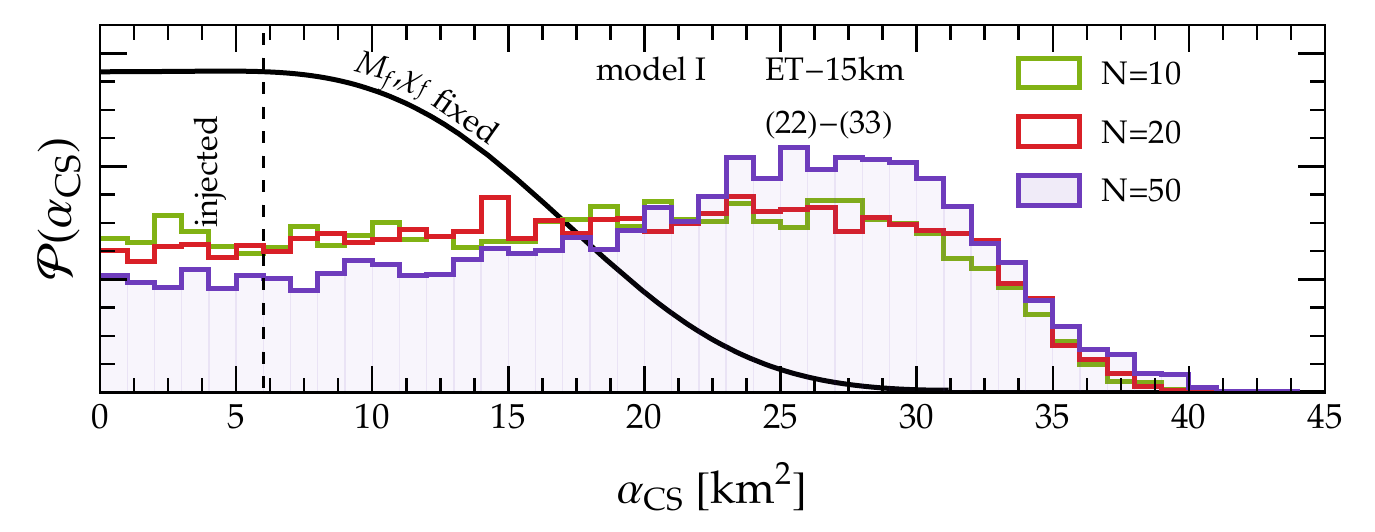}
    \includegraphics[width=8cm]{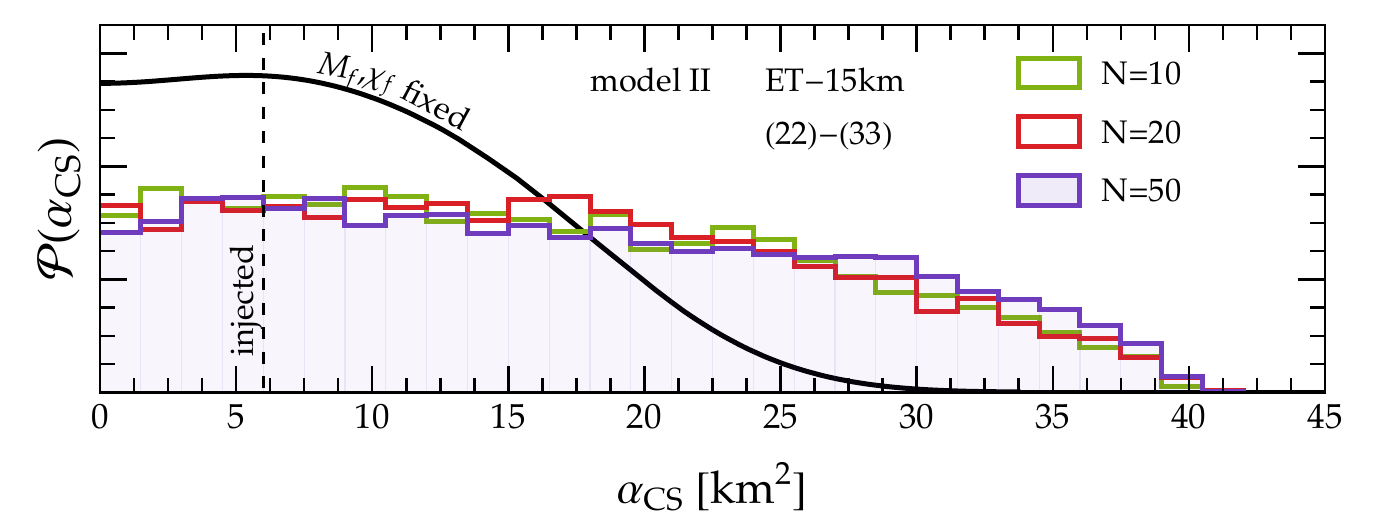}\\
    \includegraphics[width=8cm]{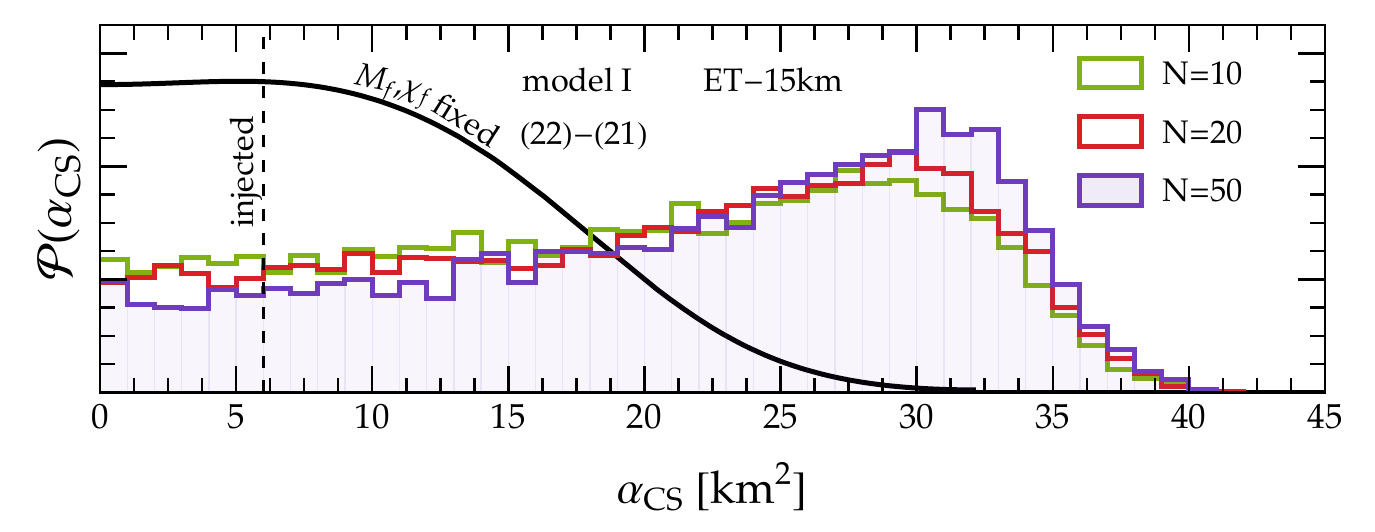}
    \includegraphics[width=8cm]{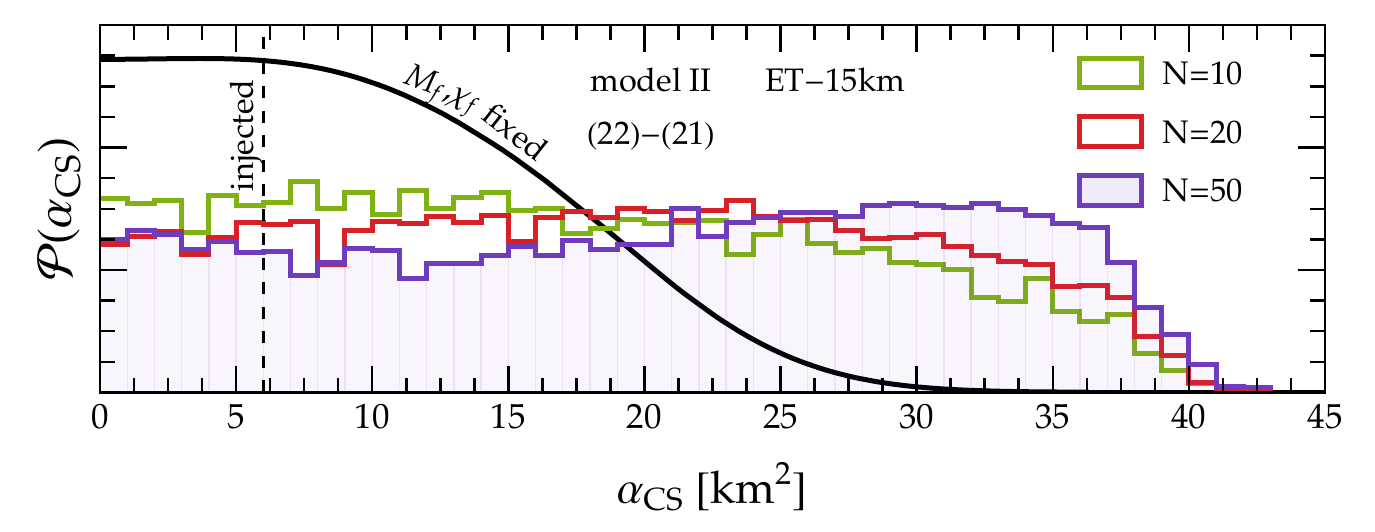}
    \caption{Posterior distributions of the dCS coupling constant $\alpha_\tn{CS}$ obtained by stacking different sets of events observed by ET in a two-mode analysis. The vertical dashed lines identifies the injected value of the coupling. The mass and spin of the remnants are sampled together with $\alpha_\tn{CS}$. For comparison we also show the posterior of the coupling obtained by stacking the largest set of events considered, and fixing $M_f$ and $\chi_f$ to their true values, with $N=50$ (solid black line). The left and right panels refer to different spin population models.}
    \label{fig:CSboundsMchi}
\end{figure*} 

In EsGB gravity, including masses and spins within the sampling parameters yields uninformative posteriors on the coupling constants. The results for dCS shown in the four panels of Fig.~\ref{fig:CSboundsMchi} are more promising. Here we follow the same strategy discussed in Sec.~\ref{sec:gbresults} in the context of EsGB gravity, and we focus on BHs with masses smaller than $50M_\odot$.
Green, red and purple histograms are the marginalized distributions of $\alpha_\tn{CS}$, obtained by stacking an ever larger set of events observed by ET. For comparison we also show (in black) the posterior obtained by fixing $M_f$ and $\chi_f$ to their true values, and assuming $N=50$ events: this posterior
is consistent with zero, but it usually peaks around the injected value of the dCS coupling (marked by a vertical dashed line).

The inferred value of $\alpha_\tn{CS}$ is quite sensitive to the spin distribution of the population and only mildly affected by choice of the secondary mode, even if dCS corrections to the $(33)$ Kerr frequencies and damping times are always one order of magnitude larger than the corresponding corrections to the $(21)$ component (compare  Fig.~\ref{fig:app_dCS}).
This confirms that the bulk of information comes from the fundamental QNM, as we discussed earlier in the context of Fig.~\ref{fig:CSbounds}.
For the \texttt{model I} population (left panels of Fig.~\ref{fig:CSboundsMchi}) the posterior distributions shift to the wrong value of $\alpha_\tn{CS}$, peaking far from the injected value of the coupling as the number of observations grows. The origin of this systematic bias can be traced to the degeneracy between the source parameters (i.e., $M_f,\chi_f$) and the beyond-GR deviation parameter (in this case, the dCS coupling) that we also observe in the agnostic two-mode analysis (cf. Sec.~\ref{sec:res2}).  Indeed, the bias in Fig.~\ref{fig:CSboundsMchi} is slightly larger when we consider the $(21)$ mode (which leads to smaller changes in the Kerr spectrum) as the secondary mode.

For the \texttt{model II} population (right panels of Fig.~\ref{fig:CSboundsMchi}) the bias is less evident. In this case, the posterior become wider that those obtained by fixing the BH masses and spins, with the upper bound on $\alpha_\tn{CS}$ being a factor two bigger.  As for EsGB gravity, the remnant mass and spin posteriors are always correctly centered around the ``true'' injected values: see Fig.~\ref{fig:AppMchi} in Appendix~\ref{sec:appmasses}.
It is also clear from Fig.~\ref{fig:AppMchi} that the mass and spin inference is generally better for large-SNR systems. These are usually large-mass systems: unfortunately, the degeneracy between mass, spin and coupling constants is less relevant precisely for those systems for which beyond-GR corrections are suppressed.

\subsubsection{Effective Field Theory}\label{sec:eftresults}

We now discuss how well ringdown observations by XG detectors could constrain the EFT gravity models of Ref.~\cite{Cano:2023jbk}. Once again, we first assume that BH masses and spins are fixed to their injected values, and only sample over the coupling constant $\alpha_q$ for each of the eight EFT models we consider. We inject signals with three different values of $\alpha_q$ such that $\gamma_{q}^\tn{cubic}=\alpha^4_q/\min[m_1,\,m_2]^4=(0.1,\,0.2,\,0.5)$ and $\gamma_{q}^\tn{quartic}=\alpha^6_q/\min[m_1,m_2]^6=(0.1,\,0.2,\,0.5)$, where $\min[m_1,\,m_2]$ is the smallest mass among all binary components that yield an observable ringdown signal within each of our population models.

\begin{figure*}[t]%
    \includegraphics[width=13cm]{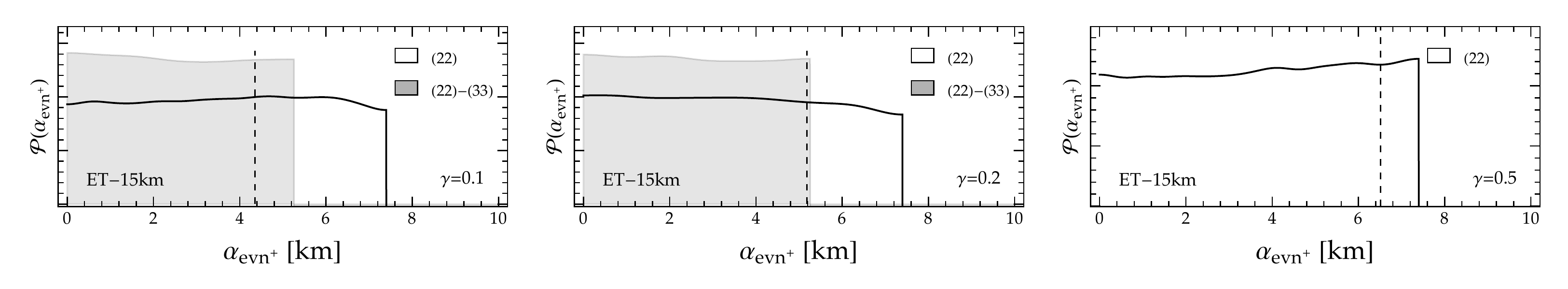}\\
    \includegraphics[width=13cm]{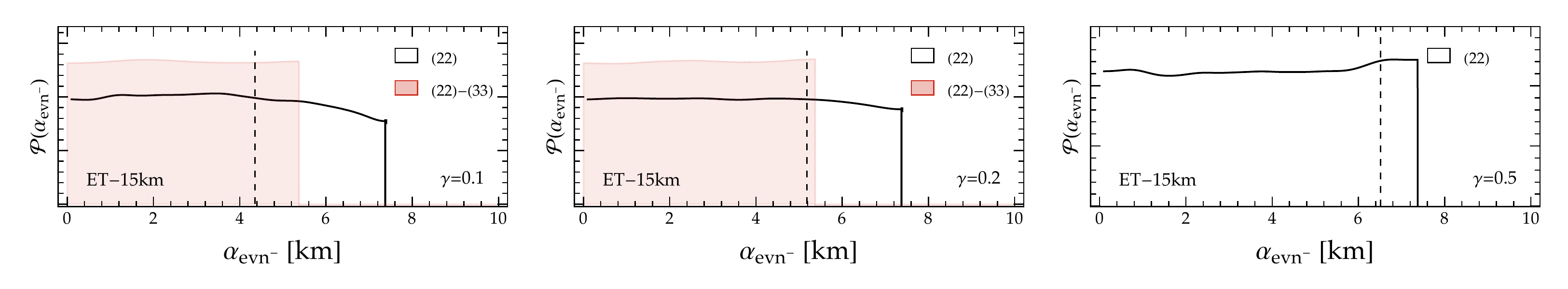}\\
    \includegraphics[width=13cm]{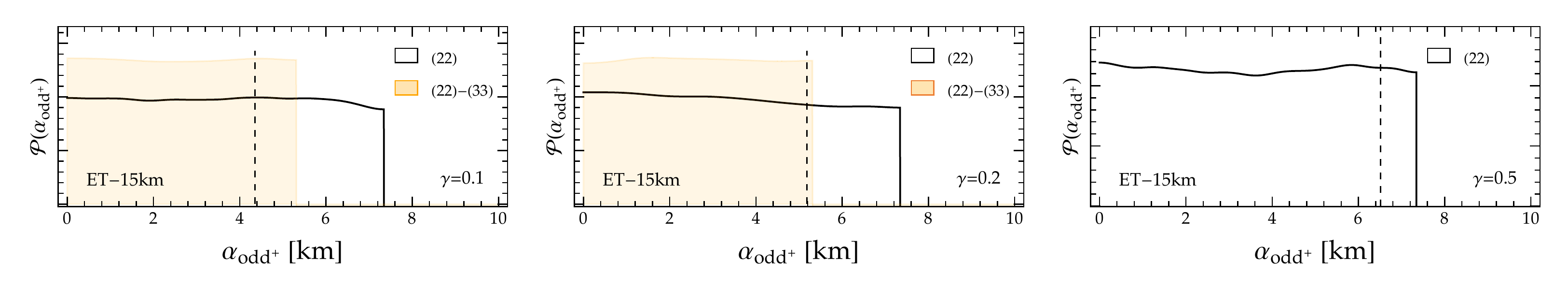}\\
    \includegraphics[width=13cm]{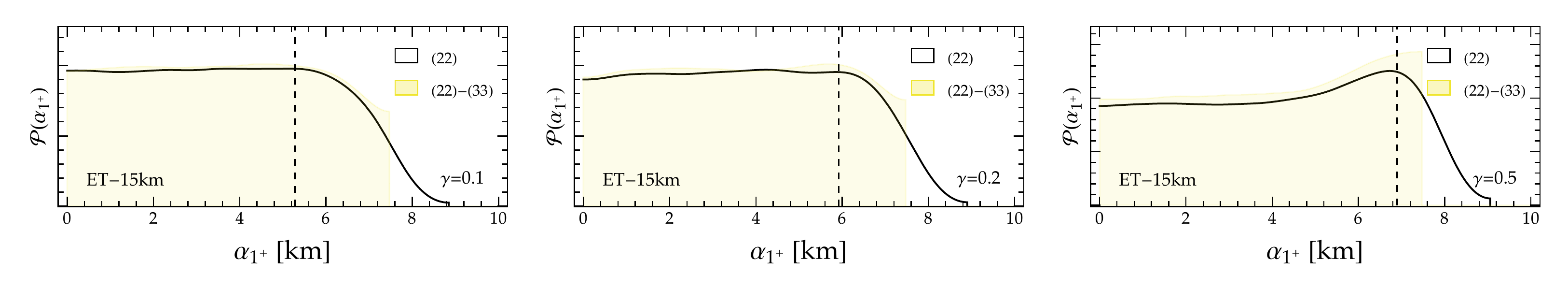}\\ 
    \includegraphics[width=13cm]{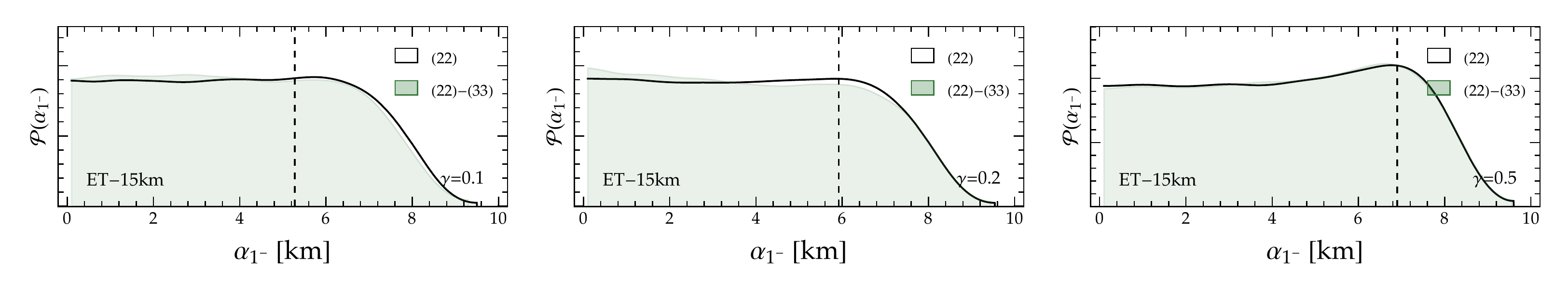}\\
    \includegraphics[width=13cm]{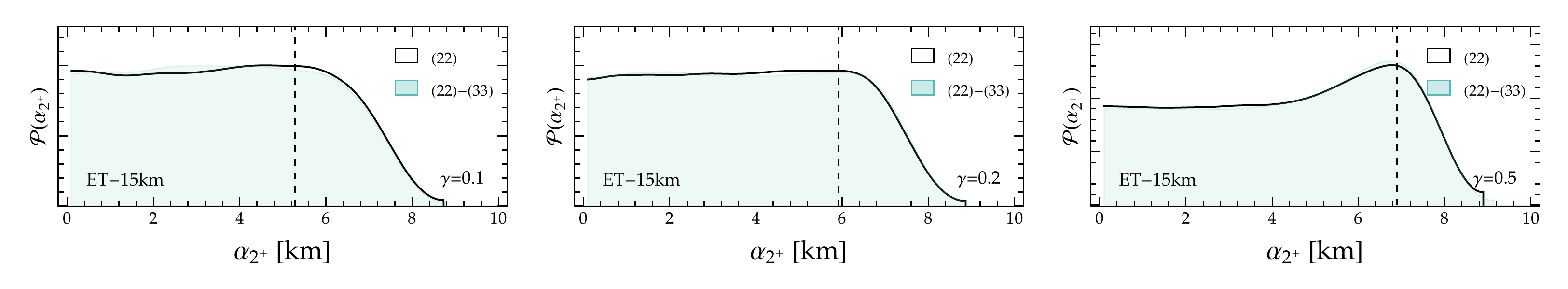}\\
    \includegraphics[width=13cm]{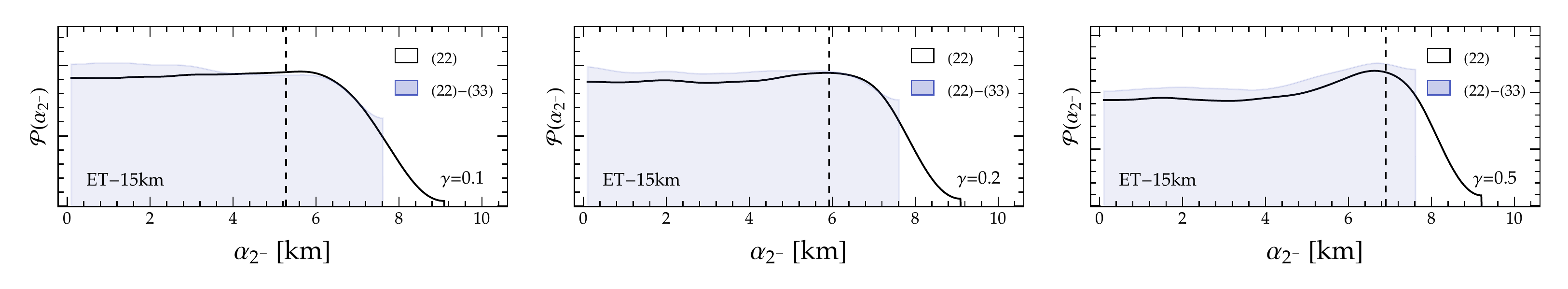}\\
    \includegraphics[width=13cm]{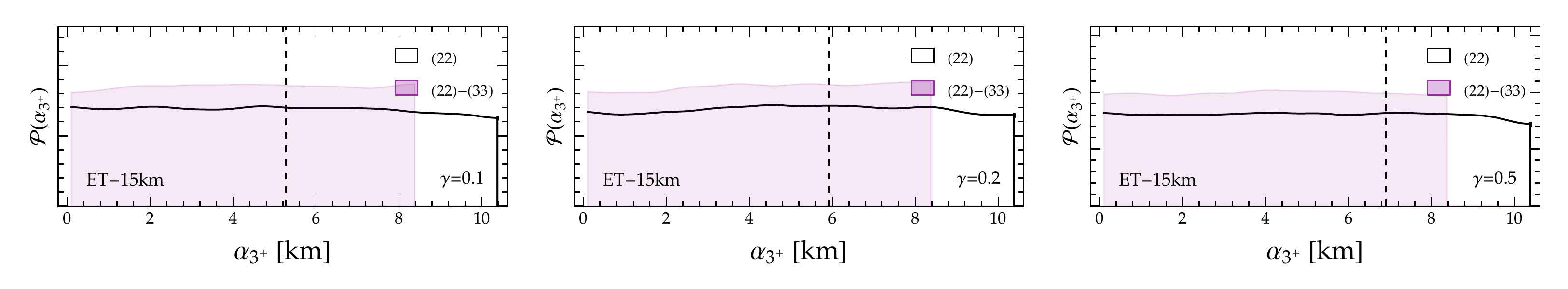}
    \caption{Posterior probability distributions for the coupling parameters of the eight EFT models we consider.  Bounds are obtained by stacking all of the \texttt{model I} signals observed by ET. Left, middle and right panels refer to a different injected value of $\alpha_q$, corresponding to $\gamma_{q}=(0.1,\,0.2,\,0.5)$ in each specific theory.  The injected values are marked by vertical lines.}
    \label{fig:EFTbounds}
\end{figure*} 

The posterior distributions for both cubic and quartic couplings are shown in Fig.~\ref{fig:EFTbounds}, where we stack all BH remnants observed by ET and we assume the \texttt{model I} spin distribution. Each panel shows results obtained by either a single-mode or a two-mode analysis.  As we observed for EsGB gravity, the hard cut in the posteriors inferred from the $(22)$-$(33)$ combination for some of the models corresponds to imposing the requirement that the \PS coefficients remain in the perturbative regime.

All of the EFT gravity theories we analyze display a common trend: for $\gamma_q\le 0.2$ (first and second column) the inferred posteriors are poorly informative, yielding upper bounds on $\alpha_q$ of the order of $\mathcal{O}(10)$~km, which depend only mildly on the mass scaling of the coupling (either $\gamma_q\propto M^{-4}$ or $\gamma_q\propto M^{-6}$) and on the specific gravity theory. All distributions rail against zero, showing support for the GR hypothesis.

At the highest value of $\gamma_q$ we consider (third column in Fig.~\ref{fig:EFTbounds}), the posteriors for some of the models narrow, peaking around the injected value of $\alpha_q$. Note however that all posterior distributions still rail against the GR value within the 90\% credible interval. These results imply that ringdown observations may be able to mildly constrain the coupling constant of these EFT gravity models only when the coupling constants are very large, possibly violating the domain of validity of the theory. For signals with $\gamma^\tn{cubic}_{q}=0.5$ (top-right panels) we only show results for the single-mode inference: in the two-mode inference, the coupling (which is outside the EFT domain of validity) is such that we always find a BH mass in our catalogue which violates the perturbative character of the \PS corrections. As usual, including the mass and spins of the remnant within the MCMC sampling introduces additional correlations among the parameters, thus widening the posteriors.

\begin{figure}[t]
    \includegraphics[width=8cm]{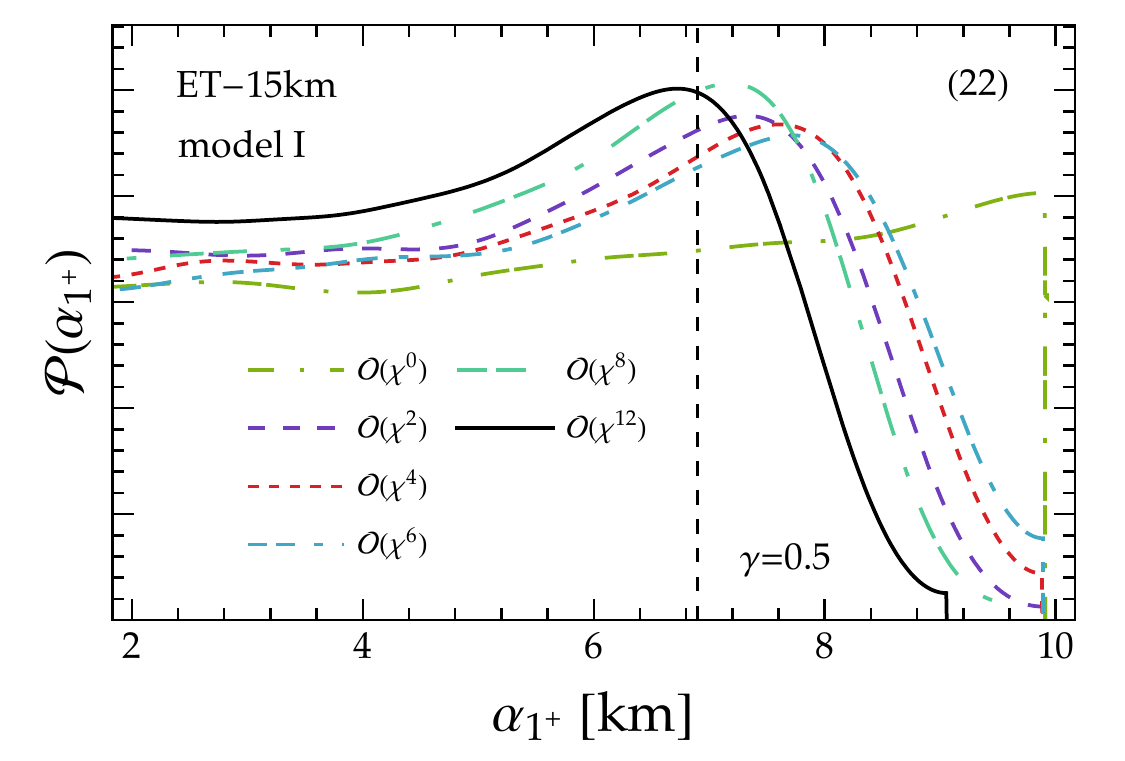}
    \caption{Posterior distribution of the EFT coupling parameter $\alpha_{1^{+}}$. Different colors correspond to recovery templates that include spin effects at different orders, as indicated in the legend. The vertical dashed line marks the injected value of the coupling.}
    \label{fig:EFTspinbounds}
\end{figure}

The high-order expansion of the QNM frequencies derived in Ref.~\cite{Cano:2023jbk} allows us to test the relevance of different contributions in the spin expansion to the QNM spectrum.
To this end, in Fig.~\ref{fig:EFTspinbounds} we repeat the previous analysis with a \PS template that includes spin terms only up to a given order. For concreteness we show the constraints on the coupling parameter of the $1^{+}$ model, and we inject signals corresponding to large coupling ($\alpha^6_{1^+}/\min[m_1,m_2]^6=0.5$).  While a zero-spin template introduces a significant bias in the posterior distribution, a sixth-order expansion in $\chi$ is sufficient to recover the ``true'' injected value of $\alpha_{1^+}$. Note that the quadratic expansion is remarkably good in this case. However the convergence of the series may well be asymptotic, and this is probably just a coincidence.
The results for the other families of EFT gravity models are similar: depending on the specific model, including between the sixth and the ninth order is typically sufficient to infer the injected coupling parameter without significant bias.

\section{Conclusions}
\label{sec:conclusions}

In this paper we employ the {\sc ParSpec} parametrization of the QNM spectrum of spinning BHs beyond GR developed by some of us in previous work~\cite{Maselli:2019mjd} to address the following question: can future ringdown observation with XG detectors detect (or constrain) deviations from the Kerr spectrum by stacking multiple observations of binary mergers? We use astrophysically motivated populations and address this question for two different families of spectroscopy tests: (i) \textit{agnostic} (null) tests, and (ii) \textit{theory-based} tests, which make use of QNM frequency calculations in EsGB gravity, dCS gravity, and various EFT-based extensions of general relativity.
When we inject beyond-GR signals in the data, we generally assume the coupling constant of specific beyond-GR theory to saturate current experimental bounds.

We find that robust inference of hypothetical corrections to GR requires pushing the slow-rotation expansion to high orders. Even when high-order expansions are available, ringdown observations alone may not be sufficient to measure deviations from the Kerr spectrum for theories with dimensionful coupling constants because the constraints are dominated by ``light'' black hole remnants, and only few of these light remnants have sufficiently high SNR in the ringdown. 

For theory-agnostic tests with dimensionless couplings $(p=0)$ which include spin corrections to the QNM spectrum we find a hierarchy among the beyond-GR parameters, with the nonrotating terms providing the tightest constraints. This confirms some of the main results of Ref.~\cite{Maselli:2019mjd}: spin corrections to the damping time may be unmeasurable, even when stacking thousands of events.

For theory-agnostic tests where couplings have dimensions of $({\rm mass})^4$ ($p=4$), we can only constrain nonrotating corrections to the QNM spectrum. The inferred bounds on the \PS parameters imply bounds on the fundamental coupling of a putative beyond-GR theory of the order of $\mathcal{O} (10)\,\textnormal{km}$.

For EsGB gravity, the constraints inferred by XG detectors are strongly dependent on the accuracy of the beyond-GR spin expansion (cf. Figs.~\ref{fig:GBbounds2} and \ref{fig:GBbounds3})).
When we include second-order terms in the small-spin expansion we can better constrain the EsGB coupling constant,, but there is still support for the GR hypothesis ($\alpha_\tn{GB}=0$). The analysis suggests that the constraints may change when we include the unknown $\mathcal{O}(\chi^3)$ terms in the beyond-GR spin expansion. Moreover, even the most optimistic constraints inferred with the inclusion of second-order spin terms are still weaker than current bounds from LIGO/Virgo/KAGRA inspiral observations (Fig.~\ref{fig:CSGBbounds}). All of the bounds discussed above are optimistic, because they are found assuming that the masses and spins of the remnants are known (e.g., from the full inspiral-merger-ringdown waveform) and fixed to their injected values. When we allow the mass and spin to vary in the MCMC, the posteriors on $\alpha_\tn{GB}$ become uninformative.

Our findings for dCS are slightly more optimistic: by stacking ringdown observations with ET and CE we may constrain the dCS coupling constant $\alpha_\tn{CS}$ with an accuracy about five times better than current bounds (Fig.~\ref{fig:CSGBbounds}), at least when we assume that the masses and spins of the remnants are known. In any case, ringdown-only constraints on $\alpha_\tn{CS}$ would still be about two orders of magnitude worse than those achievable by  XG observations of post-Newtonian corrections in the inspiral~\cite{Perkins:2020tra}.

All of the EFT models that we consider lead to uninformative posteriors for cubic and quartic GR modifications, even when we consider values of the coupling so large that they would violate the domain of validity of the EFT.  Constraining quartic models is particularly difficult because of the $1/M_f^6$ suppression of the beyond-Kerr deviations.  The QNMs computed in EFT gravity include high-order terms in the spin expansion up to $\mathcal{O}(\chi^{12})$. The inclusion of these contributions is useful to understand the convergence properties of the small-spin expansion, but it does not significantly improve the bounds.

In conclusion, it is difficult to set constraints on theories with a {\it dimensionful} coupling -- like those modifying GR in the large curvature regime -- using ringdown observations alone. This is because constraints on any dimensionful coupling constants are dominated by low-mass black holes, i.e. by higher frequency signals, for which ET and CE are less sensitive (at least, with the currently planned configuration).
For both ET and CE, we find that the bounds obtained by stacking $\sim 100$ events with $M_f\lesssim50M_\odot$ are comparable to the bounds found by stacking the full data set.

It should be possible to achieve better constraints on theories with dimensionful coupling constants by exploiting the full inspiral-merger-ringdown waveform. This is a more ambitious, ongoing research program that will require a combination of numerical simulations (see e.g.~\cite{Okounkova:2017yby,Witek:2018dmd,Okounkova:2019dfo,Okounkova:2019zjf,Ripley:2019aqj,Julie:2020vov,Witek:2020uzz,East:2020hgw,Silva:2020omi,East:2021bqk,Elley:2022ept,Ripley:2022cdh,Evstafyeva:2022rve,Corman:2022xqg}), post-Newtonian and effective-one-body models~\cite{Julie:2017pkb,Julie:2017rpw,Julie:2018lfp,Julie:2019sab,Shiralilou:2020gah,Shiralilou:2021mfl,Julie:2022huo,Julie:2022qux,vanGemeren:2023rhh}, and parametrizations of the full inspiral-merger-ringdown waveform~\cite{Brito:2018rfr,Ghosh:2021mrv,Silva:2022srr,Maggio:2022hre,Ripley:2022cdh}. 
The constraining power of ringdown observations is better for theories with a {\it dimensionless} coupling, for which suppression of beyond Kerr deviations due some energy scale is not effective. In this case the analysis we carried out in the agnostic, dimensionless scenario suggests that we may be able to set bounds at least on the leading-order (nonrotating) terms of the beyond-GR contribution to the QNM frequencies.

\acknowledgments

We thank Swetha Baghwat, Alessandra Buonanno, Mark Ho-Yeuk Cheung and Costantino Pacilio for useful discussions.
S.Y, L.R. and E.B. are supported by NSF Grants No. AST-2006538, PHY-2207502, PHY-090003 and PHY-20043, NASA Grants No. 20-LPS20-0011 and 21-ATP21-0010, and by the John Templeton Foundation Grant 62840.
S.Y. is supported by the NSF Graduate Research Fellowship Program under Grant No. DGE2139757.
A.M., L.P. and L.G. acknowledge financial support from the EU Horizon 2020 Research and Innovation Programme under the Marie Sklodowska-Curie Grant Agreement no. 101007855.
A.M. and E.B. acknowledge support from the ITA-USA Science and Technology Cooperation program,
supported by the Ministry of Foreign Affairs of Italy (MAECI).
E.B. acknowledges the support of the Indo-US Science and Technology Forum through the Indo-US Centre for Gravitational-Physics and Astronomy, grant IUSSTF/JC-142/2019. 
A.M. acknowledges financial support from the Italian Ministry of University and Research (MUR) for the PRIN grant METE under contract no. 2020KB33TP.
This work was carried out at the Advanced Research Computing at Hopkins (ARCH) core facility (\url{rockfish.jhu.edu}), which is supported by the NSF Grant No.~OAC-1920103. 

\appendix

\begin{table}[t]
	\centering
	\begin{tabular}{ c | c c | c c | c c }
		\hline
        \hline
	 $k_1$ & $\bar{\omega}^{(k_1)}_{22}$ & $\bar{\tau}^{(k_1)}_{22}$ & 
	$\bar{\omega}^{(k_1)}_{33}$ & $\bar{\tau}^{(k_1)}_{33}$ & $\bar{\omega}^{(k_1)}_{21}$ & $\bar{\tau}^{(k_1)}_{21}$ \\
		\hline
 0 & 0.37368 & 11.241 & 0.59946 & 10.787 & 0.37367 & 11.241 \\
 1 & 0.12947 & 0.98133 & 0.2065 & 0.95996 & 0.062929 & 0.4055 \\
 2 & -0.017526 & -19.094 & 0.0046294 & -19.021 & 0.049757 & -6.8293 \\
 3 & 0.80811 & 184.11 & 0.91245 & 185.08 & -0.057124 & 70.950 \\
 4 & -2.8392 & -766.01 & -3.0061 & -769.48 & 0.46321 & -292.17 \\
 5 & 4.3651 & 1434.2 & 4.1145 & 1441.7 & -1.1694 & 534.65 \\
 6 & 1.0487 & -608.63 & 2.9075 & -614.3 & 1.4399 & -182.24 \\
 7 & -8.1789 & -1337.3 & -10.535 & -1340.6 & -0.27864 & -566.17 \\
 8 & -0.21522 & 594.52 & -1.51 & 599.45 & -1.0989 & 190.92 \\
 9 & 11.546 & 1730.8 & 15.248 & 1734.2 & 0.72622 & 749.45 \\
 10 & 2.5717 & -111.70 & 4.3535 & -114.94 & 0.9947 & 4.3652 \\
 11 & -18.135 & -2473.0 & -24.505 & -2476.4 & -1.6423 & -1098.0 \\
 12 & 9.4186 & 1387.6 & 12.54 & 1390.2 & 0.71601 & 604.36 \\
		\hline
        \hline
	\end{tabular}
		\caption{Values of the GR coefficients for the \PS spin expansion for 
	the modes with $(\ell m)=(22),(33),(21)$ modes up to order $12$ in the spin.}
	\label{tab:Parspec-tab-GR}
\end{table}

\section{Quasinormal mode frequencies for modified theories of gravity}\label{app:qnmtheories}

In this appendix we map the frequencies and damping times computed in specific modified theories of gravity to the \PS parameters defined in Eq.~\eqref{eq:parspec-expa1} and \eqref{eq:parspec-expa2}.  As discussed in Sec.~\ref{sec:approaches}, we consider two models of quadratic gravity for which QNM frequencies have been computed in a slow-rotation expansion (EsGB and dCS gravity) and on eight models of EFT gravity.

The expansion coefficients for a slowly-rotating Kerr background can be found (in principle, to arbitrary order in the small-spin expansion) by fitting the numerical values computed, e.g., in Ref.~\cite{Berti:2005ys}. For consistency with the highest-order expansion computed in Ref.~\cite{Cano:2023jbk}, we fit the frequencies and damping times by a $12$th order polynomial in $\chi$ using the \texttt{NonlinearModelFit} routine in {\sc Mathematica}: 
\begin{align}
M\omega^\tn{K,s}_{\ell m}=&\sum_{k_1=0}^{12}\bar{\omega}_{\ell m}^{(k_1)}
\chi^{k_1}\ ,\label{math:kerromega}\\    
\tau^\tn{K,s}_{\ell m}/M=&\sum_{k_1=0}^{12}\bar{\tau}_{\ell m}^{(k_1)}
\chi^{k_1}\ .\label{math:kerrtau}
\end{align}
We further impose that $(\bar{\omega}_{\ell m}^{(0)},\bar{\tau}_{\ell m}^{(0)})$ are fixed to match exactly their Schwarzschild value.
The resulting fitting coefficients are listed in Table~\ref{tab:Parspec-tab-GR}.

\begin{figure}[t]%
    \includegraphics[width=4cm]{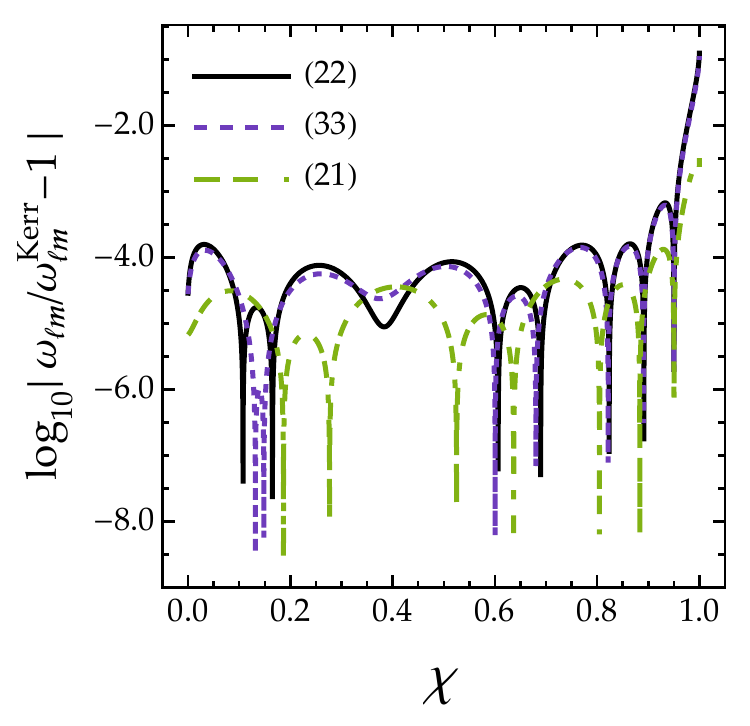}
    \includegraphics[width=4cm]{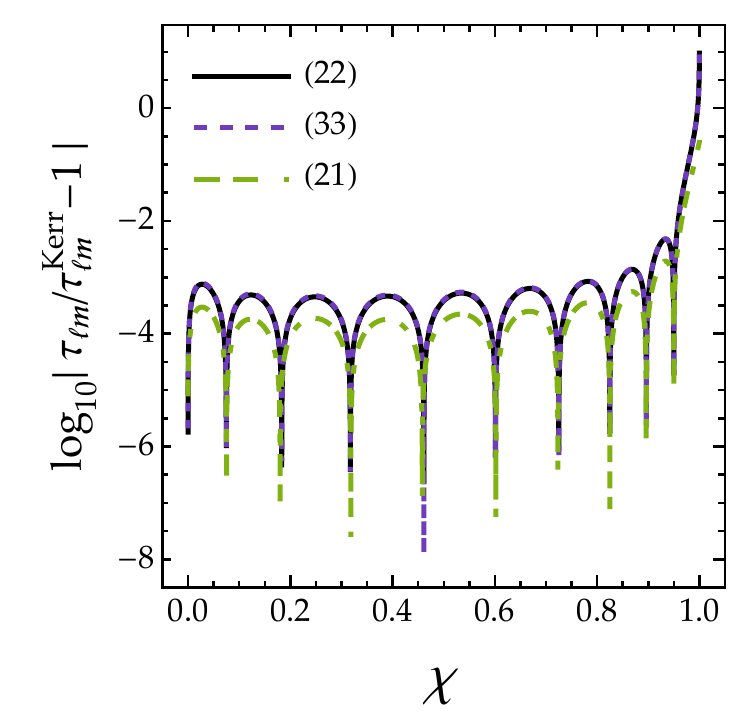}
    \caption{Relative errors between the frequencies and damping times in GR computed by solving the Teukolsky equation and their slow-rotation approximations at twelfth-order in a Taylor expansion in the spin, with coefficients given in Table~\ref{tab:Parspec-tab-GR}.}
    \label{fig:app_convGR}
\end{figure}

It would be interesting to consider alternative fitting procedures proposed in the literature (see e.g. Ref.~\cite{Carullo:2021oxn}, in particular their Fig.~5), but the Taylor series presented here is sufficiently accurate for our present purposes.
As shown in Fig.~\ref{fig:app_convGR}, the expansion reproduces the numerical QNM spectrum with a relative accuracy better than $1\%$ in both frequency and damping time, as long as $\chi\lesssim 0.95$.

\subsection{Einstein-scalar-Gauss-Bonnet gravity}

The EsGB QNM spectrum at second order in a small-spin expansion was computed in Ref.~\cite{Pierini:2022eim}. In terms of the dimensionless spin $\chi=J/M^2$ and of the coupling constant $\beta_\tn{GB}=\alpha_\tn{GB}/M^2$ it reads
\begin{align}
M \Omega_{\ell m} = &\left(\Omega_0^A + \beta_\tn{GB}^2 \Omega_0^B \right)+ m \chi \left(\Omega_1^A + \beta_\tn{GB}^2 \Omega_1^B \right)   \notag\\
& + \chi^2 \left[\left(\Omega_{2a}^A + \beta_\tn{GB}^2 \Omega_{2a}^B\right)\right.\nonumber \\
&\left.+ m^2 \left(\Omega_{2b}^A+\beta_\tn{GB}^2 \Omega_{2b}^B\right)\right] \ ,
\end{align}
where $\Omega_{\ell m}=\omega_{\ell m}-i/\tau_{\ell m}$ and the subscripts ``A'' and ``B'' refer to the GR and beyond-GR contributions, respectively.
We can recast this expression into the \PS framework by identifying $\gamma =\alpha_\tn{GB}^2/M^4=\beta_\tn{GB}^2$, such that:
\begin{align}
\omega_{\ell m} = & \frac{1}{M} \left[\bar{\omega}^{(0)}_{\ell m} \left(1+ \beta_\tn{GB}^2 
\delta \omega_{\ell m}^{(0)}\right)+ \chi \bar{\omega}_{\ell m}^{(1)} \right.  \notag\\
&  \left. \left(1   +   \beta_\tn{GB}^2 \delta \omega_{\ell m}^{(1)}\right) + \chi^2 \bar{\omega}_{\ell m}^{(2)} \left(1 +\beta^2_\tn{GB} \delta \omega_{\ell m}^{(2)}\right) \right]+\nonumber\\
&+\frac{1}{M}\sum_{k_1=3}^{12}\bar{\omega}_{\ell m}^{(k_1)}\label{eq:expa-parspec-edgb-1}\\
\tau_{\ell m} = & M \left[\bar{\tau}_{\ell m}^{(0)} \left(1+ \beta_\tn{GB}^2 \delta \tau_{\ell m}^{(0)}\right)+ \chi \bar{\tau}_{\ell m}^{(1)} \right. \notag\\
& \left.  \left(1+ \beta_\tn{GB}^2 \delta \tau_{\ell m}^{(1)}\right) + \chi^2 \bar{\tau}_{\ell m}^{(2)} \left(1+ \beta_\tn{GB}^2\delta \tau_{\ell m}^{(2)}\right) \right] +\nonumber\\
&+M\sum_{k_1=3}^{12}\bar{\tau}_{\ell m}^{(k_1)}\ ,
\label{eq:expa-parspec-edgb-2}
\end{align}
where the quantities with an overbar identify the GR coefficients up to $12$th order in the spin. We determine the beyond-GR parameters in Eqs.~\eqref{eq:expa-parspec-edgb-1} and \eqref{eq:expa-parspec-edgb-2} by fitting them against the analytical expressions provided in Ref.~\cite{Pierini:2022eim} up to $\mathcal{O}(\chi^2)$ and $\mathcal{O}(\alpha_\tn{GB}^6)$.
We perform the fit in a range of spins and coupling constants where the QNM frequencies are well described by Taylor expansions according to Ref.~\cite{Pierini:2022eim}, i.e., $\chi \in [0,\,0.2], \alpha_\tn{GB}/M^2 \in [0,\,0.2]$.

\begin{figure}[t]%
    \includegraphics[width=8cm]{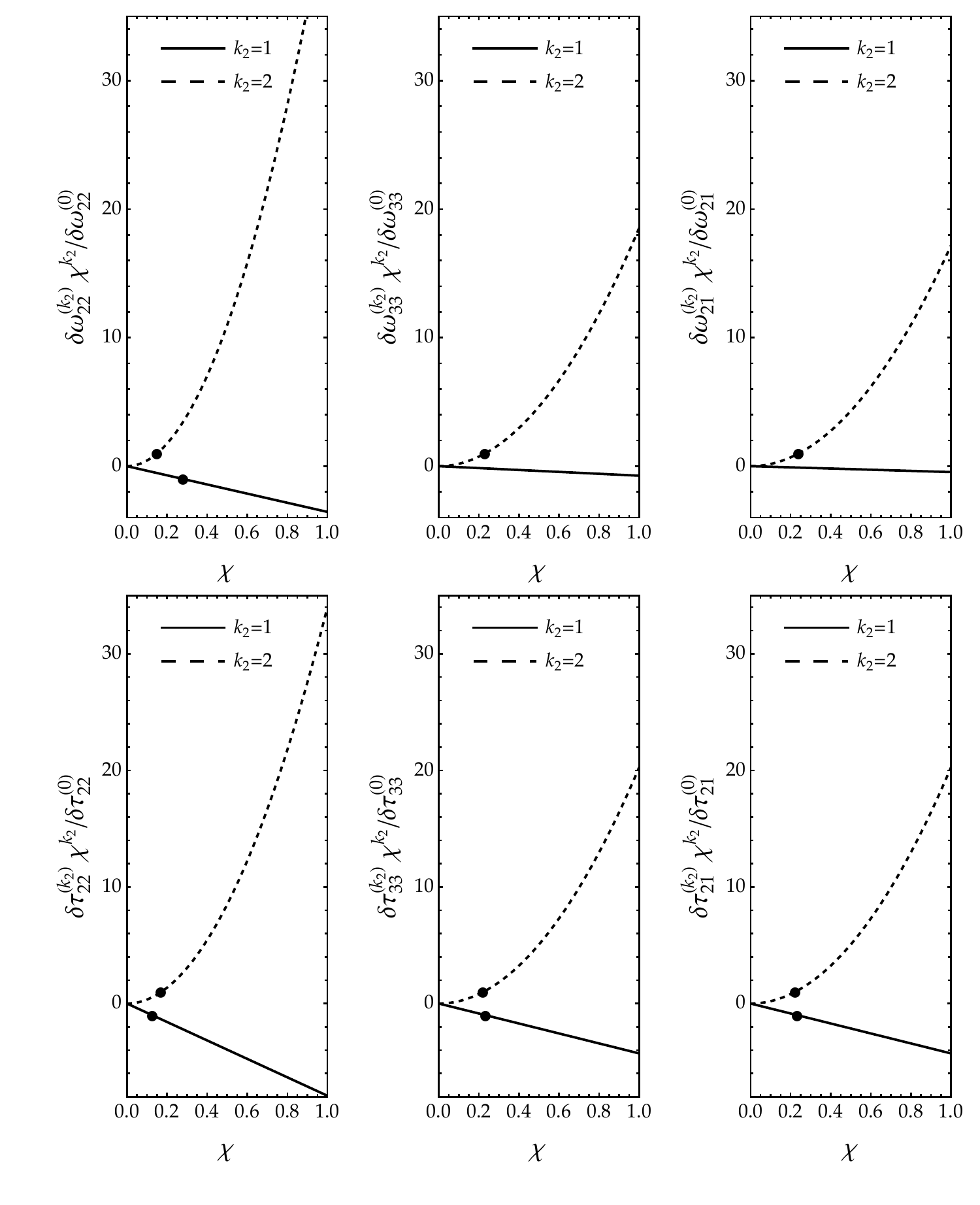}
    \caption{Linear ($k_2=1$) and quadratic ($k_2=2$) corrections to the EsGB QNM frequency (top) and damping time (bottom) as a function of $\chi$, normalized to the nonrotating term, for selected values of $(\ell m)$. 
    Bullets mark values of $\chi$ for which the linear or quadratic correction becomes larger (in absolute value) than the nonrotating contribution.}
    \label{fig:app_esgborder}
\end{figure}

The values of the EsGB fitting coefficients found in this way are listed in Table~\ref{tab:Parspec-tab-EdGB}.
In Fig.~\ref{fig:app_esgborder} we plot the linear and quadratic spin corrections to the frequency and damping time of the $(22)$, $(33)$ and $(21)$ modes, normalized to the nonrotating term, as functions of $\chi$. Bullets mark values of $\chi$ for which the linear or quadratic correction becomes larger (in absolute value) than the nonrotating contribution. For example, the second-order corrections to the frequency and damping time of the $(22)$ mode become larger than the nonrotating contribution for $\chi\gtrsim 0.2$.  This highlights the importance of pushing the spin expansion to higher orders.

\begin{figure}[t]%
    \includegraphics[width=4.2cm]{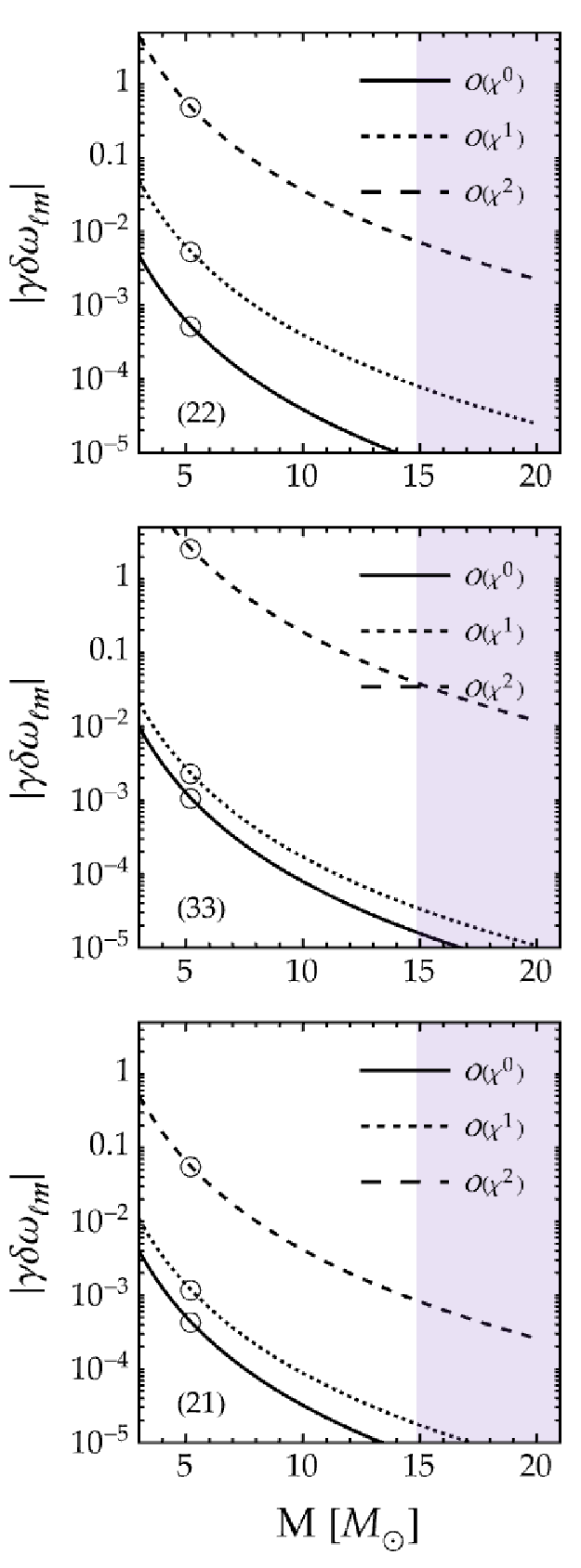}
   \includegraphics[width=4.2cm]{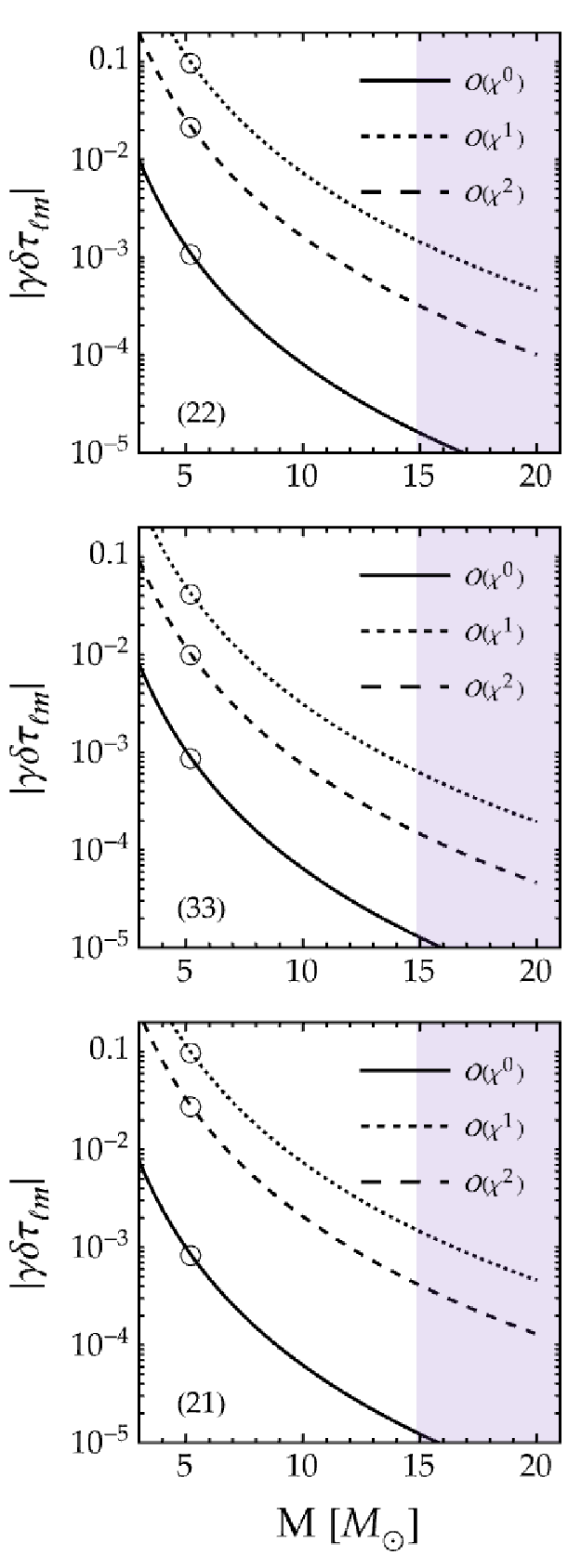}
   \caption{\PS coefficients $\gamma(\delta\omega^{k_2}_{\ell m},\delta\tau^{k_2}_{\ell m})$, with $\gamma=\beta_\tn{GB}^2=\alpha^2_\tn{GB}/M^4$ for EsGB gravity, as a function of the BH mass. Solid, dashed and dotted curves refer to nonrotating, linear and quadratic coefficients. Empty dots correspond to the smallest observable binary component mass in our population models. Vertical shaded regions identify the lightest remnant BH mass among all events with an observable ringdown. The top, center and bottom rows show values of the coefficients for the $(22)$, $(33)$ and $(21)$ mode, respectively.}
    \label{fig:esgbparameters}
\end{figure}

To further clarify the relevance of spin corrections, we study the mass dependence of the \PS coefficients $\gamma \delta\omega^{k_2}_{\ell m}$ and $\gamma \delta\tau^{k_2}_{\ell m}$ for EsGB gravity.  In Fig.~\ref{fig:esgbparameters} we plot the nonrotating, linear and quadratic spin terms as a function of $M$, assuming $\sqrt{\alpha_\tn{GB}}\simeq2.5\,\tn{km}$, as discussed in Sec.~\ref{sec:theoryspec}. Empty circles in each panel correspond to the minimum observable component mass among all binary BHs present in our catalogs. For EsGB gravity, both frequencies and damping time coefficients remain much smaller than one in the whole range of BH \textit{remnant} masses.

\begin{table}[t]
	\centering
	\begin{tabular}{ c | c c | c c | c c }
		\hline
		\hline
		 $k_2$ & $\delta \omega^{(k_2)}_{22}$ & $\delta \tau^{(k_2)}_{22}$ 
		& $\delta\omega^{(k_2)}_{33}$ & $\delta\tau^{(k_2)}_{33}$ & $\delta\omega^{(k_2)}_{21}$ & $\delta\tau^{(k_2)}_{21}$ \\
		\hline
		0 & $-0.05062$ & $-0.10606$ & $-0.10432$ & $-0.08524$ & $-0.04227$ & $-0.08179$ \\
		1 & $0.51999$ & $9.6119$ & $0.22466$ & $4.0986$ & $0.11534$ & $9.7028$ \\
		2 &  $47.388$ & $2.1226$ & $-250.52$ & $0.97981$ & $-5.4561$ & $2.7290$\\
		\hline
        \hline
	\end{tabular}
		\caption{Values of the EsGB coefficients for the \PS spin expansion of the $(\ell m)=(22)$, $(33)$, $(21)$ modes. }
	\label{tab:Parspec-tab-EdGB}
\end{table}

\begin{figure}[t]%
    \includegraphics[width=8.5cm]{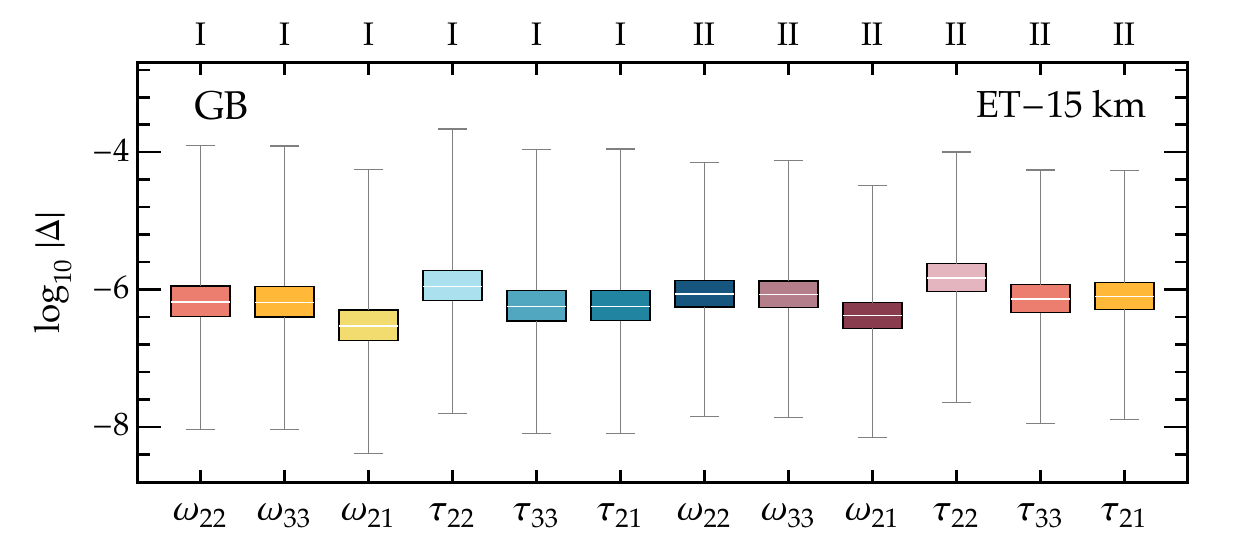}
    \includegraphics[width=8.5cm]{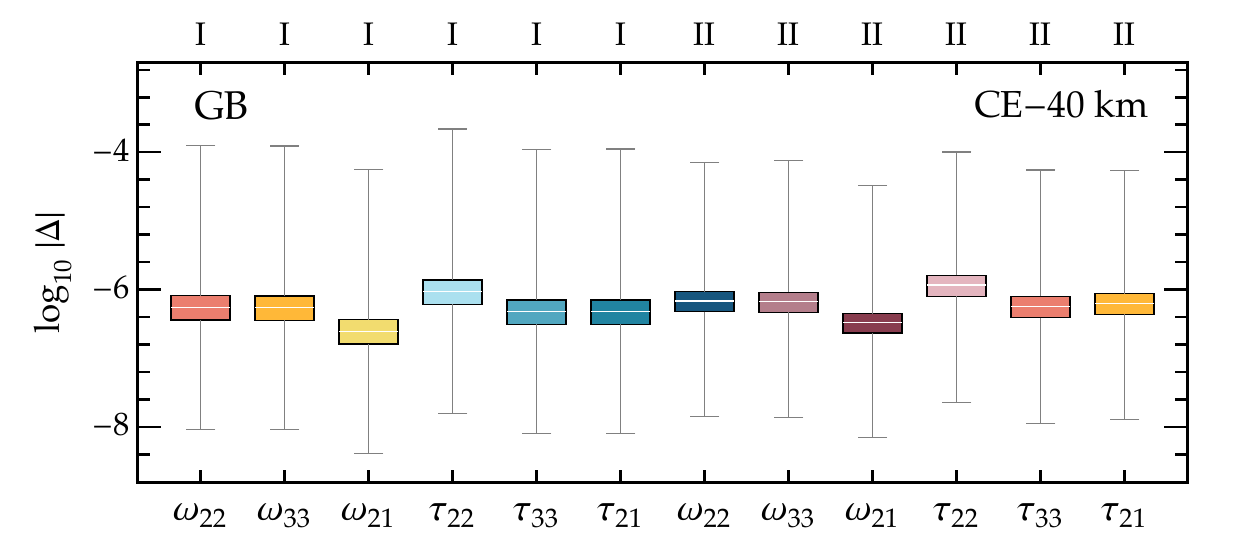}
    \caption{Distributions of the relative change \eqref{math:shiftGBome}-\eqref{math:shiftGBtau} in frequencies and damping times induced by EsGB gravity with respect to GR. White horizontal lines in each colored box mark the median over all events with ringdown observable by ET (top) and CE (bottom); the edges of the box identify the upper and lower quartiles, while the ends of the whiskers are the maximum and minimum value in the distribution. Labels in the top x-axis of the panels refer to the spin model used in the astrophysical population (\texttt{model I} or \texttt{model II}). The value of the EsGB coupling constant is 
    fixed to $\sqrt{\alpha_\tn{GB}}\simeq 2.5$~km~\cite{Lyu:2022gdr}.}
    \label{fig:app_EsGB}
\end{figure} 

In Fig.~\ref{fig:app_EsGB} we show the distributions of the relative change with respect to the slow-rotation expansion of the Kerr QNM frequencies and damping times $(\omega_{\ell m}^\tn{K,s},\tau_{\ell m}^\tn{K,s})$, given by Eqs.~\eqref{math:kerromega}-\eqref{math:kerrtau}, due to the EsGB corrections:
\begin{align}
\Delta\omega^\tn{GB}_{\ell m}=&\frac{\beta_\tn{GB}^2}{\omega_{\ell m}^\tn{K,s}}\left[\bar{\omega}^{(0)}_{\ell m} 
\delta \omega_{\ell m}^{(0)}+ 
\chi \bar{\omega}_{\ell m}^{(1)}\delta \omega_{\ell m}^{(1)} + \chi^2 \bar{\omega}_{\ell m}^{(2)} \delta \omega_{\ell m}^{(2)}\right]\ ,\label{math:shiftGBome}\\
\Delta\tau^\tn{GB}_{\ell m}=&\frac{\beta_\tn{GB}^2}{\tau_{\ell m}^\tn{K,s}}\left[\bar{\tau}^{(0)}_{\ell m} 
\delta \tau_{\ell m}^{(0)}+ 
\chi \bar{\tau}_{\ell m}^{(1)}\delta \tau_{\ell m}^{(1)} + \chi^2 \bar{\tau}_{\ell m}^{(2)} \delta \tau_{\ell m}^{(2)}\right]\ ,\label{math:shiftGBtau}
\end{align}
As discussed before we fix the EsGB coupling to 
$\sqrt{\alpha_\tn{GB}}\sim2.5$~km and we evaluate $\Delta\omega^\tn{GB}_{\ell m}$ and $\Delta\tau^\tn{GB}_{\ell m}$ for all BHs of the two populations (\texttt{model I} and \texttt{II}) that have a detectable ringdown in ET, i.e., those for which $(22)$ mode SNR $\rho\geq 12$. The fundamental $(22)$ mode leads to the largest deviations, but deviations in the $(33)$ and $(21)$ components are comparable in order of magnitude. Note however that the SNR, and therefore the uncertainties on frequency and damping times, depend on the energy released in each mode, $\epsilon_\tn{RD}$ (or, equivalently, on the mode amplitudes).  As discussed in Sec.~\ref{sec:results} (see in particular Fig.~\ref{fig:hist_amp}), the values of $\epsilon_\tn{RD}$ for the $(22)$ QNM are more than one order of magnitude larger than those of the $(33)$ and $(21)$ modes, and therefore the $(22)$ mode is the most important for Bayesian inference.

\subsection{Dynamical Chern-Simons gravity}
\label{app_sub:dCS}

For dCS gravity we use the analytical fits provided in Ref.~\cite{Wagle:2021tam}, which are valid up to first order in the spin and up to second order in the coupling constant $\alpha_\tn{CS}$\footnote{Ref.~\cite{Srivastava:2021imr} provides terms at leading order in the dCS coupling constant which were missing in Ref.~\cite{Wagle:2021tam}. However Ref.~\cite{Srivastava:2021imr} does not list values for $l=3$. Therefore, for consistency, we only use results from Ref.~\cite{Wagle:2021tam}.}.  For simplicity we focus on the axial sector, which is more strongly affected by the dCS modification with respect to the polar sector.

At this order, defining $\gamma=\alpha_\tn{CS}^2/M^4= \beta_\tn{CS}^2$, the \PS mapping becomes
\begin{align}
\omega_{\ell m} = & \frac{1}{M} \left[\bar{\omega}_{\ell m}^{(0)} \left(1+ \beta_\tn{CS}^2 \delta \omega_{\ell m}^{(0)}\right) \right. \notag\\
& \left. + \chi \bar{\omega}_{\ell m}^{(1)} \left(1+ \beta_\tn{CS}^2\delta \omega_{\ell m}^{(1)}\right) \right]+M\sum_{k_1=2}^{12}\bar{\tau}_{\ell m}^{(k_1)}\ ,\label{eq:parspec-dCS-expaA}\\
\tau_{\ell m} = & M \left[\bar{\tau}_{\ell m}^{(0)} \left(1+ \beta_\tn{CS}^2\delta \tau_{\ell m}^{(0)}\right) \right.\nonumber\\
& \left. + \chi \bar{\tau}_{\ell m}^{(1)} \left(1+ \beta_\tn{CS}^2\delta \tau_{\ell m}^{(1)}\right) \right]+M\sum_{k_1=2}^{12}\bar{\tau}_{\ell m}^{(k_1)}\ .
\label{eq:parspec-dCS-expaB}
\end{align}

We find the beyond-GR parameters with a procedure analogous to the EsGB case. We fit Eqs.~\eqref{eq:parspec-dCS-expaA} and \eqref{eq:parspec-dCS-expaB}, with the GR coefficients fixed to the values listed in Table~\ref{tab:Parspec-tab-GR}, to the expressions provided in Ref.~\cite{Wagle:2021tam} in the range where the authors find their expansion to be reliable, i.e., $\chi \in [0,\,0.04], \alpha_\tn{CS}/M^2 \in [0,\,0.1]$. The dCS axial coefficients of the \PS expansion determined in this way are listed (up to first order in the spin) in Table~\ref{tab:Parspec-tab-dCS-AX}.

\begin{figure}[t]
    \includegraphics[width=4.2cm]{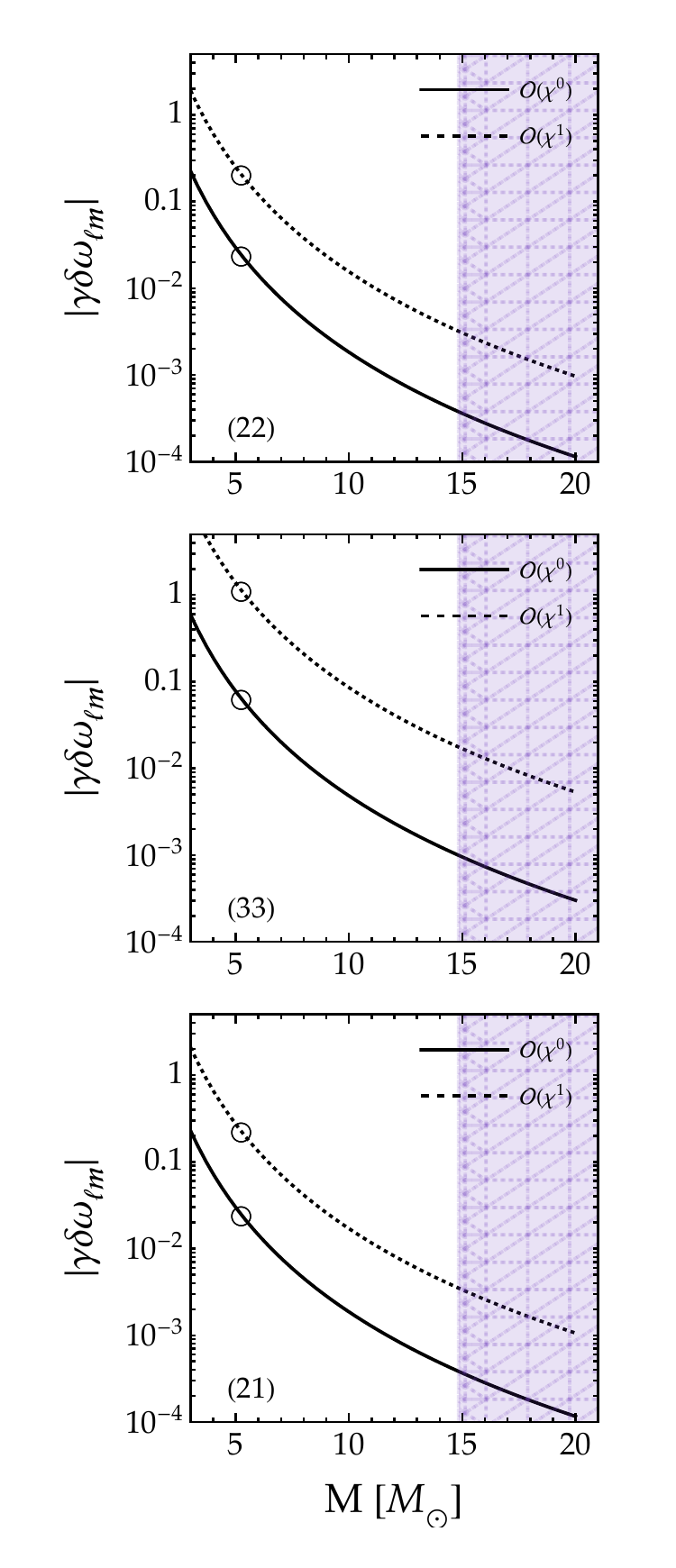}
    \includegraphics[width=4.2cm]{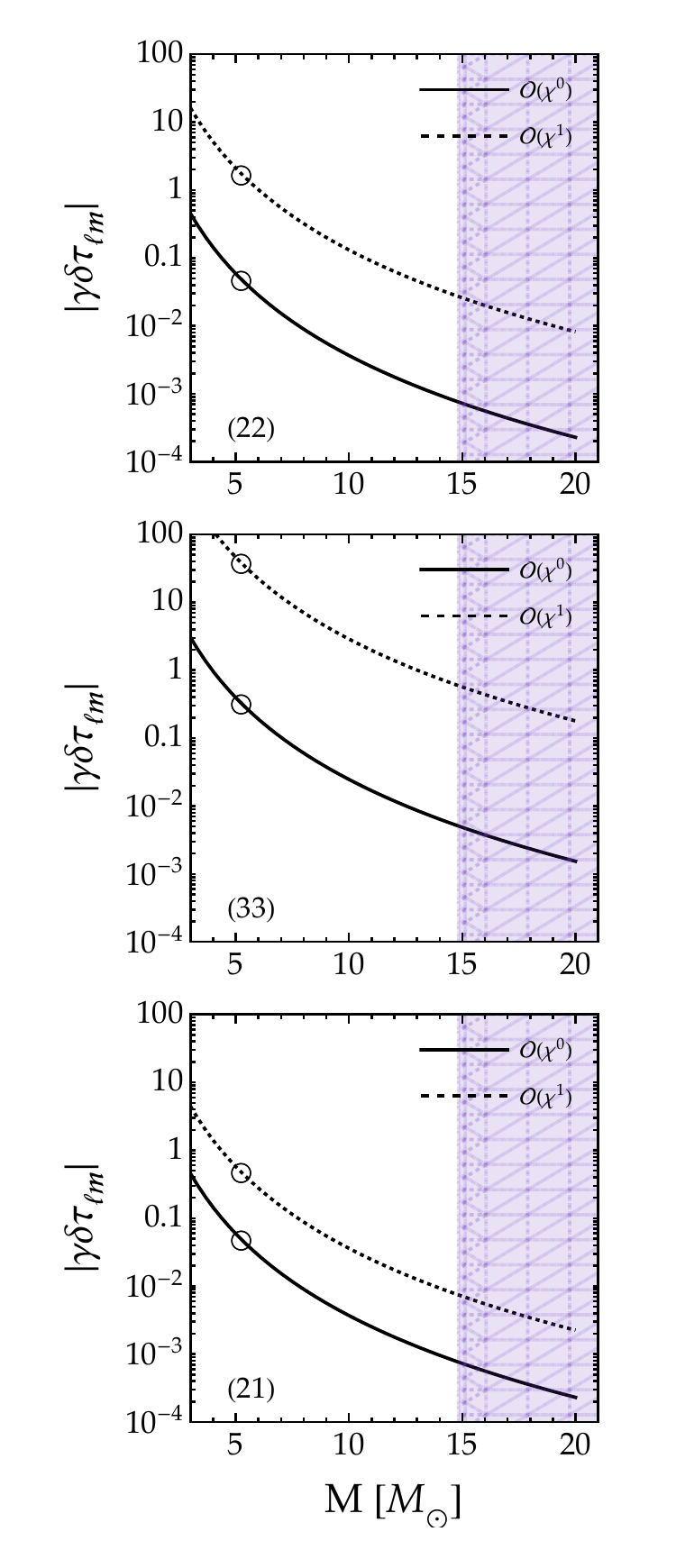}
    \caption{Same as Fig.~\ref{fig:esgbparameters}, but for dCS gravity.}
    \label{fig:dcsparameters}
\end{figure}

In Fig.~\ref{fig:dcsparameters} we show the behavior of $\gamma \delta\omega^{k_2}_{\ell m}$ and $\gamma \delta\tau^{k_2}_{\ell m}$ as a function of the BH mass, for $\sqrt{\alpha_\tn{CS}}\simeq2.5\,\tn{km}$.  Compared to EsGB gravity (Fig.~\ref{fig:esgbparameters}) the values of the dCS parameters are, in general, larger.

\begin{figure}[t]
    \includegraphics[width=8.5cm]{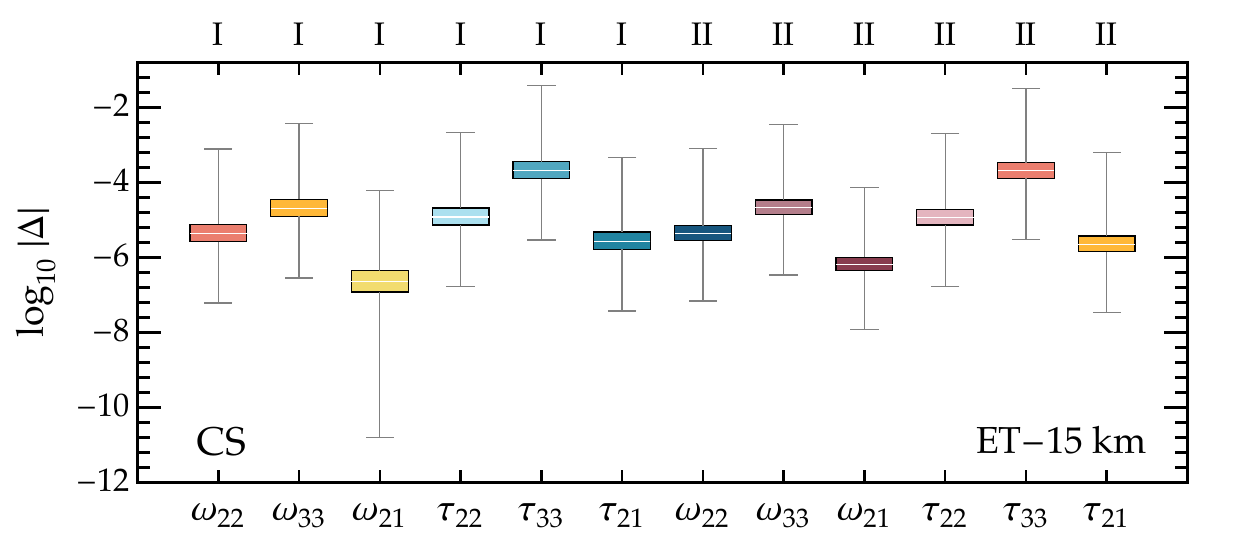}
    \includegraphics[width=8.5cm]{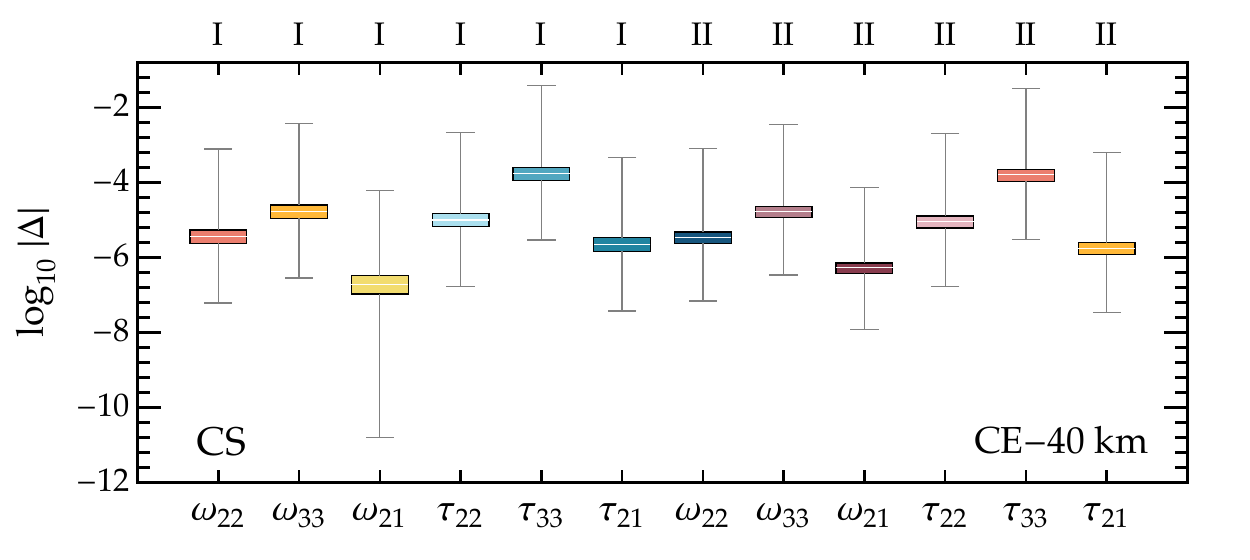}
    \caption{Same as Fig.~\ref{fig:app_EsGB}, but for 
    dCS gravity. We assume the maximum value of the dCS coupling constant allowed by GW observations, $\sqrt{\alpha_\tn{CS}}\simeq 2.5$~km~\cite{Lyu:2022gdr}.}
    \label{fig:app_dCS}
\end{figure}

Figure~\ref{fig:app_dCS} is the dCS equivalent of Fig.~\ref{fig:app_EsGB} for EsGB gravity. Most of the considerations we made for EsGB gravity apply also in this case. Note, however, that deviations in the $(33)$ component are now always larger than deviations in the $(22)$ and $(21)$ modes.

\begin{table}[t]
	\centering
	\begin{tabular}{ c | c c | c c | c c }
        \hline
		\hline
		 $k_2$ & $\delta \omega^{(k_2)}_{22}$ & $\delta \tau^{(k_2)}_{22}$ 
		& $\delta\omega^{(k_2)}_{33}$ & $\delta\tau^{(k_2)}_{33}$ & $\delta\omega^{(k_2)}_{21}$ & $\delta\tau^{(k_2)}_{21}$ \\
		\hline
		0 & $2.4182$ & $4.8493$ & $6.3807$ & $-32.535$ & $2.4569$ & $4.8768$ \\
		1 & $20.499$ & $173.56$ & $113.02$ & $-3792.9$ & $-22.421$ & $-47.561$ \\
		\hline
        \hline
	\end{tabular}
		\caption{Values of the beyond-GR coefficients for the \PS spin expansion for the axial modes with $(\ell m)=(22)$, $(33)$, $(21)$ in dCS gravity, as computed from the fits provided in Ref.~\cite{Wagle:2021tam}.}
	\label{tab:Parspec-tab-dCS-AX}
\end{table}

\subsection{Effective Field Theory}

For EFT gravity we use the numerical fits provided in Ref.~\cite{Cano:2023jbk}, where beyond-Kerr QNM corrections are written in the form:
\begin{equation}
M\omega=M\omega_{\ell m}^\tn{kerr}+\alpha_q
\delta\omega^\tn{fit}\ ,\label{math:canofit}
\end{equation}
where 
\begin{equation}
\delta\omega^\tn{fit}_{\ell m}=\sum_{n=0}^{12}c_n\chi^n\ .
\end{equation}
The numerical coefficients $c_n$ for the $(22)$ and $(33)$ QNMs -- but not for the $(21)$ QNM, which therefore is not included in this analysis -- are given in Tables~I and II of Ref.~\cite{Cano:2023jbk}.

\begin{figure*}[t]%
    \includegraphics[width=8cm]{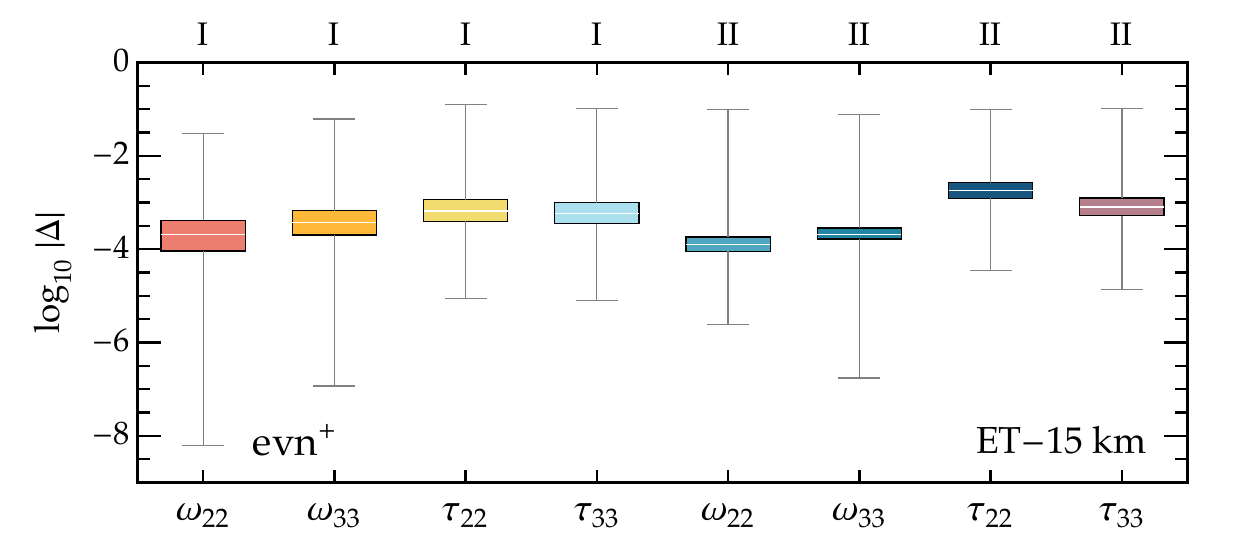}
    \includegraphics[width=8cm]{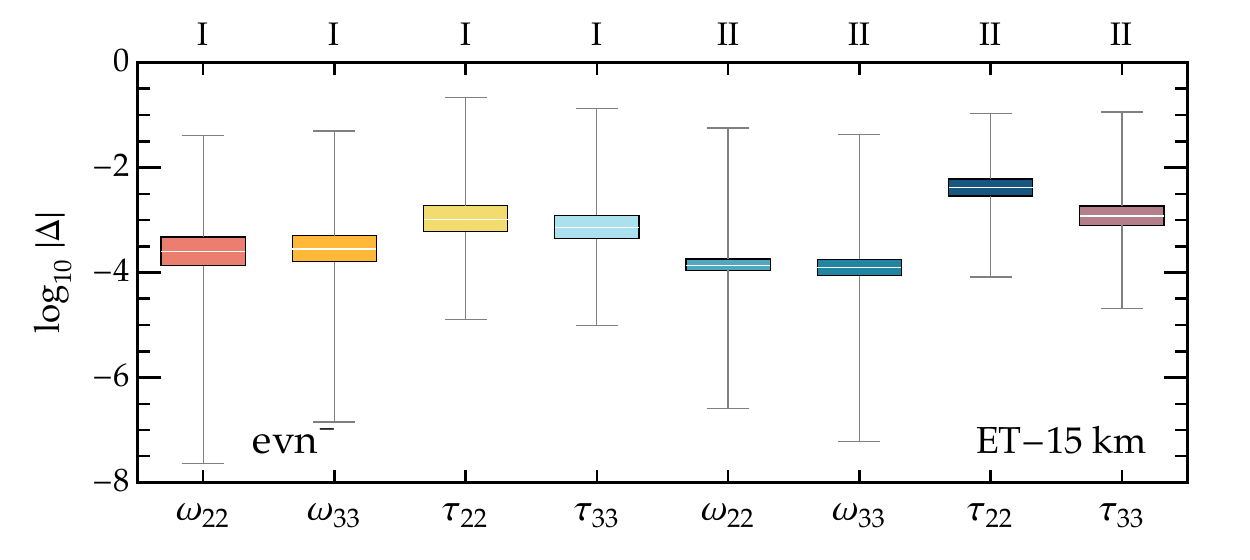}\\
    \includegraphics[width=8cm]{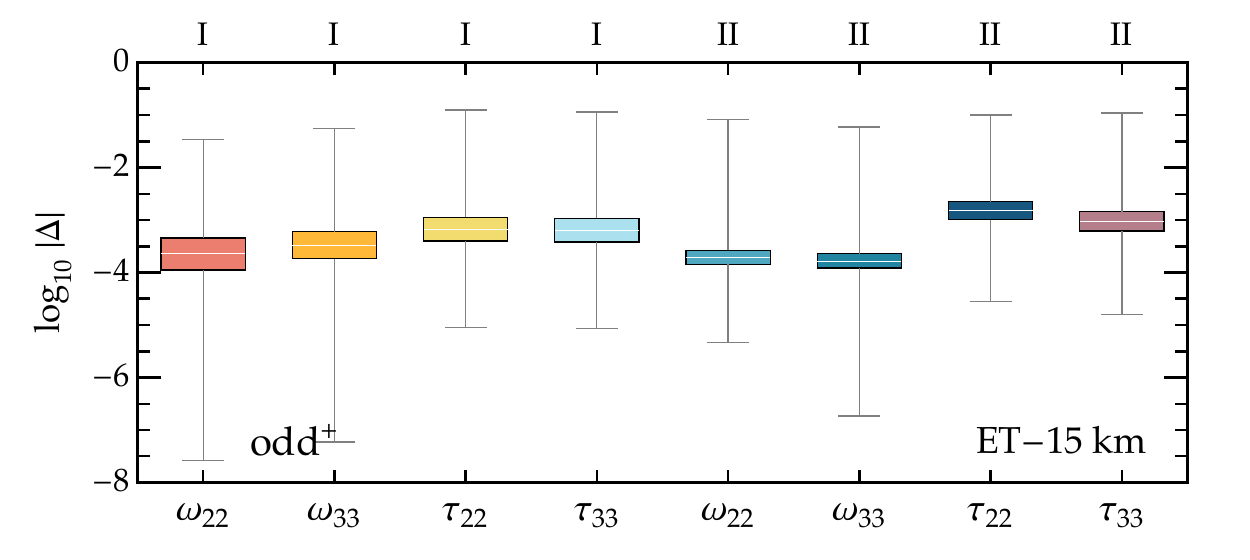}
    \includegraphics[width=8cm]{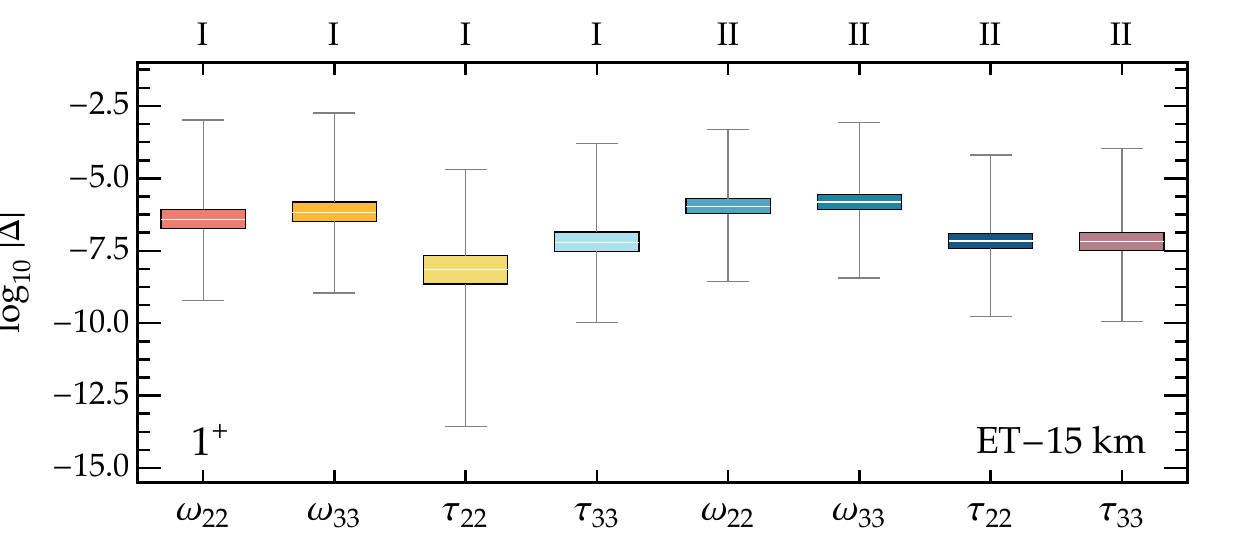}\\    
    \includegraphics[width=8cm]{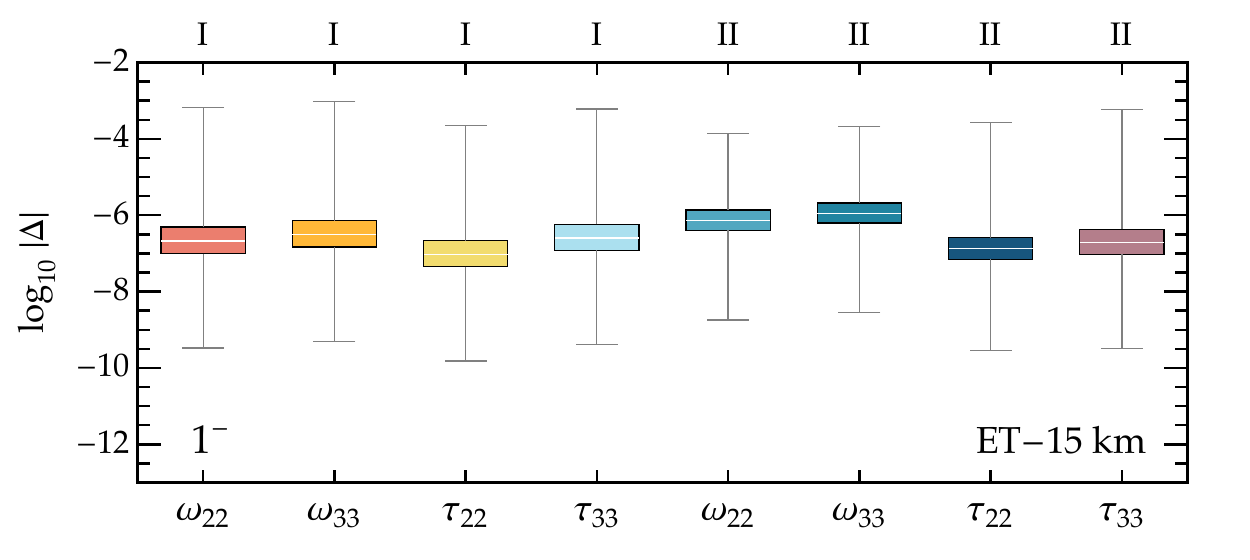}
    \includegraphics[width=8cm]{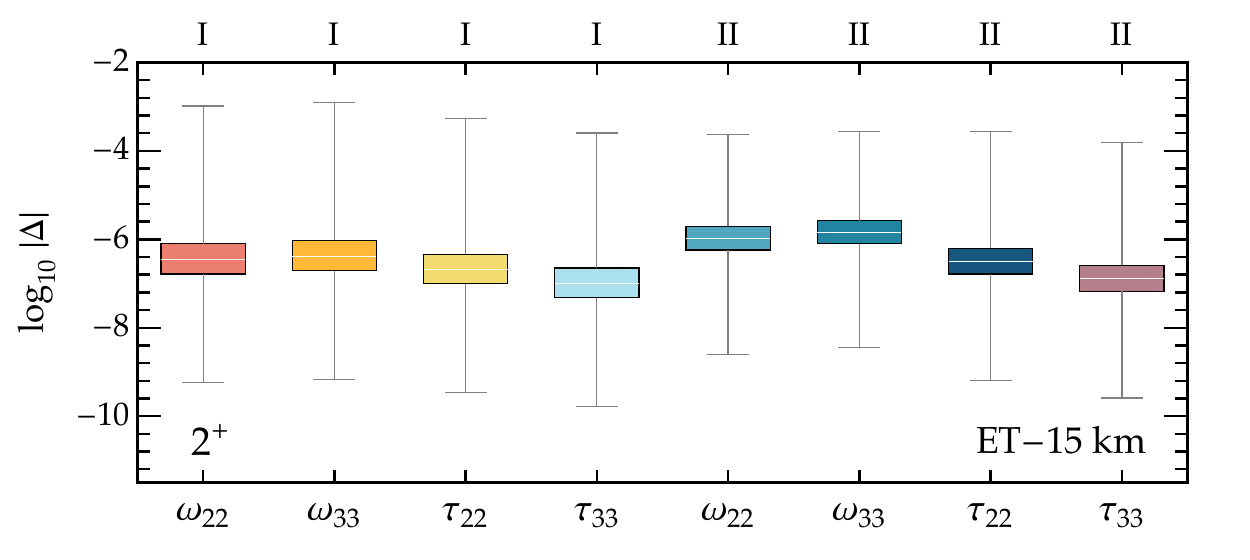}\\
    \includegraphics[width=8cm]{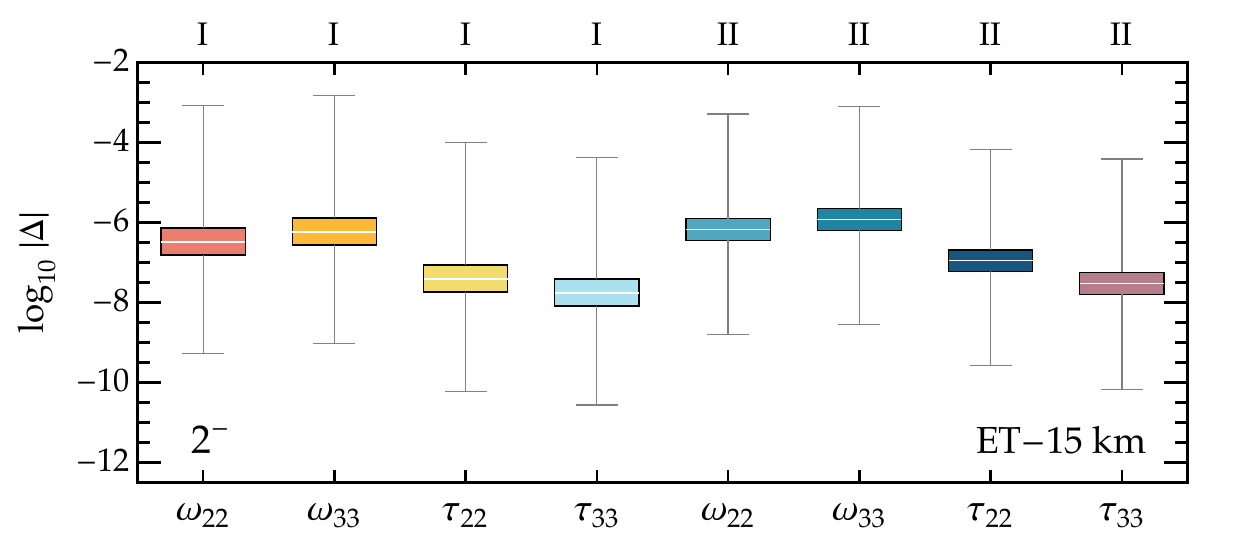}
    \includegraphics[width=8cm]{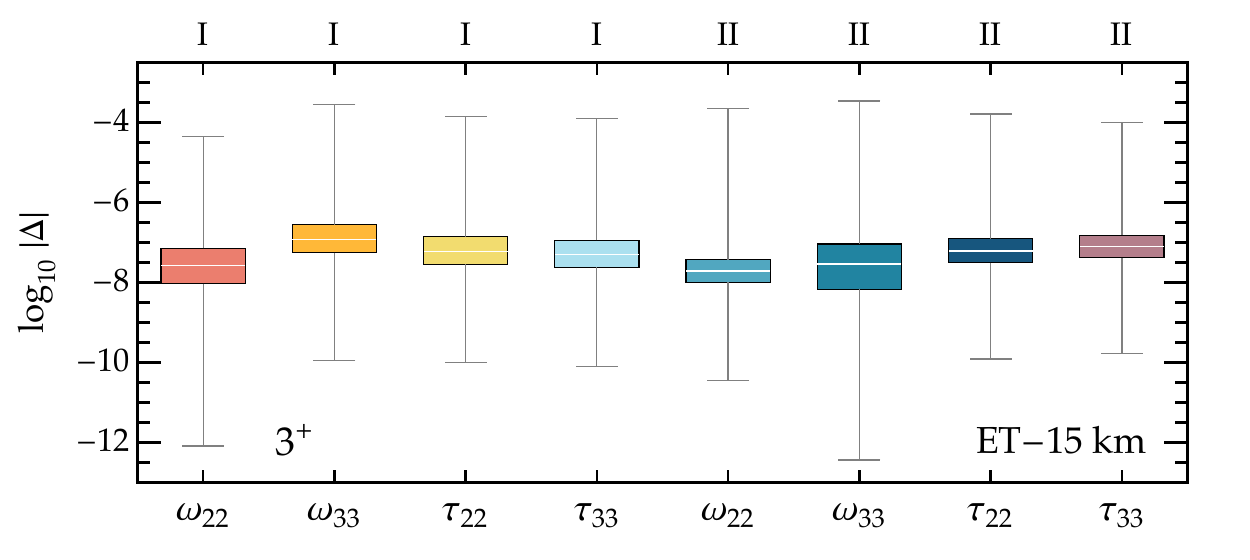}
    \caption{Same as Figs.~\ref{fig:app_EsGB} and \ref{fig:app_dCS}, but for the eight EFT gravity models of Ref.~\cite{Cano:2023jbk} considered in this work. We only show ringdown signals detectable by ET. The results for CE are qualitatively similar.}
    \label{fig:app_eft}
\end{figure*}

Equation~\eqref{math:canofit} can be straightforwardly mapped to the \PS coefficients in Eqs.~\eqref{eq:parspec-expa1}-\eqref{eq:parspec-expa2} once we are given the GR part of the expansion, Eqs.~\eqref{math:kerromega}-\eqref{math:kerrtau}.  Note that for EFT gravity the beyond-GR component of the spin expansion is known to much higher order ($n_1=n_2=12$) than for EsGB and dCS gravity. The distributions of deviations from the Kerr frequencies and damping times in the eight EFT models we consider are shown in Fig.~\ref{fig:app_eft}. In all panels we assume a value of the coupling such that $\gamma_q=0.1$, and (for brevity) we only consider ringdown observations with ET. 

\section{Black hole mass and spin measurements}\label{sec:appmasses}

\begin{figure*}[t]
    \includegraphics[width=15.5cm]{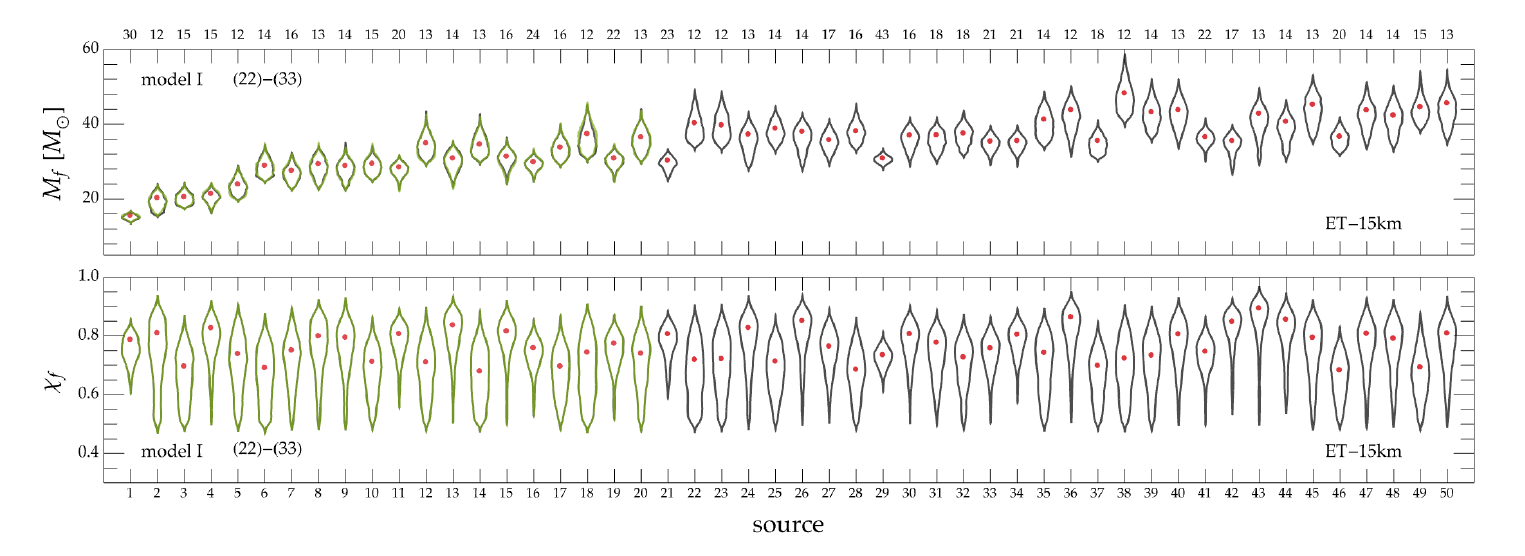}\\
    \includegraphics[width=15.5cm]{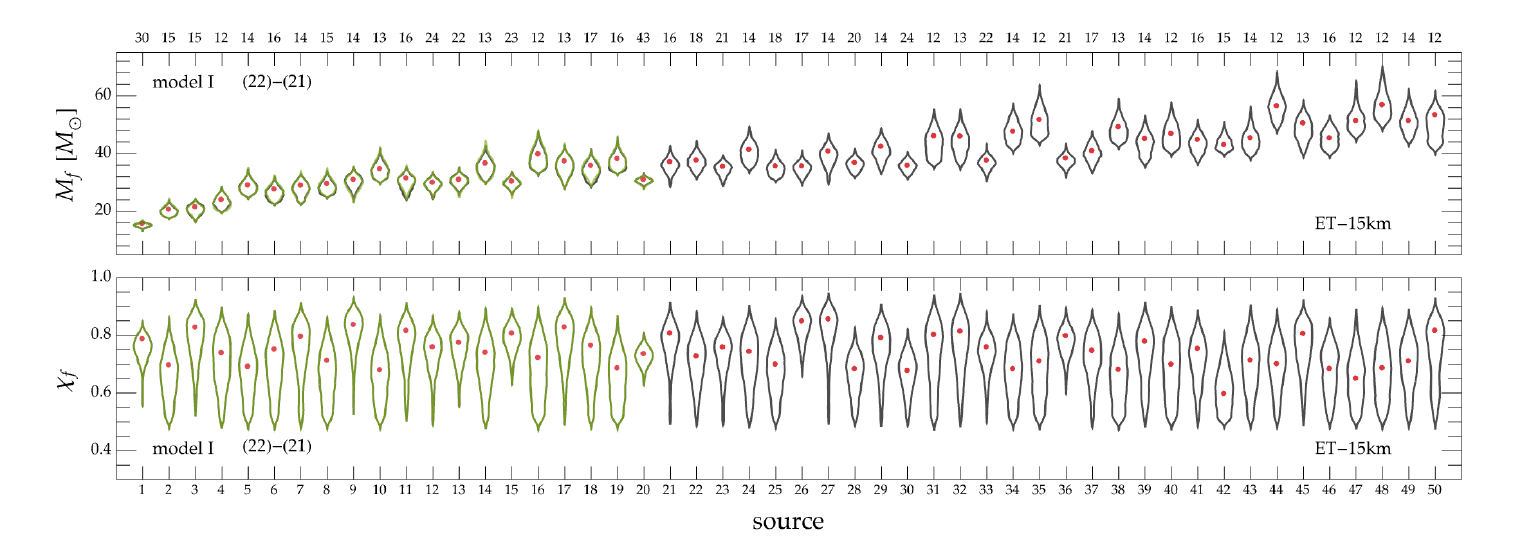}
    \includegraphics[width=15.5cm]{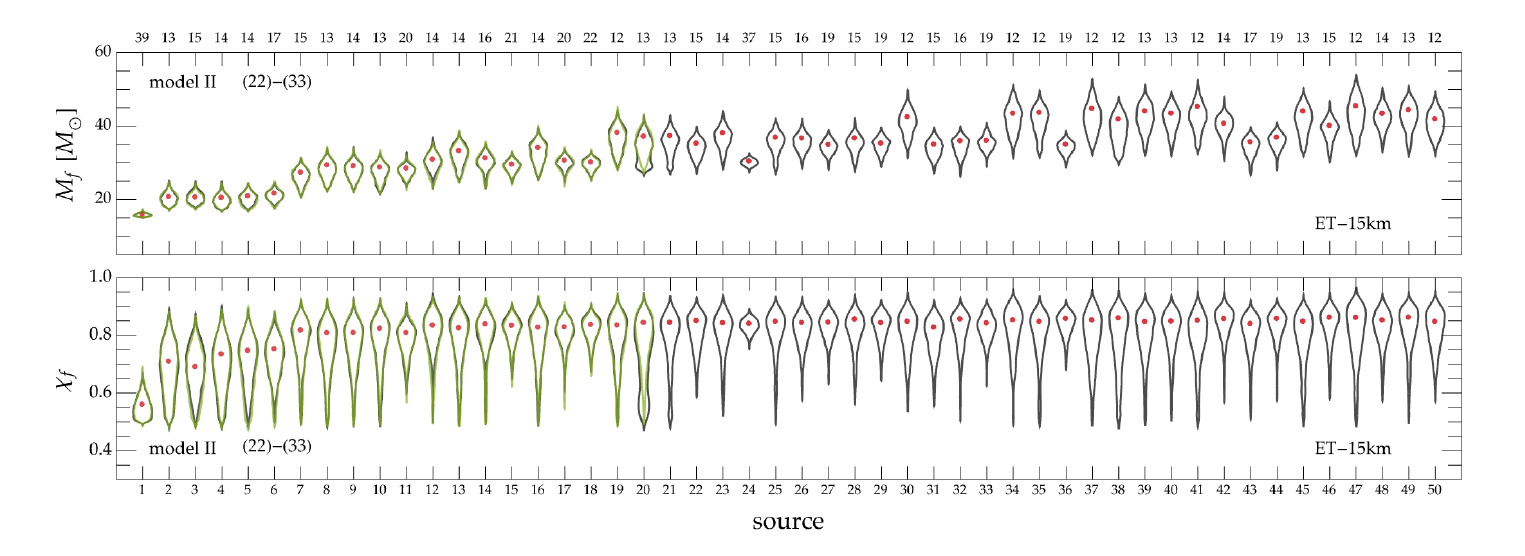}\\
    \includegraphics[width=15.5cm]{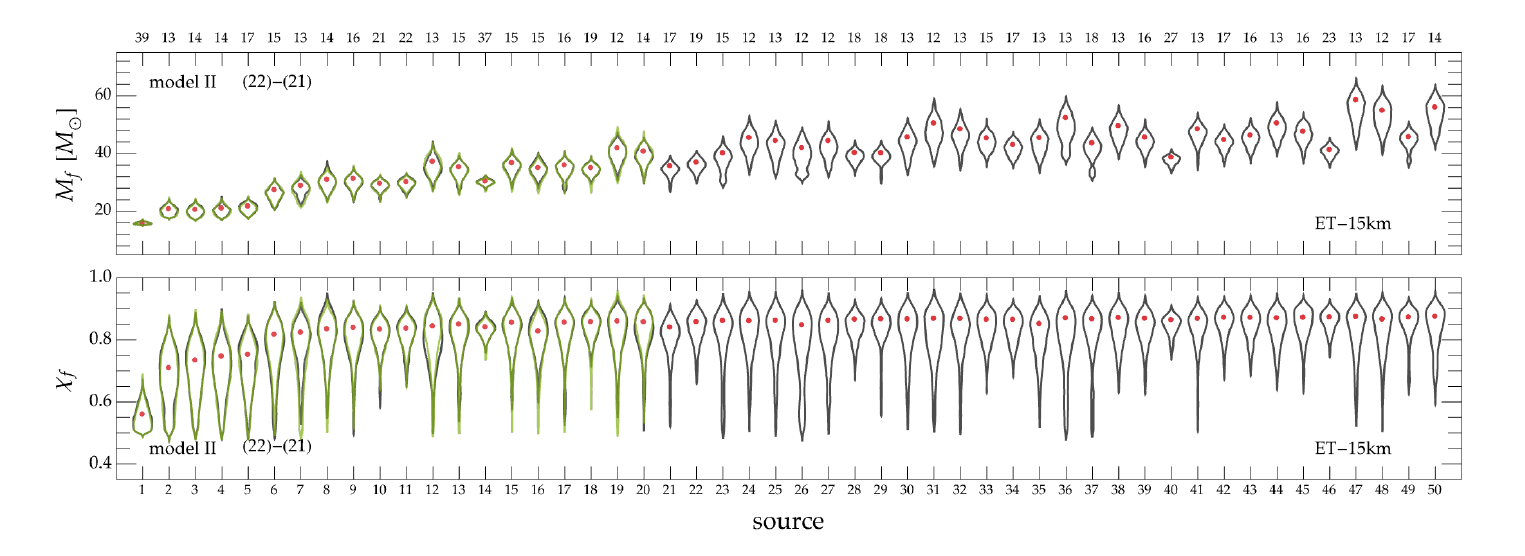}
    \caption{Posterior densities for the BH masses and spins inferred from a $(22)$-$(33)$ and $(22)$-$(21)$ analysis for dCS gravity.  All constraints are inferred from ringdown observations made by ET for BHs following either the \texttt{model I} or \texttt{model II} spin distributions.  Black (green) violins are obtained by stacking the largest (second largest) set of observations available in the population model (see also Fig.~\ref{fig:CSboundsMchi}). Red bullets are the injected values. Labels on the top horizontal axes identify the values of the SNR of the fundamental mode.}
    \label{fig:AppMchi}
\end{figure*}

The main purpose of this appendix is to show that the remnant BH's mass and spin are generally recovered with remarkable accuracy. For illustration, in Fig.~\ref{fig:AppMchi} we plot the probability distributions for the masses and spins inferred from a two-mode analysis for dCS gravity. Black and green contours correspond to constraints derived by stacking together two different sets of ringdown events (compare Fig.~\ref{fig:CSboundsMchi}), while red dots identify the injected values of $M_f$ and $\chi_f$.

The constraints are strongly dependent on the SNR of the fundamental mode, shown on the top horizontal axis of each panel.  The figure illustrates four key features common to the $(22)$-$(33)$ and the $(22)$-$(21)$ analysis: (i) the constraints on $M_f$ and $\chi_f$ are almost independent of the number of events that we stack; (ii) masses are always better measured than spins; (iii) measurement accuracies are slightly worse for \texttt{model II}, which exhibits long tails in the posterior distributions of the spins.

\section{SNR and detector configurations}\label{sec:snrconf}

In this appendix we analyze the dependence of the ringdown SNR on the remnant BH mass $M_f$ and on the progenitor binary mass ratio $q$.  We also study how the SNR changes for ET and CE detectors with different armlengths. Note that in this case we consider a \textit{single} ET, at variance with the network configuration of two L-shaped detectors used in the rest of the paper.  To compute the SNR we uniformly sample the primary mass in the range $m_1\in[10,\,100]M_\odot$ and the mass ratio in the range $q=m_1/m_2\in[1,\,3]$. For simplicity we assume the component spins to be equal and aligned, with magnitude $\chi_1=\chi_2=0.6$. This choice leads to BH remnants with spins in the range $\chi_f\sim[0.78,\,0.86]$, but we have checked that our conclusions are only mildly dependent on the magnitude of the progenitor BH spins. For concreteness, and without loss of generality, we assume all sources to be located at luminosity distance $d_\tn{L}=1$ Gpc.  We compute the amplitude of the $(22)$ mode and the resulting SNR as described in Sec.~\ref{sec:setup}.

\begin{figure}[t]
    \includegraphics[width=9cm]{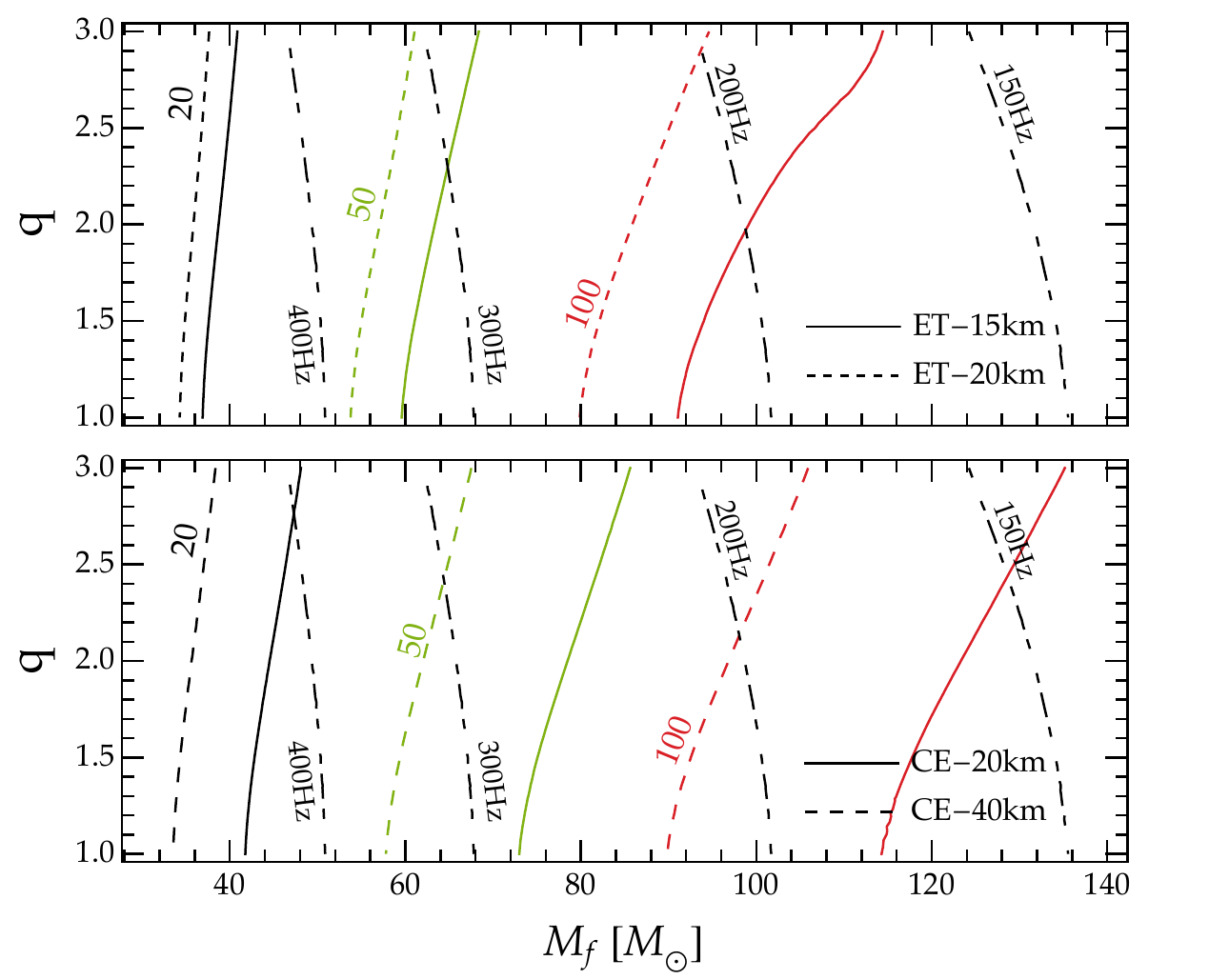}
    \caption{Curves of constant SNR for the $(22)$ mode as a function of the remnant BH mass $M_f$ and of the binary progenitor's mass ratio $q$ for ET (solid) and CE (dashed).  Dot-dashed curves identify configurations with a specific $(22)$ frequency.}
    \label{fig:AppSNR}
\end{figure} 

In Fig.~\ref{fig:AppSNR} we show constant-SNR contours for ET and CE computed in this way. The two configurations we considered in this paper, the 15 km ET and the 40 km CE, have similar performances, with the latter yielding slightly larger SNRs. A 20-km armlength CE would yield lower values of the ringdown SNR, with differences with respect to ET increasing as the remnant BH mass grows.

To further clarify what frequency range is most relevant for ringdown tests, we have computed the accumulated SNR for ET (single detector) and CE as a function of the upper integration limit $f_\tn{max}$ in Eq.~\eqref{math:snr3}. We focus on BH remnants from four selected binary systems with primary mass $m_1=(10,50,80,200)M_\odot$, mass ratio $q=2$, aligned spins $\chi_1=\chi_2=0.6$, and luminosity distance $d_\tn{L}=1$~Gpc.

\begin{figure}[t]
    \includegraphics[width=8.5cm]{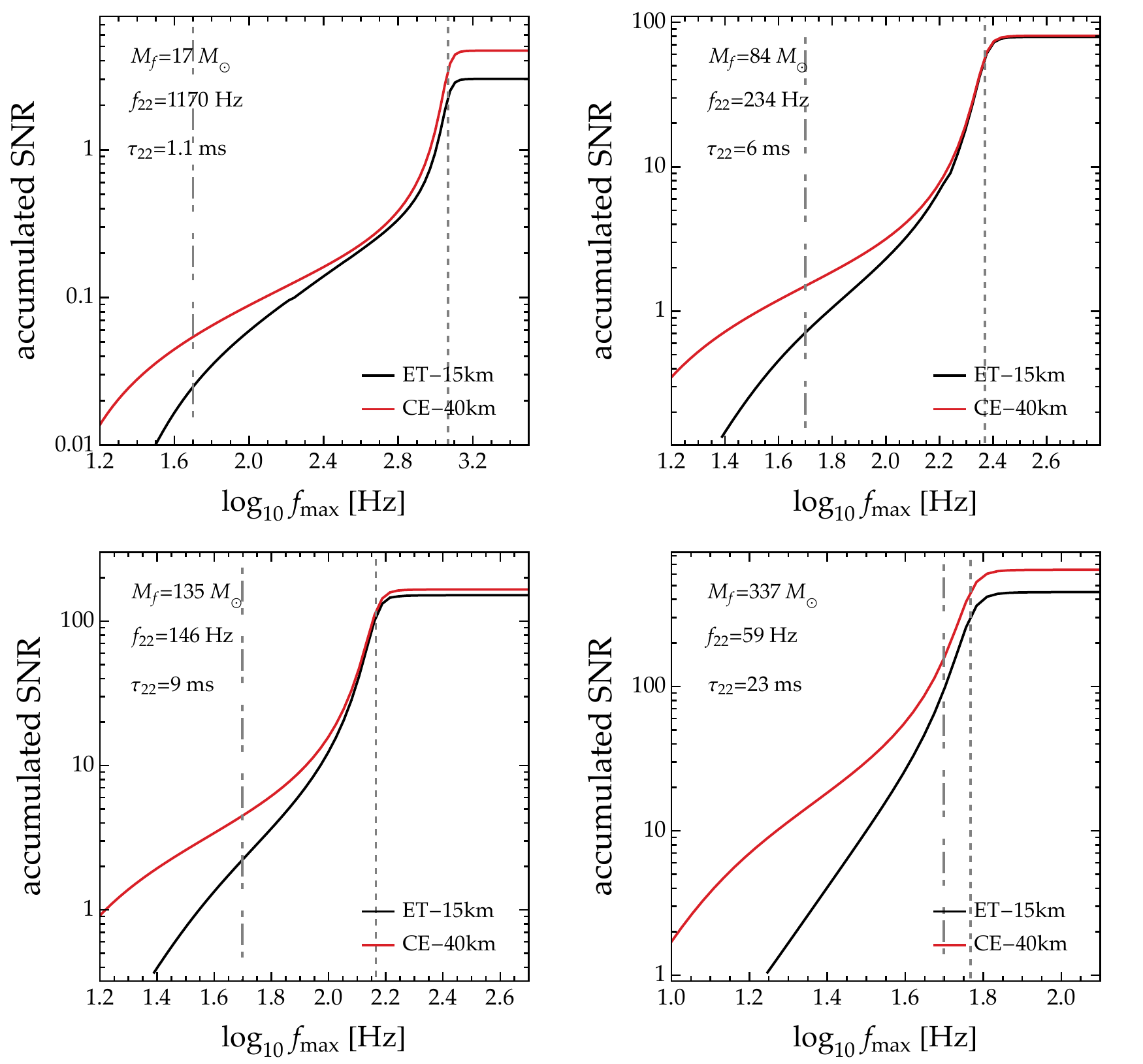}
    \caption{Accumulated SNR as a function of the maximum integration
    frequency $f_\tn{max}$ for four different BH binaries, all at 
    luminosity distance $d_\tn{L}=1$~Gpc. (see text). The vertical 
    dashed and dot-dashed lines in each panel mark the frequency 
    of the fundamental $(22)$ mode and a reference frequency of 
    50~Hz, respectively.}
    \label{fig:AppSNR2}
\end{figure}

In the four panels of Fig.~\ref{fig:AppSNR2} (one for each binary) we see some common trends: the SNR for CE is always larger than for ET when $f_\tn{fmax}\lesssim50$~Hz, since the PSD of CE is better between 3 and 50 Hz. As $f_\tn{max}$ increases, ET ``catches up'' and yields an overall similar value of the SNR. Indeed, for $f\gtrsim 50$Hz, the ratio of the two PSDs is similar, with CE being less than a factor two better than ET. Note also that for the mass range of interest, i.e., for BHs with $M_f\sim 100 M_\odot$, the two interferometers yields almost identical results.

\section{Fisher matrix components}\label{sec:fishercomp}

Here we list explicit expressions for  the components of the FIM~\eqref{math:FisherM}, averaged over the source orientation in the sky. The FIM is computed from the ringdown waveform model of Eqs.~\eqref{math:hcf}--\eqref{math:totalh}. For a given multipolar component $(\ell m)$ it depends on the parameters $\vec{\theta}=(\bar{{\cal A}}_{\ell m},\phi_{\ell m},\omega_{\ell m},\tau_{\ell m})$, where $\bar{{\cal A}}_{\ell m}=M{\cal A}_{\ell m}/r$ is the effective amplitude. Denoting partial derivatives by commas, the ten independent components of the FIM are given by:
\begin{widetext}
\begin{align}
\Gamma_{\bar{{\cal A}}_{\ell m}\bar{{\cal A}}_{\ell m}}=&\frac{b_+^2+b_-^2}{10\pi}\quad\ ,\quad\Gamma_{\phi_{\ell m}\phi_{\ell m} }=\frac{\bar{{\cal A}}^2_{\ell m}}{10\pi}\left[b_+^2+b_-^2\right]\quad\ ,\quad \Gamma_{\omega_{\ell m}\omega_{\ell m} }=\frac{\bar{{\cal A}}^2_{\ell m}}{10\pi}\left[b_{+,\omega_{\ell m}}^2+b_{-,\omega_{\ell m}}^2\right]\ ,\\
\Gamma_{\tau_{\ell m}\tau_{\ell m} }=&\frac{\bar{{\cal A}}^2_{\ell m}}{10\pi}\left[b_{+,\tau_{\ell m}}^2+b_{{-},\tau_{\ell m}}^2\right]\quad\ ,\quad
\Gamma_{\bar{{\cal A}}_{\ell m}\phi_{\ell m} }=\frac{\bar{{\cal A}}_{\ell m}}{10\pi}\left[-ib_+^2+ib_-^2\right]\ ,\\
\Gamma_{\bar{{\cal A}}_{\ell m}\omega_{\ell m} }=&\frac{\bar{{\cal A}}_{\ell m}}{10\pi}\left[b_+b_{+,\omega_{\ell m}}+b_-b_{-,\omega_{\ell m}}\right]\quad\ ,\quad
\Gamma_{\bar{{\cal A}}_{\ell m}\tau_{\ell m} }=\frac{\bar{{\cal A}}_{\ell m}}{10\pi}\left[b_+b_{+,\tau_{\ell m}}+b_-b_{-,\tau_{\ell m}}\right]\ ,\\
\Gamma_{\phi_{\ell m}\omega_{\ell m} }=&\frac{i\bar{{\cal A}}^2_{\ell m}}{10\pi}\left[b_+b_{+,\omega_{\ell m}}-b_-b_{-,\omega_{\ell m}}\right]\quad\ ,\quad
\Gamma_{\phi_{\ell m}\tau_{\ell m} }=\frac{i\bar{{\cal A}}^2_{\ell m}}{10\pi}\left[b_+b_{+,\tau_{\ell m}}-b_-b_{-,\tau_{\ell m}}\right]\ ,\\
\Gamma_{\omega_{\ell m}\tau_{\ell m} }=&\frac{\bar{{\cal A}}_{\ell m}^2}{10\pi}\left[
b_{+,\omega_{\ell m}}b_{+,\tau_{\ell m}}+b_{-,\omega_{\ell m}}b_{-,\tau_{\ell m}}
\right]\ .
\end{align}
\end{widetext}

\bibliography{bibliography}
\end{document}